\documentclass[9pt,journal,onecolumn]{IEEEtran}
\usepackage[table]{xcolor}
\usepackage [T1]{fontenc}
\usepackage[utf8]{inputenc}
\usepackage{mathtools,amssymb,stmaryrd,amsmath}
\usepackage{pdfpages}
\usepackage{bibentry}
\nobibliography*
\usepackage{booktabs}
\usepackage{graphicx}
\usepackage{wrapfig}
\usepackage{footnote}
\usepackage[graphicx]{realboxes}

 \usepackage[normalem]{ulem}
 \useunder{\uline}{\ul}{}

\ifCLASSOPTIONcompsoc \usepackage[caption=false,font=normalsize,labelfon
t=sf,textfont=sf]{subfig}
\else
\usepackage[caption=false,font=footnotesize]{subfig}
\fi
  \usepackage[margin=10pt,font=small,labelfont=bf]{caption}
\usepackage{macros}

\usepackage{url}
\usepackage[yyyymmdd,hhmmss]{datetime} % for running heads
\usepackage{xspace}

\AtBeginDocument{}

\title{Three Decades of Formal Methods in Business Process Compliance: A Systematic Literature Review}
\author{Hugo A. López\thanks{H. A. López is with the Technical University of Denmark. Email: hulo@dtu.dk} and Thomas T. Hildebrandt\thanks{T.T. Hildebrandt is with the University of Copenhagen. Email: hilde@di.ku.dk}
}

\begin{document}
% \linenumbers  % line numbers (for reviewing)
\maketitle

\newcommand{\todo}[1]{\upshape[\textbf{TODO:} #1]}

\begin{abstract}
Context: Digitalization endeavors face a dichotomy: while on the one hand business processes should be effective and efficient in the achievement of their goals, on the other hand, they also need to abide by the set of laws that rule the organization. The term business process compliance refers to the alignment of business processes and their regulations. The literature reports several frameworks to achieve process compliance, with the earliest citation in 1981. We focus on rigorous frameworks using formal methods to verify or ensure business process compliance.  A work that collects, abstracts, and synthesizes an in-depth summary of the current literature will help develop research in this field. Objective: We conducted a systematic literature review (SLR) focusing on process compliance frameworks based on formal models in the existing literature to understand better the state of the research on process model compliance, and identify the gaps and opportunities for future research. Method: We used automated search engines to collect studies between 1981 and the establishment of GDPR using standardized electronic libraries. Out of 5018 publications retrieved initially, we selected 46 primary studies for thorough analysis. We identified, analyzed, and categorized compliance frameworks based on their phases, process and compliance languages utilized, and reasoning techniques employed. We also analyzed evidence provided by the applicability of each framework: what case studies compliance frameworks have been tested on, what kind of users have utilized each framework, and what competencies are required to achieve compliance. We assessed the maturity level for each framework based on their phases, and users considered.   Results: Our analysis identifies a major agreement regarding verification techniques at the core of process compliance, but less so in phases before or after this. Concerning the techniques used, model checking is the de facto technique, but the nature of compliance and process languages utilized have changed. Finally, most works presented only conceptual results with prototypical implementations, do not consider actual users coming from compliance departments (e.g.: legal specialists), and most of the works do not consider changes in laws.  Conclusions: The existing research calls for comprehensive empirical studies to contribute to a better understanding of the anatomy and maturity of regulatory compliance frameworks and for extensive evaluation methods where compliance frameworks could be benchmarked. Our study is a comprehensive source for researchers working on the compliance of process models against laws and related fields and a useful guide for practitioners aiming to generate compliant process models.

\end{abstract}

%%% Local Variables:
%%% mode: latex
%%% TeX-master: "main"
%%% End:

% \input{changelog}
\section{Introduction}

Business Processes are considered the heart of organizations. Companies achieve business objectives through processes, coordinate and optimize resources, and comply with regulatory requirements. Legal compliance has become a substantial task for all organizations, independent of whether they are in the public sector or operate privately. Compliance affects all stages in the business process lifecycle: it poses requirements on the activities that must be included, it affects the times for execution of a trace and it brings queries that need to be analyzed in auditing phases.  Moreover, it is a continuous effort: both business processes and laws change continuously, and a change in legislation potentially impacts vertically the execution of business processes across the organization. 

The dream of \textit{compliance by design} is not new, and two dates are critical. The first was in 1986 when Sergot et al. published an encoding of parts of the British nationality act as Prolog rules~\cite{sergot1986british}. This work sets a baseline for formal methods techniques in reasoning about laws, and the computational support for legal decision-making. The second date is (25th of May) 2018 when the European Parliament and Council of the European Union declare the General Data Protection Regulation (GDPR) effective date. While previous efforts such as the EU directive on the protection of data existed, the impact of GDPR on non-compliant behavior incurred heavier penalties, the clear definition of necessary IT processes, and definitions of responsible roles for the execution (and monitoring) of business processes following privacy and data security regulations. It is natural to question, given 32 years of research in rigorous approaches for compliance, \textit{To which extent can formal methods support the compliance of business processes against laws?}

Answering this question is key for multiple reasons: first, compared to other approaches in compliance using data science or machine learning, the advantage of formal methods is the reliability of their answers. Equipped with formal semantics, queries to a model evidence mathematical properties that a process exhibits. Applying formal methods in this way requires laws to be formalized as mathematical objects, processes to be specified, and algorithmic techniques that align these two artifacts. Second, techniques based on formal methods look at all possible variations and not only those with statistical significance. This will be important in the case of rare violations with high impact in an organization, for instance, financial fraud. Third, formal techniques are not prone to the hallucination typical of machine learning models, so compliance or violations do not depend on the probabilities in the underlying model used as a reference. Finally, it equips organizations with explainable techniques, a key requirement when automated decision-making grants or denies rights to citizens, and is contemplated in Recital 71 of the GDPR.

% \subsubsection{Question focus}

% In this article, we aim to identify frameworks, case studies, and experience reports in software engineering, business process management, and formal methods where formal verification techniques are applied to achieve regulatory
% compliance of process-oriented technologies.

This paper examines the state of the art in business process compliance right on the verge of the adoption of GDPR as a European legislation. Our focus is to identify frameworks, case studies, and experience reports in software engineering, business process management, and formal methods where formal verification techniques are applied to achieve regulatory
compliance in business process technologies. The main objective of the work is to build a body of knowledge of the techniques we consider part of business process compliance, from early extraction of requirements to implementation and post-mortem techniques. In addition, we would like to contribute to the state of the art by identifying use cases (in terms of laws and processes) that can help define different compliance frameworks. Finally, we would like to assess the maturity of existing frameworks, identifying possible areas of improvement for next-generation compliance techniques. 

% \subsection{Question quality and amplitude}
% \subsubsection{Research objective}

The main objective is to characterize the initiatives in formal verification applied to compliance with business processes, covering the entire spectrum of uses from the design phase
of systems and processes, the process execution phase, and the
post-execution phase of processes. Moreover, we would like to assess the maturity of research initiatives for a possible industrial application and identify the research gaps. In particular, this SLR has the
following specific objectives:

% \paragraph{Specific objectives}
\begin{enumerate}
\item To identify what formal technologies aim at achieving regulatory compliance of
  process-oriented technologies exist, what are their uses, and what
  is the maturity level of the research solutions presented.
  \label{obj2-1}
\item To identify what gaps are limiting the adoption of formal verification technologies concerning the requirements in regulatory frameworks.\label{obj3-1}
\item To propose a research roadmap for bridging the
gaps
identified in objective \ref{obj3-1}, raising the maturity level of research in formal methods and technologies for regulatory compliance.
%\item To propose a research %roadmap that could integrate %previous
%  attempts identified in Objective \ref{obj2-1} and the products
%  resulting from solutions to the gaps identified in objective \ref{obj3-1}.
\end{enumerate}

To accomplish this goal, we performed a Systematic Literature Review (SLR) following methodologies from Kitchenham \cite{kitchenham_systematic_2013}, which is described in detail in Section \ref{sec:methodology}. We collected 5.018 works from some of the biggest scholar databases, reducing the set of primary studies to 46 primary studies after a double round of screenings.

Compared to previous surveys~\cite{sackmann_classification_2008,pourmirza_systematic_2017,shamsaei_systematic_2011,ghanavati_systematic_2011,hashmi_are_2018,ly_compliance_2015,becker_generalizability_2012,casanovas_legal_2017,hashmi_norms_2017,fellmann_state---art_2014}, the scope aims at helping adopters of compliance technologies with a mapping of the available techniques, their maturity and their possible pitfalls. On the other hand, it updates previous results with an updated view of compliance frameworks, as well as providing a critical discussion on their affordances and possibilities for extensions based on novel technologies such as LLMs.

We used automated search engines to collect studies between 1981 and the establishment of GDPR using standardized electronic libraries. Out of 5018 publications retrieved initially, we selected 46 primary studies for thorough analysis. We identified, analyzed, and categorized compliance frameworks based on their phases, process and compliance languages utilized, and reasoning techniques employed. We also analyzed evidence provided by the applicability of each framework: what case studies compliance frameworks have been tested on, what kind of users have utilized each framework, and what competencies are required to achieve compliance. We assessed the maturity level of each framework based on their phases, evidence on the technologies and use cases generated, deployment in
operational settings, and users considered.

\textbf{Contributions} Our analysis identifies a major agreement regarding verification techniques at the core of process compliance, where most works situate their concerns at a process design level, and a small representation of works considering runtime and post-mortem phases. Concerning techniques, model checking, and compliance-by-design are the most common techniques, with the emergence of conformance-checking techniques in recent years. Our observations show no agreement about a common specification language for laws or processes.
%, although standards such as %BPMN provide a good baseline. 
The use cases identified are within areas where compliance is particularly relevant, including finance, commerce, healthcare, insurance, and public sectors. Our observations regarding the laws
utilized by our primary studies are not conclusive regarding the depth and breadth of the formalization and showcases the difficulties in comparing different compliance frameworks. For competencies, we identified that most compliance frameworks rely on manual labor for specification extraction, including document selection, filtering, ambiguity resolution, and refinement as important challenges that need to be resolved for compliance frameworks to scale. Among the gaps not covered, most frameworks
document empirical validation, data compliance, and extensions for runtime monitoring as their most pressing needs. At the same time, it also uncovered uncommon requirements such as novel
languages capable of expressing specific legal terms (e.g.: mandates, power structures), missing tasks in compliance frameworks (e.g. extraction of specifications from natural language), and the introduction of compliance drifts. Finally, our assessment of technology readiness levels situates most works in lower levels, with few reaching operational levels tested in relevant environments over multiple cases.

\textbf{Impact}
Our mapping formed a baseline for the development of a second-generation set of tools for compliance considering some of the gaps identified in the mapping. %The existing research calls for comprehensive 
The project involved researchers spanning from formal methods to
empirical studies and ethnographic studies of case work in practice, and of technology providers and local government agencies. A concrete outcome of the project is the DCR technology and tool suite for offered by the spin-off company DCR Solutions (DCRSolutions.net). The technology is now widely used in Danish governmental organizations and its use is starting to spread outside Denmark. 
We believe that much more of such interdisciplinary research and development is needed to contribute to a better understanding of the anatomy and maturity of regulatory compliance frameworks. There is also a need for evaluation methods where compliance frameworks can be benchmarked. We hope our survey can serve as a
comprehensive source for researchers working on the compliance of process models against laws and related fields and a useful guide for practitioners aiming to generate compliant process models.

\textbf{Document Structure}
Section \ref{sec:methodology} presents the research methodology to construct the SLR. The data collection phase is documented in Section~\ref{sec:dataExtraction}. Section~\ref{sec:results} presents the summary of the results. Our interpretations of the state of the art and a comparison to current trends are given in Section ~\ref{sec:interpretation}. Section~\ref{sec:threatsValidity} documents our threats to validity. Related work is documented in~\ref{sec:relatedWork}. We conclude in Section  ~\ref{sec:conclusions}.

\section{Research Methodology} \label{sec:methodology}

To retrieve, select, and catalog primary studies for this
review, we have conducted a Systematic Literature Review following the
guidelines from Kitchenham~\cite{kit_cha_2007}. The definition of our research questions follows the following PICOC
criteria~\cite{keele2007guidelines}:

\begin{itemize}
\item \textbf{Population:} This experiment will include research in formal verification, business process management, and digitization of law.
\item \textbf{Intervention:} We have selected requirements specification, formal
verification, and compliance assistance as the main technologies. 
\item\textbf{Comparison:} No comparison criteria were defined a priori in this experiment. 
\item \textbf{Outcome:} The experiment aims at answering what are the current initiatives of regulatory
compliance assistance of business processes using formal methods. The outcome of this study will be an indicator of
the maturity level of current solutions. 
\item \textbf{Context:} Both academic and industrial contexts have been included in this experiment.
\end{itemize}

\subsubsection{Research questions}

The main aim is to assess the maturity of research initiatives in formal verification
technology to achieve regulatory compliance of
process-oriented technologies, covering the entire spectrum of uses from the design phase
of systems and processes, the process execution phase, and the
post-execution phase of processes. In particular, this SLR presents
the following questions:

\begin{description}
\item[RQ1] What are the common elements that conform a \textbf{formal
  regulatory compliance framework} for \textbf{process-oriented technologies}?
\item[RQ2] How do \textbf{current technologies} \textbf{formalise real regulatory documents}?
  % in a way that is \textbf{understandable} and \textbf{satisfactory} for \textbf{policy-makers}?
%   \todo{revise policy-makers} \\
%   \todo{Policy makers are those writing the regulations after the
%     laws, or proposing changes in the legislation.\\
%   Policy is what will enlighten the adoption of a law or its
%   modification.\\
%   but the law is adopted by the parliament and the MPs (Members of the
% Parliament).}
  \begin{description}
    \item[Q2.1] What \textbf{formal languages} are used to \textbf{extract,
      represent and organize} requirements from regulations?
  \item[Q2.2] What \textbf{case studies} of \textbf{formal regulatory compliance frameworks} exists?
  \item[Q2.3] What \textbf{process models} have been used for achieving 
    \textbf{conformance} against \textbf{law}? 
  \item[Q2.4] What are the \textbf{competences} required to \textbf{translate} a \textbf{law text} into
    a \textbf{formal model}? 
  \item[Q2.5] What are the \textbf{gaps} not covered in current formal regulatory compliance frameworks?
  \end{description}
  \item[RQ3] What \textbf{compliance techniques} using formal methods exists? 
\item[RQ4] How possible is to accommodate \textbf{changes} in \textbf{laws} in current
  \textbf{regulatory compliance frameworks}?
  \item[RQ5] How \textbf{mature} is the method/tooling available, from the
    evidence present in research literature?
\end{description}

The answers to these questions can be derived from the output of one
single general question, that can be automated in this SLR:

\begin{center}
\textbf{RQ} What are the current technologies in the \textbf{formal
  development} of \textbf{regulatory compliance frameworks} for \textbf{process models}?
\end{center}

%%% Local Variables:
%%% mode: latex
%%% TeX-master: "main"
%%% End:

\subsection{Search strategy}
\label{sec:search-strategy}
\begin{figure}[t]\centering
  \includegraphics[width=0.6\textwidth]{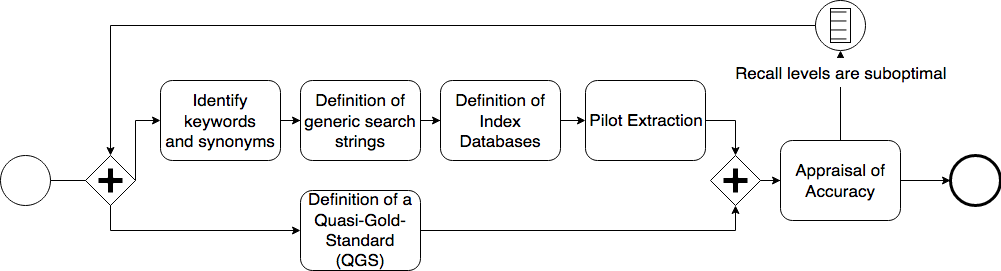}
  \caption{The search strategy protocol}
  \label{searchProcedure}
\end{figure}

Before starting our SLR, we produced a review protocol \cite{SRLProtocol-TSEN} which is summarised in this section. Figure \ref{searchProcedure} details the hybrid search strategy followed in this study. We combined both an automated search over digital libraries as well as the construction of a manual Quasi-Gold-Standard (QGS) Dataset. An automatic search procedure retrieves a large
dataset of candidate studies for review. This search is validated against a QGS including representative studies that were selected following a systematic procedure. Results after the appraisal can either show that the search strategy was too narrow to identify a good
percentage of the primary studies in the QGS, or that the
search is too wide to make the manual selection feasible in a timespan. 

  The first
step consists of identifying the keywords and generic search strings.

\subsubsection{Keywords and synonyms} \label{keywords}

The following is the list of main terms composing our research
question, as well as their synonyms. These terms have been used during
the review execution phase. We will use operators  \{ \}
to denote character grouping, commas to denote atomic search units, as well as the use of the operator ? to
describe the optional matching of a group in the search query.

\begin{description}
  \item[KW1] = \{formal method, formal model, formalism, formal language\}.
\item[KW2]= \{Compliance, conformance, accordance, coherence,
  enforcement\}.
\item[KW3] =\{Framework, method\{ology\}?\}.
\item[KW4]= \{Business Process\{es\}?, workflow\{s\}?, case management\}.
\item[KW5]= \{Regulatory text\{s\}?, regulation\{s\}?, law\{s\}?, legislation\{s\}?,
  legal document\{s\}?, polic\{y,ies\}, legal\}.
\end{description}

\subsubsection{Generic search string} \label{searchString}

We have used a single search string to perform a query that
captures the requirements of the research questions
in the previous section.  The terms starting with KW refer to the
keywords identified in section \ref{keywords}, and it has been
instantiated using OR connectives (e.g.: KW3 stands for ``Framework OR
method OR methodology'' ). The generic search string has been then: 

\begin{center}
  \textbf{Q1}:   KW1 \textbf{AND}  KW2 \textbf{AND} KW3 \textbf{AND} KW4 \textbf{AND} KW5
\end{center}

Given that questions \textbf{RQ1--RQ5} considered a subset of the
primary studies considered in \textbf{RQ}, we have opted for having
one single automated query, producing the remaining answers as a
a by-product of the data analysis of the sample retrieved with \textbf{RQ}.

% \paragraph{Sources}

We have limited our review to primary studies available online,
published in peer-reviewed conferences and journals,  and that appear indexed in reputable
citation databases (c.f.: section \ref{sources}). We only considered
primary studies that allowed full
access to the contents published. Failure to provide full access was a
sufficient condition to consider a source discarded. 

Given the multidisciplinary nature of this systematic review, we have
decided to combine subject-specific databases in computer
science, as well as meta-indexes covering computer science, business
and management, and law.  Table \ref{table:selectedDatabases}
summarises the databases considered.  We have restricted our source selection to publishing houses that
provide advanced keyword search through their web interfaces.
Google Scholar provides an exceptional case. This search
database indexes most academic content on the web; however, its
indexing criteria also includes articles that have not been
peer-reviewed. We have refrained from the use of Google Scholar for this reason. Similarly, we have initially considered additional search
engines such as DBLP\footnote{\url{http://www.dblp.org/search/index.php}}, FDBLP\footnote{\url{http://dblp.l3s.de}} and CiteseerX\footnote{\url{http://citeseerx.ist.psu.edu/index}}, but they have been further
discarded since their respective query language do not  
allowed us to
construct complex queries using AND or OR operators at the time of experimenting.

\begin{table}[t] \footnotesize \centering
\begin{tabular}{ m{3cm} m{0.5cm} m{5.4cm} m{6.4cm}  }
 \hline
\rowcolor[rgb]{ .357,  .608,  .835} \textcolor[rgb]{ 1,  1,  1}{ \textbf{Name}} & \textcolor[rgb]{ 1,  1,  1}{ \textbf{Code}} &\textcolor[rgb]{ 1,  1,  1}{ \textbf{Database Type}} & \textcolor[rgb]{ 1,  1,  1}{ \textbf{Source}}\\
 \hline \hline
EBSCO Discovery Service    &      EDS  & Meta Search                               &\url{http://search.ebscohost.com/login.aspx?authtype=shib&custid=ns215211&profile=eds}\\
\rowcolor[rgb]{ .867,  .922,  .969}  Scopus &      SCO  & Meta Search  &\url{https://www.scopus.com/}\\  
  Web of Science &      WOS  & Meta Search   &\url{http://www.webofknowledge.com}\\  
\rowcolor[rgb]{ .867,  .922,  .969}  JSTOR    &     JST  & Meta Search  &\url{https://www.jstor.org}\\  
Science Direct    &      SD  & Publisher-specific Search   &\url{http://www.sciencedirect.com}\\  
\rowcolor[rgb]{ .867,  .922,  .969}   SpringerLink    &      SPL  & Publisher-specific Search  &\url{http://www.springerlink.com}\\  
IEEE Xplore    &     IEEE  & Publisher-specific Search  &\url{http://ieeexplore.ieee.org}\\  
\rowcolor[rgb]{ .867,  .922,  .969}   ACM Digital Library    &      ACM  & Publisher/Subject Specific (Computer Science)  &\url{https://dl.acm.org/}\\  
Business Source Premier    &      BSP  & Subject Specific (Business \& Law)   &\url{http://search.ebscohost.com/login.aspx?authtype=ip,uid&profile=ehost&defaultdb=bth}\\
  \hline
\end{tabular}
\caption{Selected source databases}
\label{table:selectedDatabases}
\end{table}

  % \subsection{Time period}
  We have included a period as an inclusion criterion. In this, we
  have Sergots' study on the formalization of the Naturalization Act \cite{sergot1986british} as the starting
  date for our search. The automated search has covered all articles published until
  September 2017.

\subsection{Pilot extraction}

An initial pilot extraction on selected search engines was
performed. The generic search string was adapted to the formats
accepted by each of the indexes. Its results have been collected in
Table \ref{Table:Pilot}. The exact queries are presented in Appendix
\ref{appendix:pilot}.

\begin{table}[t!]
  \centering
    \begin{tabular}{lr}
    \rowcolor[rgb]{ .357,  .608,  .835} \textcolor[rgb]{ 1,  1,  1}{\textbf{Index}} & \multicolumn{1}{l}{\textcolor[rgb]{ 1,  1,  1}{\textbf{Number of Hits}}} \\
    \rowcolor[rgb]{ .867,  .922,  .969} EBSCO Discovery Service &1.418 \\
    Web of Science & 17 \\
    \rowcolor[rgb]{ .867,  .922,  .969} Scopus & 375 \\
    JSTOR & 87 \\
    \rowcolor[rgb]{ .867,  .922,  .969} Science Direct & 765 \\
    SpringerLink & 2.331 \\
    \rowcolor[rgb]{ .867,  .922,  .969} Xplore & 13 \\
    ACM Digital Library & 10 \\
      \rowcolor[rgb]{ .867,  .922,  .969} Business Source Premier & 2
      \\
      \midrule \midrule
      \textbf{Total Hits} & \textbf{5.018} \\
    % \bottomrule
    \end{tabular}%
    \caption{Results of the automated search}
        \label{Table:Pilot}%
\end{table}%

The variance in the number of hits retrieved by the indexes consulted
can be explained by the diversity of the criteria that each of them
implement. The meta searchers consulted (EDS, SCO, WOS, JST) provided
specific filters for searches on title, abstract, and keywords,
limiting the search space to content that is most relevant to the
SLR. This functionality was not present in some of the indexes (e.g.:
SPL), increasing the sample with articles that contain the keywords in
their full-text versions but whose topics do not correspond to the
interest in this SLR. The total set of articles retrieved contained
duplicate entries as some meta-searchers index the same
journals than publisher-specific databases (e.g.: SD and SCO are both owned
by Elsevier, and the journals indexed in SD are also present in SCO). We have decided to allow
this as meta-searchers are likely to include hits in areas not
considered in our standard search procedures (computer science).

% \subsection{Search method}
% For
% automatic searches, we will start by using controlled subject terms
% (title, abstract and indexed keywords). The advantage of using
% controlled subject terms is that the sources identified will be closer
% to the main topic of each search. Its drawback is that it will
% potentially give few results. In case the index database does not
% provide subject term search, or the resulting dataset is not
% considerable (e.g. corresponding to less than 10 sources) we will
% expand our search method into considering full text searches.

\subsection{Inclusion and exclusion criteria}
\label{sec:inclusionCriteria}

This SLR restricted sources to primary
formal verification and business process management studies. No
statistical methods have been applied in the research.
The studies in consideration must present initiatives to ensure
compliance of business processes against a regulatory framework. Both
theoretical frameworks and research experiences of using formal
methods in the  formalisation of laws will be considered. In
particular, we have considered the following set of inclusion criteria:

\begin{description}
  \item[IN1] The study is concerned with compliance of business processes
    against regulatory frameworks using formal methods. 
  \end{description}  
The criterion has been assessed by reading the title and abstract.
After the initial set of included studies has been selected, we will
apply further exclusion criteria. Our exclusion criteria considered the
following parameters:

\begin{description}
  \item[EX1] The study does not propose a compliance
    model/framework.
    
  \item[EX2] The study does not make use of a formal model to ensure
    the correctness of its approach.
    
  \item[EX3] The study does not place considerations regarding
    compliance of legal texts.

  \item[EX4] The language of the primary study is not English.

  \item[EX5] The application area of the paper does not refer to process-oriented technologies. 
  \item[EX6] This entry is not a primary study, for instance, other
    reviews.
    \item[ExNP] This entry is not a full paper
    (only abstract, summary, etc.). 
\end{description}

Moreover, we have applied the following subsumption criteria:
% to determine whether
% certain primary study  is similar to other in the set of included
% studies, yet it does not provide a
% significant contribution. A study is considered subsumed if any of the
% following criteria apply:

\begin{description}
    \item[S1] There exists a more recent, or longer version of the same
    study, from the same set of authors (i.e.: a conference version
    is subsumed by a journal version).
    \item[S2] It does not propose a significant modification from
      another documented study by the same set of authors.
    \item[S3] The study refers to the implementation whose method
        has been previously
        documented in another study.
\end{description}

The original SLR protocol considered forward snowballing in the case
that the resulting dataset contained less than 20 primary studies. This
strategy was abandoned since the resulting set amounted to 46 primary
studies.

\subsection{Control Set} \label{sources}

\begin{figure}[t]\centering
  \includegraphics[width=12cm]{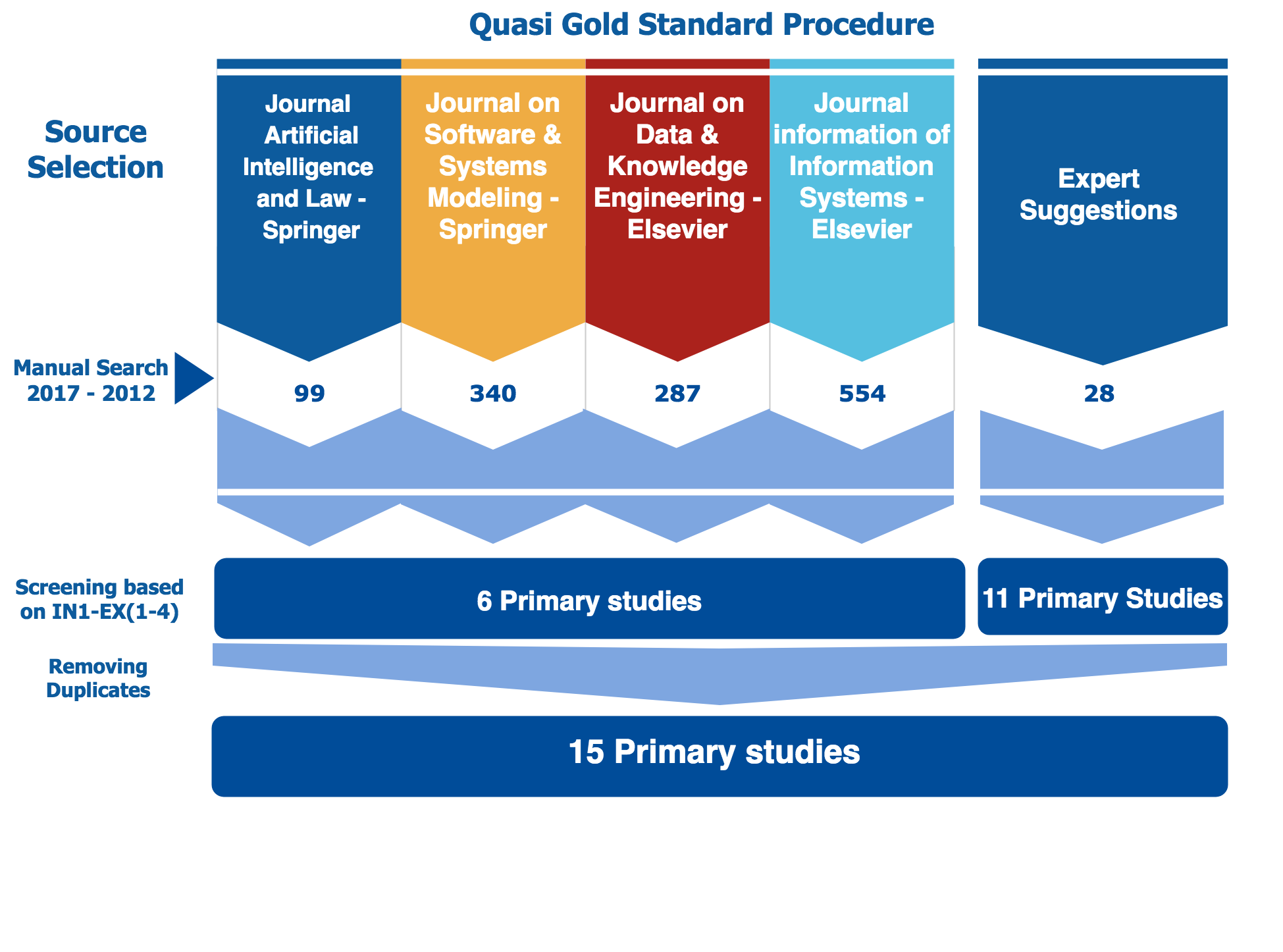}
  \caption{Construction of the Quasi Gold Standard Dataset}
  \label{fig:QGSPilot}
\end{figure}

In the development of this SLR, we have built a Quasi-Gold-Standard
(QGS) dataset \cite{zhang2011identifying}. This involved the construction of a pilot study and the collection of expert
suggestions, as described in Figure \ref{fig:QGSPilot}. The pilot
was built on a manual search of all relevant studies in the 
following journals published between 2012 and 2017:

\begin{itemize}
\item Springer's Journal of Artificial Intelligence and Law \cite{JAIL}.
\item Springer's Journal on Software \& Systems \cite{sosym}.
\item Elsevier's Data \& Knowledge Engineering \cite{DKE}.
  \item Elsevier's Journal of Information Systems \cite{JIIS}.
\end{itemize}

In the construction of the QGS dataset, we screened a total of
979 articles. After applying the inclusion and exclusion criteria
defined in Section \ref{sec:inclusionCriteria}, we identified a total
of 5 primary studies that were used as part of the control set.

In addition to the set of primary studies identified through the
pilot, the authors identified an additional set of 28 articles, considering of
relevance in the SLR. After applying the inclusion and exclusion
criteria from Section \ref{sec:inclusionCriteria}, a set of 11 primary
studies were identified. The final Quasi-Gold-Standard included the
studies identified by the pilot as well as the ones derived from
expert suggestions and it is included in the Appendix 
\ref{Appendix:QGS}.

\subsection{Validity of the search strategy (internal validity)}

The selection 
of sources in Section \ref{sources}  considered the inter-disciplinary nature of this field by ensuring that
journals in Computer Science,
Information Systems, Business Processes and Law were included. The selection of the
papers included in the QGS comply with the inclusion/exclusion
criteria defined, and it was verified by at least two of
the authors. The search process detailed in Section \ref{sec:search-strategy} was validated against the QGS. To test the validity of the search we have used
the recall measure \cite{zhang2011identifying}:

{\small
\[
\textsl{Recall} = \frac{\text{Number of relevant studies in the QGS
    found by automated search}}{\text{Number of papers in the QGS}}
\times 100\%
  \]
}
In the execution of the initial pilot extraction, we  
recorded  5.018 hits. As suggested by \cite{zhang2011identifying}, an
extraction will be considered valid if the recall rate reaches over
80\%. In our case, 14 out of 15 studies in the QGS were
retrieved by the automated search, leading to a 93.33\% % recall
validating the search query. 

% \subsubsection{Keyword refinement}

Automatic searches could filter out some of the relevant studies
identified by the QGS. After analyzing the results of the papers
omitted, we could deduce that the articles filtered corresponded to
those that, while being subscribed in the area of formal verification,
did not use the keyword ``formal'' in their title, abstract, or set
of keywords. A further attempt for data extraction was performed, this
time modifying building a new query \textbf{Q2} = KW1' \textbf{AND}
KW2 \textbf{AND} KW3 \textbf{AND} KW4 \textbf{AND} KW5, where  KW1'
refers to the set of terms that exclude an explicit mention of the
``formal'' keyword:
\begin{align*}
  &\text{KW1'} & =&&\{\text{method, model, language}\}.\\
  &\text{KW2--KW5} & =&& \text{as before}\\
\end{align*}

Table \ref{table:KeywordComparison} compares the
size of the automatic search for the search queries
used. For query \textbf{Q2}, the recall level reaches 100\%, but the
dataset retrieved by \textbf{Q2} is also 69,28 times the size of the
one retrieved by \textbf{Q1}. We have decided to continue using the
original search query, as the increase in recall did not justify the
increased dimensions of the studied dataset.

\begin{table}[t]
    \centering 
    \begin{tabular}{|c r r|} \hline
      \rowcolor[rgb]{ .357,  .608,  .835} \textcolor[rgb]{ 1,  1,  1}{
      \textbf{Search string}} & \textcolor[rgb]{ 1,  1,  1}{
        \quad \textbf{Number of Hits}} &
         \textcolor[rgb]{ 1,  1,  1}{ \quad
        \textbf{Recall}}\\
      \rowcolor[rgb]{ .867,  .922,  .969} Q1& 5.018 & 93.33\% \\
      Q2&             347.666 & 100.00\% \\ \hline
    \end{tabular}
    \caption{Comparison of number of hits by vs. search string}
    \label{table:KeywordComparison}
  \end{table}

%%% Local Variables:
%%% mode: latex
%%% TeX-master: "main"
%%% End:

\subsection{Selection of Primary Studies}

\begin{figure}[t]\centering
  \includegraphics[width=5cm]{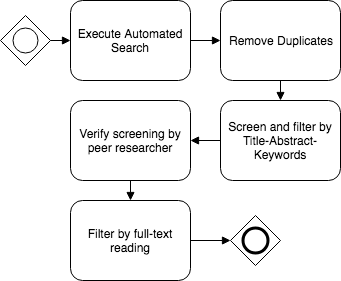}
  \caption{The search and selection protocol}
  \label{fig:searchSelectionProcedure}
\end{figure} 
Figure \ref{fig:searchSelectionProcedure} details the search and selection procedure in this SLR. 
First, the search strings have been adapted and executed at the selected
databases (c.f. Appendix \ref{appendix:pilot}). The aggregated dataset was further pruned to
eliminate duplicate results. The remaining sources were first screened over title, abstract, and keywords, and those that corresponded to the exclusion criteria were filtered out from the sample. To validate consistency, uniformity, and validity in the application of the inclusion/exclusion instrument, this sample was controlled by at least one different researcher in the team. The results were collated and any disagreements were discussed until an agreement was reached. After this, the remaining set of candidate studies was controlled by performing a full-text analysis. Those studies that fulfilled the inclusion criteria after this stage were considered primary studies. We performed the collection of data for this final set of primary studies, and the data extraction forms have been collated and verified for completeness by at least one author different from the data extractor.

%\subsubsection{On the selection process}

Table~\ref{table:results-process} documents the primary study selection. Starting from an automated search, we identified $45,5\%$ duplicated entries that were removed. The title, abstract, and keyword filtering filtered $4,56\%$ from the potential studies, which were used in the second screening applying further selection criteria. The result included $88,46\%$ of the filtered studies, that went for a full review. In this last process, we filtered $50\%$ of the studies. The selection of primary studies is presented in Table \ref{table:primaryStudies}. 

\begin{table}[t]
    \centering 
    \begin{tabular}{|m{4cm}r|} \hline
\rowcolor[rgb]{ .867,  .922,  .969} \textbf{Phase}&                                          \textbf{Selected Studies}\\
Automated Search&      5.018\\
\rowcolor[rgb]{ .867,  .922,  .969} Duplicate Removal&      2.283\\
Title, Abstract \& Keyword filtering  &      104\\
\rowcolor[rgb]{ .867,  .922,  .969} Second Screening&      92\\
Full-text filtering&      46\\
\hline
\end{tabular}
\caption{Results of the automated search and filtering phases}
\label{table:results-process}
\end{table}

\begin{table}[t] \scriptsize \centering
\begin{tabular}{@{}m{0.15cm}m{0.35cm}m{0.25cm}m{5cm}m{9cm}m{0.8cm}@{}}
\toprule
\textbf{PId} & \textbf{Id.} & \textbf{Year} & \textbf{Author}                                                                                                                  & \textbf{Title}                                                                                                                       & \textbf{Source}                                                                                                                                                        \\ \midrule
S01                 & PS002              & 2016          & Elgammal,  Turetken, Heuvel, Papazoglou               & Formalizing and appling compliance patterns for business process compliance                                                          & \cite{elgammal_formalizing_2016}                                                                                                                          \\ 
S02                 & PS003              & 2011          & Awad,  Weidlich, Weske                                   & Visually specifying compliance rules and explaining their violations for business processes                                          & \cite{awad_visually_2011}                                                                                                                            \\
S03                 & PS004              & 2011          & Nešković, Paunović,  Babarogić                                  & Using Protocols and Domain Specific Languages to Achieve Compliance of Administrative Processes with Legislation                     & \cite{neskovic_using_2011}                                                                                                      \\
S04                 & PS005              & 2012          & Ly, Rinderle-Ma, Göser, Dadam                          & On enabling integrated process compliance with semantic constraints in process management systems                                    & \cite{ly_enabling_2012}                                                                                                                \\
S05                 & PS006              & 2016          & Hashmi,  Governatori, Wynn                                  & Normative requirements for regulatory compliance: An abstract formal framework                                                       & \cite{hashmi_normative_2016}                                                                                                                                         \\
S06                 & PS008              & 2010          & D’Aprile, Giordano,  Gliozzi,  Martelli, Pozzato, Dupré    & Verifying Business Process Compliance by Reasoning about Actions                                                                     & \cite{daprile_verifying_2010}                                                        \\
S07                 & PS009              & 2010          & Governatori, Rotolo                            & Norm Compliance in Business Process Modeling                                                                                         & \cite{governatori_norm_2010}    \\
  S08                 & PS010              & 2017          & Ribino, Lodato, Cossentino                                 & Modeling Business Rules Compliance for Goal-Oriented Business Processes                                                              & \cite{ribino_modeling_2017}
  \\
S09                 & PS011              & 2015          & Jiang, Aldewereld, Dignum, Wang, Baida                    & Regulatory compliance of business processes                                                                                          & \cite{jiang_regulatory_2015}                                                                                                                \\
S10                 & PS014              & 2012          & Hashmi, Governatori, Wynn                                  & Business Process Data Compliance                                                                                                     & \cite{hashmi_business_2012}                                    \\
S11                 & PS016              & 2012          & Ramezani, Fahland, van der Aalst                                           & Where Did I Misbehave? Diagnostic Information in Compliance Checking                                                                 & \cite{ramezani_where_2012}                                                                                                                  \\
S12                 & PS017              & 2014          & Ramezani,  Gromov, Fahland, van der Aalst                      & Compliance Checking of Data-Aware and Resource-Aware Compliance Requirements                                                         & \cite{taghiabadi_compliance_2014}                                                                                                                  \\
S13                 & PS018              & 2015          & Knuplesch, Reichert,  Kumar                                & Visually Monitoring Multiple Perspectives of Business Process Compliance                                                             & \cite{knuplesch_visually_2015} \\
S14                 & PS020              & 2007          & Ghose, Koliadis                                           & Auditing Business Process Compliance                                                                                                 &  \cite{ghose_auditing_2007}                                                                          \\
S15                 & PS025              & 2011          & Islam, Mouratidis, Jürjens                            & A framework to support the alignment of secure software engineering with legal regulations                                               & \cite{islam_framework_2011}                                                                                                                                          \\
S16                 & PS028              & 2010          & Schleicher, Anstett, Leymann, Schumm                       & Compliant Business Process Design Using Refinement Layers                                                                            & \cite{schleicher_compliant_2010}                                                                                          \\
S17                 & PS029              & 2013          & Ramezani,  Fahland, Dongen, van der Aalst              & Diagnostic Information for Compliance Checking of Temporal Compliance Requirements                                                   &  \cite{taghiabadi_diagnostic_2013}                                                                                                               \\
S18                 & PS030              & 2009          & Schleicher, Anstett, Leymann, Mietzner                       & Maintaining Compliance in Customizable Process Models                                                                                &\cite{schleicher_maintaining_2009}                                                              \\
S19                 & PS031              & 2009          & Saeki,  Kaiya, Hattori                                & Checking Regulatory Compliance of Business Processes and Information Systems                                                         & \cite{saeki_checking_2009}                                                      \\
S20                 & PS033              & 2008          & Arbab, Kokash, Meng                                 & Towards Using Reo for Compliance-Aware Business Process Modeling                                                                     & \cite{arbab_towards_2008}                                                         \\
S21                 & PS035              & 2011          & Bruni,  Corradini, Ferrari, Flagella, Guanciale, Spagnolo        & Applying Process Analysis to the Italian eGovernment Enterprise Architecture                                                         & \cite{bruni_applying_2011}                                                                                                               \\
S22                 & PS047              & 2017          & Knuplesch, Reichert                                           & A visual language for modeling multiple perspectives of business process compliance rules                                            & \cite{knuplesch_visual_2017}                                                        \\
S23                 & PS048              & 2017          & Guarda, Ranise, Siswantoro                                          & Security analysis and legal compliance checking for the design of privacy-friendly information systems                               & \cite{guarda_security_2017}                                                      \\
S24                 & PS053              & 2014          & Elgammal,  Sebahi, Turetken, Hacid, Papazoglou, Van Den Heuvel                       & Business process compliance management: An integrated proactive approach                                                             & \cite{elgammal_business_2014} \\
S25                 & PS055              & 2013          & Knuplesch, Reichert, Pryss, Fdhila, Rinderle-Ma                               & Ensuring compliance of distributed and collaborative workflows                                                                       &  \cite{knuplesch_ensuring_2013}                                                    \\
S26                 & PS056              & 2013          & Knuplesch, Reichert, Fdhila, Rinderle-Ma                               & On enabling compliance of cross-organizational business processes                                                                    & \cite{knuplesch_enabling_2013}                            \\
S27                 & PS060              & 2012          & Elgammal, Turetken, Van Den Heuvel                                        & Using patterns for the analysis and resolution of compliance violations                                                              & \cite{elgammal_using_2012}                                                            \\
S28                 & PS061              & 2011          & Turetken, Elgammal, Van Den Heuvel, Papazoglou                                 & Enforcing compliance on business processes through the use of patterns                                                               & \cite{turetken_enforcing_2011}                                                                              \\
S29                 & PS062              & 2011          & Papazoglou                                                                        & Making business processes compliant to standards \& regulations                                                                      & \cite{papazoglou_making_2011}                                                               \\
S30                 & PS063              & 2010          & Schumm, Turetken,  Kokash, Elgammal, Leymann,  Van Den Heuvel                            & Business process compliance through reusable units of compliant processes                                                            & \cite{schumm_business_2010}       \\
S31                 & PS064              & 2010          & Elgammal,  Turetken, Van Den Heuvel,  Papazoglou                                  & Root-cause analysis of design-time compliance violations based on property patterns                                           & \cite{elgammal_root-cause_2010}            \\
S32                 & PS067              & 2007          & Liu, Müller,  Xu                                      & A static compliance-checking framework for business process models                                                                   & \cite{liu_static_2007}                                                                                    \\
S33                 & PS071              & 2017          & Koetter, Kintz, Kochanowski,  Wiriyarattanakul,  Fehling,  Gildein,  Wagner,  Leymann,  Weisbecker & An universal approach for compliance management using compliance descriptors                                                         & \cite{koetter_universal_2017}                                                \\
S34                 & PS072              & 2015          & Governatori                                                                        & The regorous approach to process compliance                                                                                          & \cite{governatori_regorous_2015}               \\
S35                 & PS074              & 2014          & Hashmi, Governatori, Wynn                                          & Normative Requirements for Business Process Compliance                                                                               & \cite{hashmi_normative_2014}                                                        \\
S36                 & PS079              & 2009          & Awad, Smirnov, Weske                                            & Towards resolving compliance violations in business process models                                                                   & \cite{awad_towards_2009}\\
S37                 & PS080              & 2009          & Governatori, Sadiq                                         & The journey to business process compliance                                                                                           &                                                                      \cite{governatori_journey_2009}          \\
S38                 & PS081              & 2009          & Governatori, Rotolo                                     & How do agents comply with norms?                                                                                                     & \cite{governatori_how_2009}                  \\
S39                 & PS082              & 2008          & Sackmann, Kähmer                                            & ExPDT: A policy-based approach for automating compliance & \cite{sackmann_expdt:_2008}                                                                                                                                                 \\
S40                 & PS085              & 2006          & Milosevic, Sadiq, Orlowska                                       & Translating business contract into compliant business processes                                                                      & \cite{milosevic_translating_2006}                                                     \\
S41                 & PS089              & 2016          & Burattin, Maggi,  Sperduti                            & Conformance checking based on multi-perspective declarative process models                                                           & \cite{burattin_conformance_2016}                   \\
S42                 & PS092              & 2013          & Lohmann                                                                        & Compliance by design for artifact-centric business processes                                                                         & \cite{lohmann_compliance_2013}                                                                                                            \\
S43                 & PS093              & 2012          & Awad, Goré, Hou, Thomson, Weidlich                   & An iterative approach to synthesize business process templates from compliance rules                                                 &  \cite{awad_iterative_2012}                                                                                                   \\
S44                 & PS095              & 2010          & Tamisier, Didry,  Parisot,  Wax,  Feltz                   & A Model-Driven Architecture for Implementing Business Reasoning Maintenance Systems                                                  & \cite{tamisier_model-driven_2010}                                                                                     \\
S45                 & PS101              & 2017          & Ghooshchi,  van Beest, Governatori, Olivieri, Sattar                            & Visualisation of Compliant Declarative Business Processes                                                                            & \cite{ghooshchi_visualisation_2017}                                                              \\
S46                 & PS104              & 2007          & Sadiq,  Governatori, Namiri                                    & Modeling control objectives for business process compliance                                                                          & \cite{sadiq_modeling_2007}                                                                                                                \\ \bottomrule
\end{tabular}
\caption{List of selected primary studies}
\label{table:primaryStudies}
\end{table}

%\subsubsection{Demography of Primary Studies}

\begin{figure}[t]
  \centering
    \includegraphics[width=0.7\textwidth]{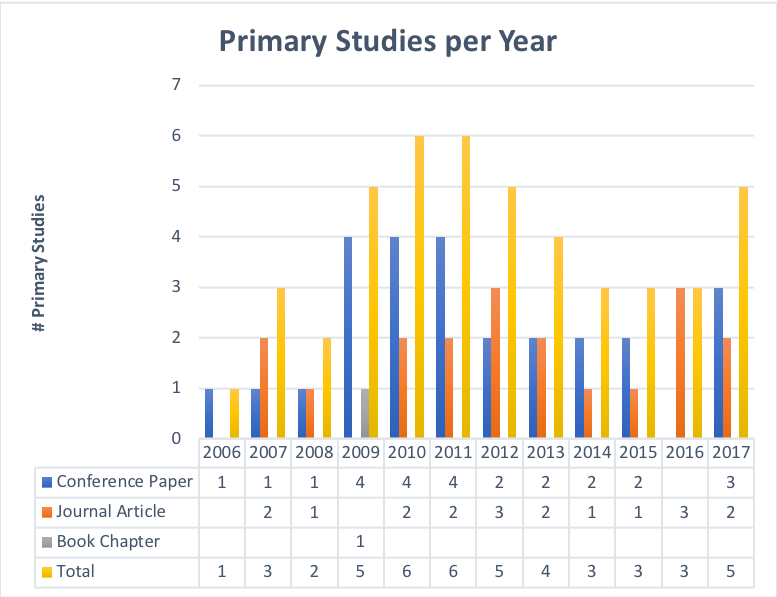}
    \caption{Primary Studies: Distribution per year/type of publication}
    \label{fig:summary-studies}
  \end{figure}

The results in Table \ref{Table:Pilot} show how most of the entries correspond to outlets in databases from computer science and information systems, while very few entries correspond to legal databases. The outlets of the primary studies show that while the selection criteria included more than 30 years of research, the studies focused on compliance frameworks only appeared in the second part of the 2000s.  See Figure \ref{fig:summary-studies}. 26 studies ($56,52\%$ of the total) appeared as conference/workshop publications, 19 studies ($41,3\%$) corresponded to journal publications and only 1 study ($2,17\%$) corresponded to a book chapter. 

%\hugo{Do we need to say more? the section is quite short, perhaps we can merge it with the next one (calling it results)}

\subsection{Quality Assessment}

The study adopted the quality controls suggested in \cite{kit_cha_2007}. The SLR protocol, including the sources used in the automated search, and the 
selected journals to construct QGS were evaluated
 by two external reviewers with expertise in SLR and Business Process Management. We acknowledge the work of 
Dr. Paolo Tell (IT University of
Copenhagen) and Prof. Barbara Weber (University of Saint Gallen) in these aspects. The datasets resulting from the phases of screening and data extraction were re-evaluated by different researchers than the data extractor, and conflicts were reported using email and electronic forums.

%%% Local Variables:
%%% mode: latex
%%% TeX-master: "main"
%%% End:

%\input{qualityAssesment}
\section{Data extraction}
\label{sec:dataExtraction}  
Data regarding standard bibliographic information (i.e.: authors, title, publication type, date of publication, abstract, ISSN/ISBN, and publication details) was extracted using automated tools (i.e.: Zotero). The data required to answer the research questions was manually extracted by one reviewer. A random sample accounting for 10\% of the
primary studies was checked by another author to verify repeatably and accuracy in the extraction.
Conflicts were discussed and resolved among the authors.  The following sections describe the data collected and their relation to the research questions:

\subsection{Research Question 1: Elements conforming Compliance Frameworks} \label{sec:guidelines-rq1}

To answer this question we categorized the studies according to their abstraction levels and the activities that they included.  A framework might combine several abstraction levels: for instance, it might include stages to elicit compliance specifications (strategic level), compare them against an existing business process model (process level), and use the results in the execution of the process in a process engine (execution level). 

\subsubsection{\textbf{Abstraction Levels}}

In particular, we used the following abstraction phases:

\begin{itemize}
    \item \textbf{Strategic level}: A framework that considers this level includes concepts regarding initial requirement engineering abstractions. The concerns of this level included  definition of the goals and business objectives and the definition, analysis, and catalog of organizational models in which processes \& regulations operate.
    \item \textbf{Process level}: A framework that considers how business goals are operationalized through process models. It includes the phases of process model creation, analysis, and verification of business process models.
    \item \textbf{Execution level}: A framework that considers this phase includes the refinement of business process models into executable specifications, such as the Web Service Business Process Execution Language (WS-BPEL) \cite{jordan2007web} or the XML Process Definition Language (XPDL),  \cite{Palmer2009}. In addition, we consider at this level all frameworks including process monitoring against compliance requirements. 
    \item \textbf{Post-mortem level}: In this level, we considered frameworks that included analysis of the sequences of events generated by the execution of business processes. These might include, for instance, the construction of process models from event logs (process mining) or the post-mortem analysis of event logs against compliance rules (process conformance).
\end{itemize}

Our categories roughly correspond to those proposed by El Kharbili  \cite{el_kharbili_business_2012}. However, frameworks including process monitoring were categorized as part of the execution level, instead of post-mortem levels.

\subsubsection{\textbf{Framework phases}}
We collected information regarding the type of activities included in each framework. While there is a correlation between certain phases and the specific abstraction levels (e.g.: the extraction of compliance requirement happens at the strategic level only), other activities happen across different abstraction levels (e.g. compliance analysis). In particular, we use the following categories.

\begin{itemize}
    \item \textbf{Extraction of regulatory models}:
        In this phase, a framework performs the interpretation of legal texts in their original form (typically natural language) and generates a model for it. The language used to express such models will be referred to as \emph{compliance language} hereafter. 
    \item \textbf{Modelling}: This phase is centered on the description and structural analysis of processes using business process description languages, such as BPMN, Workflow/Petri Nets, etc. 
    \item \textbf{Compliance checking}: This phase pertains the process of deciding whether a compliance requirement (typically defined using a compliance language) is satisfied by the executions of a process model. 
    \item \textbf{Compliance analysis}: This phase refines the decision problem in compliance checking, making an emphasis in \emph{how is the compliance model fulfilled}. A framework that considers a compliance analysis phase typically generates documentation, for example, event traces that do not comply with the regulatory requirements, or quantitative information on how many logs do not comply with the requirements.
    \item \textbf{Compliance enactment}: This phase contemplates proactive and reactive actions taken to preserve compliance against regulatory requirements in the execution of process models. This phase may involve the identification of source events that led to compliance violation and the implementation of corrective actions to reestablish compliance (i.e.: \emph{compliance resolution}), or the implementation of runtime mechanisms to react and compensate violations (i.e.: \emph{compliance recovery}).
    \item \textbf{Runtime analysis}: This phase considers the checking of compliance rules against the data generated at runtime. In this phase both predictive (how this activity may lead to violation of compliance) and coercive (how to apply compensation activities to maintain compliance) are considered.  
    \item \textbf{Postmortem analysis}: also known as \emph{process auditing}. It proceeds to analyze whether and how finished event logs have deviated from compliance rules.
\end{itemize}

\subsection{Research Question 2: Formalization of regulatory documents}\label{sec:guidelines-rq2}

To answer how compliance frameworks formalize regulatory documents, we have collected the following data:

\subsubsection{\textbf{Compliance Language}} Regulatory documents are written by personnel with training in law, and their language of choice is typically the language of the country legislating. Here we collect the formal language used to interpret regulatory documents. 

\subsubsection{\textbf{Process Language}} Here we collect which languages a framework supports when defining working with process descriptions.

\subsubsection{\textbf{Verification Techniques}} The verification technique used for a compliance framework heavily depends on the compliance and process languages used. Here we summarize each framework's verification technology used.

\subsubsection{\textbf{User support}} The framework is usable if users have access to it. We have collected evidence on the kind of users supported by each framework. Since several frameworks were only in seminal stages, this category included ``none'' as a possible answer. The remaining categories were built after data collection and included:
\begin{itemize}
    \item \emph{Compliance Experts}: In this category, we included lawyers, compliance consultants, compliance officers, law experts, legal experts, and business users with knowledge of the law.
    \item \emph{Domain Experts}: It includes process modelers, process analysts, domain specialists, business specialists, process owners, business analysts, business process domain experts, and process designers.  
    \item \emph{Others}: These included users involved in the compliance framework outside the categories mentioned above. For instance, IT security experts.
\end{itemize}

\subsubsection{\textbf{Limitations of the framework}} To answer what were the gaps not considered by the tool, we extracted information regarding the future work of each of the primary studies. 
From the data collected, we have selected only the comments that relate general challenges of regulatory frameworks and excluded specific features/improvements for a given tool.

\subsection{Research Question 3: Compliance Technologies}
\label{sec:guidelines-rq3}

The data gathered information regarding specific details on the compliance technologies used, and whether there was evidence in the study regarding translations of regulatory documents to formal specifications. In particular, we collected answers to the following questions:

\subsubsection{\textbf{Compliance checking technologies}} \label{def.compliance-checking-technologies}

Here we collected evidence at which stage compliance checking was performed. It included approaches at design-time (i.e.: \emph{compliance verification}), run-time (i.e. \emph{compliance monitoring/enforcement}), and post-mortem (i.e. \emph{compliance audit}). Some studies included compliance-checking technologies that did not map precisely to these categories. For instance, pre-processing stages regarding the verification of consistency between different regulations \cite{jiang_regulatory_2015}, and the conformance between laws and bylaws \cite{neskovic_using_2011}. Those approaches that could not be clustered were treated specially. 

\subsubsection{\textbf{Compliance analysis performed}}
\label{def.compliance-analysis-technologies}
This category collected information regarding the kind of analysis performed in compliance verification, if any. It includes the output of documentation regarding the type and 
size of the experiments performed (\emph{Compliance reporting}).  This could encompass traces checked for compliance, and aggregated measurements of the overall result of compliance checking. Other frameworks might provide \emph{explanation} on how to replicate compliance violation, in terms of counterexamples and deviations from compliant behavior. This later category will be referred to as \emph{compliance explanation}.

\subsubsection{\textbf{Compliance enactment performed}} 
\label{def.compliance-enactment-technologies}
This item collected information regarding how the framework provided support for compliance enactment technologies, if any. It included methods for automatically engaging and correcting process models at design time when they were non-compliant (\emph{Violation Resolution}) and, runtime mechanisms allowing the framework to react dynamically to possible violations in a trace, in the form of compensation activities that bring led event traces back to compliance. This last category is known as  \emph{compliance recovery}.

\subsubsection{\textbf{Translation from legal texts into formal specifications}} Even after search and selection filtered many frameworks that did not correspond to regulatory documents, many of the frameworks collected did not show evidence of actual translation of regulatory documents. In this category, we collected such evidence.

\subsection{Research Question 4: Flexibility}
\label{sec:guidelines-rq4}

To answer how flexible compliance frameworks are, we consider two types of flexibility: reactions when the laws change, and support for legal reasoning in dynamic contexts. We explain them below:

\subsubsection{\textbf{Changes in laws}} Legislations are seldom monolithic documents, and they are constantly in the process of change. In this item, we have collected evidence on whether the primary study considered how to handle changes in regulations in the compliance framework presented.

\subsubsection{\textbf{Legal reasoning techniques}} Not all laws have the same decisive power and or can be applied in all cases. An annulment rule might remove a law from the knowledge base, and an abrogation might make a law valid within a certain time frame. Moreover, laws are promulgated based on jurisdiction and content, and accordingly, \emph{defeasible conditions} define when a rule can be applied or not. In this item, we collected whether the framework considered law resolution techniques. In particular, we collected information on whether the study provides methods for reasoning and interpretation of laws. Three types of reasoning were collected: the priority of laws ruling specific subjects versus general laws (i.e.: \emph{lex specialis}), the priority of laws according to their jurisdiction (i.e.: \emph{lex superior}), and the priority of laws according to their time of enactment (i.e.: \emph{lex posterior}). 

\subsection{Research Question 5: Maturity of the Frameworks}
\label{sec:guidelines-rq5}

Part of the objectives of this survey is to analyze how far each of the frameworks considered is from providing a product that could be used in industrial sectors. Our evaluation gathered data regarding the case studies considered, the implementations, and the data sets available. This information was used in parallel with the rest of the research questions, to categorize each of the frameworks on a maturity level, explained below. 
\subsubsection{\textbf{Case studies}}

Primary studies considering theory and implementation studies where equally considered, therefore, and the data extraction collected information regarding the nature of each study. The information regarding the case study(ies) considered included (a). the number of case studies tested, (b). the regulations considered, (c). the business process(es) checked for compliance, and (d). the enterprise contexts where the framework has been studied.  

% \begin{itemize}
%     \item Is there a case study?
%     \item Number of case studies?
%     \item regulations considered
%     \item Enterprise contexts
%     \item Business processes
% \end{itemize}

\subsubsection{\bf Accessibility}
In this excerpt, we collected information regarding the accessibility of each of the frameworks. We collected information regarding (a) the accessibility of the source code, (b) the data sets of each of the case studies,  and (c) whether the prototype was publicly available at the time of data extraction.
% \begin{itemize}
%     \item Code
%     \item data sets
%     \item Prototype
% \end{itemize}

\subsubsection{\bf Readiness Level}

Part of our objectives was to identify the level of maturity of the compliance frameworks analyzed. Each
    of the levels describe the level of maturity of the technology in
    question:
    \begin{enumerate}
    \item Basic principles observed - We will consider a
      primary study in level A if it points out a theoretical 
      problem and suggests possible solutions.
    \item Technology concept formulated - We will consider a primary study in level B if it refers to a theory paper that formulates the solution to a theoretical problem.
    \item Experimental proof of concept - We will consider a primary study in level C if it refers to a theory paper with an
      algorithm implementing a solution proposed in level B.
    \item Technology validated in lab - We will consider a primary study in level D if it refers to a paper with an algorithm
      and test runs on academic examples.
    \item Technology validated in a relevant environment
      (industrially relevant environment in the case of key enabling
      technologies) - We will consider a
      primary study in level E if it refers to a paper with an algorithm tested on real data, but not with real users. 
    \item Technology demonstrated in relevant environment (industrially relevant environment in the case of key enabling
      technologies) - We will consider a
      primary study in level F when it refers to a paper on prototype systems tested on real data for real end-users, but not in real work situations.
    \item System prototype demonstration in operational environment - We will consider a primary study in level G when the primary study is a paper on a system tested on real data for real end-users in real work situations.
    \item System complete and qualified - We will consider a primary study in level H when it refers to a paper on a finished system tested on real data for real end-users in real work situations.
    \item Actual system proven in operational environment (competitive manufacturing in the case of key enabling technologies; or space) - We will consider a
      primary study in level K when it refers to a paper on a finished system validated
      in practice over time on real data for real end-users in real
      work situation.
    \end{enumerate}
The categories proposed correspond to levels 1--9 in the Technology Readiness Level (TRL) \cite{trlscale}. Our data extraction grouped readiness levels in pairs, to take into
account some uncertainty. Thus, \textbf{TRL 1--2} will be used if the primary
study in question is a purely theoretical paper with no sign of
experiments. \textbf{TRL 3--4} will be used for a theory with validation on
artificial data (e.g. made-up examples). \textbf{TRL 5--6} will include primary
studies that perform validation on real data, but not operational. \textbf{TRL
7--8} refers to primary studies reporting tests in operational
environments. Finally, \textbf{TRL 9} refers to primary studies that report
systems that have been in used for longer period and validated. 

The calculation of readiness levels used the information collected in points 1 and 2 of this section, paired with individual assessment by data collectors. These maturity levels are indicative, and they do not pretend to replace a systematic approach performed with standardized tools, such as the U.S. Air Force Technology Readiness Level calculator \cite{nolte2003technology}.

%%% Local Variables:
%%% mode: latex
%%% TeX-master: "main"
%%% End:

%\section{Summary of the results}

%All studies accepted for %inclusion were analyzed %qualitatively, first %individually and then as a %whole. The accompanying %material includes summary %tables of study %characteristics, study quality, %and results will be presented %to the reviewers. In this %section, we will present a %narrative synthesis. 

\section{Results}
\label{sec:results}

In this section, we will answer the research questions based on the analysis. We first consider research question 1, considering the anatomy of formal compliance frameworks
\begin{center}
\textit{Research Question RQ1: What are the common elements that conform to a formal
  regulatory compliance framework for process-oriented technologies?    }
\end{center}

RQ1 was answered from the point of view of which abstraction levels the study considered, and which phases in the modeling life cycle were covered.

\subsection{Abstraction Levels}

We can characterize the different activities included in compliance frameworks into four main abstraction levels: strategic, design, execution, and post-mortem phases. 

\begin{figure}[t]
    \centering
    \includegraphics[width=0.50\columnwidth]{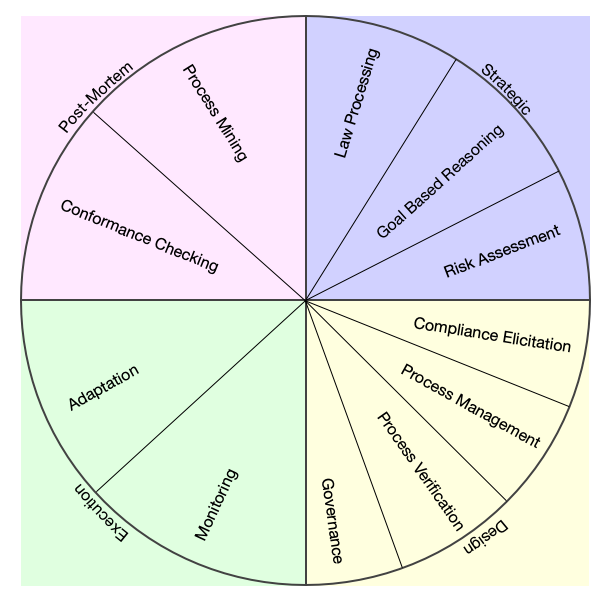}
    \caption{Abstraction Levels}
    \label{fig:abstractionLevels}
\end{figure}

\subsubsection{Strategic Level} Compliance documents, such as legal texts and guidelines, include information that does not necessarily refer to compliance laws, or that will not be prioritized in the early stages of development. Works in strategic/goal modeling aim at identifying what are the objectives of legislation. In addition, it might include the creation of compliance repositories, the analysis of the impact of the satisfaction/violation of a compliance rule, and their prioritization. 

Among strategic modeling, different techniques are applied:
\begin{itemize}
    \item \textbf{Law processing:} This phase pertains to the extraction of a formal specification from the legal text. It is mostly a manual information extraction process, whose output is the representation of the law in terms of a formal language. 
    This phase included the identification of law fragments (for instance, articles or paragraphs) that are amenable for formalization in a compliance language. Here, the expressiveness of the compliance language plays a big factor. For instance, the work in PS048 only considers the extraction of access-control policies from the EU Data Protection Directive (EU DPD), as their formal language only allows the description of such policies.  It is noted that, in our dataset, no work claimed a full formalization of a given law. Other examples of law processing in the frameworks studied appear in PS071. 
    
    Once law fragments have been identified, a further phase regarding the implementation of a law in a given context can be considered. For example, a general law such as the Administrative Procedures Act (APA) may vary when implemented in construction and importation bylaws. Similarly, the administrative procedure to release construction permits may vary in implementation depending on each municipality. Such levels are considered in  PS004 when extracting choreographies from administrative procedure acts. 
    
    To illustrate this step, we present a mapping from compliance requirements into Linear Temporal Logic (LTL), extracted from PS093. A compliance requirement in natural language can be expressed as 
         \textit{``Before opening an account the risk associated with that account must be low. Otherwise, the account is not opened.} Assuming a set of actions $Act =\{og, rl, od, rh\}$ with the following meaning:
         \begin{itemize}
             \item $og $ is "grant a request to open an account",
             \item $rl $ is "risk was assessed as low.",
             \item $od $ is "deny a request to open an account", and 
             \item $rh $ is "risk was assessed as high".
         \end{itemize}
    Then the formula $\textbf{G}(og \Rightarrow rl) \land \textbf{G}(od \Rightarrow rh)$ says that in all cases, the opening of an account is done only if its risk is low, and that in all cases, the denial implies a high risk.  
    Finally, the formalization of the law might not be consistent, and no business process will satisfy the specification. Consistency-checking techniques might be used to consider whether the specification leads to a set of rules that are contradicting, such as the ones presented in PS011.
    
    \item \textbf{Goal-based reasoning:} The ultimate objective of a business process is to achieve a business goal, but the achievement of goals can be hindered by the laws in context. The work in PS010 considers the phases of \emph{deliberation --} deciding what goals to achieve, and \emph{means-end reasoning}-- how to achieve these goals.  Goals can also be subject to interrelations: In a multi-agent system, each agent has a set of objectives to achieve, and the activities performed by a business process might block different agents from achieving their goals. Works such as PS025 use the Secure Tropos methodology~\cite{mouratidis2007secure} to identify sets of goals, agents, and legal rights, to reason whether the goals can be achieved in the presence of attacker agents. 
    \item \textbf{Normative process models:} One common approach is to identify an abstract business process to be used as a template. Abstract business processes are not complete models (they do not have enough information regarding how to execute certain tasks). Rather, they define all the important activities that a compliant process model should have, leaving the rest of the activities underspecified. Abstract business processes need to be completed in different ways (each implementation might be different). Implemented processes need to respect the traces defined by the abstract process to be considered compliant. The creation of abstract processes is still a manual activity, and it might not necessarily involve correspondence with legal documents. Some examples of normative process models are given in the form of templates (e.g.: PS028 and PS030), as well as artifact-centric models (e.g.: PS092).
    \item \textbf{Risk Assessment:} A risk is the likelihood that an event occurs, in such a way that its occurrence influences the achievement of certain goals, with consequences in the organization’s business model, loss of reputation, and/or financial condition. Risk is usually measured as a combination of impact and probability of occurrence \cite{lainhart2000cobit}. The risk assessment phase in strategic modeling focuses on the identification of (internal) controls to mitigate the risks and to ensure effective implementation of the compliance requirements. A control pairs risks with a set of process constraints. Failure to satisfy the constraints for a control increases the likelihood of a (compliance) risk to materialize. Compliance officers use controls to prioritize the satisfaction of constraints according to their impact on controls. Examples of risk assessment frameworks are presented in PS061.
\end{itemize}

\subsubsection{Design Level} Overall, this was the only level considered by all the primary studies. Some of the activities considered at the process level included the extraction of formal descriptions of compliance rules, process modeling, compliance checking, and analysis of non-compliant process models. These activities are described in more detail when describing the framework phases. 

\subsubsection{Execution Level}
Few primary studies evidenced a relationship between compliance models and executable process languages used in a business process management system (BPMS). Among the studies collected, only the works of the COMPAS project (PS002, PS053, PS060, PS064) provided a relation between compliance rules (expressed in eCRG or LTL) and specifications in the Web-Service Business Process Execution Language (WS-BPEL) \cite{jordan2007web}. These works were mainly to identify whether existing process logs conform to compliance rules, so it could also be considered a post-mortem analysis. Other works (e.g. PS002) consider process execution among the phases in compliance but do not provide evidence of implementation in their frameworks. Surprisingly, despite a trend in compliance-by-design approaches, there was no evidence that such approaches lead to compliant executions of processes. 

\subsubsection{Post-mortem Level}
Some compliance rules cannot be verified unless there is contextual information regarding a process instance. The typical example is that of compliance rules that describe links between process information (e.g. the execution of an activity) and an existing value in the event log. Primary studies considered compliance at different abstraction levels. Here frameworks consider different technologies:

\begin{itemize}
    \item \textbf{Process Monitoring:} This technique receives as inputs a set of compliance rules, and a process model. The monitor keeps track of the evolution in the states of each process instance and compares it against the set of compliance rules. As an output, the monitor provides information on whether each instance state has violated any compliance rule, as well as it might provide statistical information on the aggregated performance of all instances to the compliance rules. Process monitoring has been included in the Seaflows framework (PS005), eCRG (PS018, PS053), and outlined as part of the ongoing work on Compliance Descriptors (PS071).
    \item \textbf{Process Discovery:} This technique is useful when paired with an information system with existing logs. Event logs are processed via a process discovery algorithm to generate a structured process model, which can then be checked for compliance using standard techniques (LTL model checking). This is the approach followed by Compliance Descriptors (PS071).
    \item \textbf{Conformance Checking:} While the process mining technique presented above requires the generation of a process model, works on conformance checking examine the compliance rules directly over event logs. Such an approach is implemented in the consistency and compliance checker framework (CCCF) (PS011), eCRG (PS047), and ProM's packages for Compliance Frameworks (PS017, PS029), ConformanceToBPMN (PS035), and Multi-Perspective Declare (PS089).
\end{itemize}

\subsection{Compliance Phases}

\subsubsection{Compliance elicitation techniques} In this category we summarise the techniques used for compliance elicitation, namely selection, catalog, and assessment of compliance requirements. Formalisation of compliance requirements into compliance rules, and verification of the sound interpretation of the laws, in isolation or composition.

\begin{enumerate}
    \item Compliance elicitation: selection, catalog, and assessment of compliance requirements. Formalization of compliance requirements into compliance rules, and verification of the sound interpretation of the laws, in isolation or composition.
    \item Process management: selection/catalog/mining of processes, verification of structural constraints (soundness).
    \item Process verification: Design-time verification. Does the process model respect the compliance rules?
    \item Process Monitoring/Enactment: Design time verification is expensive to test (state explosion problem). This is evident with the introduction of compliance rules based on data, where compliance verification might need to instantiate the process to every possible data instance, which is not feasible in real-time. Process monitoring allows for compliance checking of partial traces based on the current data instances. Enactment allows for the healing of processes that are candidates of a compliance violation to be compensated with activities that keep the satisfiability of compliance rules.
    
    \item Process Auditing: It can be done in absence of a process description (process mining), or as an after-the-fact analysis to identify general patterns in the behavior admitted by the process. Compliance policies at this level can describe trace-specific properties (e.g.: \emph{Product delivery is always preceded by the charge of a credit card}), or aggregated policies (e.g.: \emph{``20\% of all the cases should be subject to an additional auditing phase by an external advisor''}. 
    
    \item Change management: Both regulations and processes may change, and every time one of them change compliance might be compromised. Frameworks must keep track of how changes in the regulations affect old processes, or if process optimizations still are compliant with laws. 
\end{enumerate}

    % \begin{wrapfigure}[20]{r}[\dimexpr\columnwidth+\columnsep\relax]{\textwidth}
    \begin{figure}[t]\centering
\includegraphics[width=0.45\textwidth]{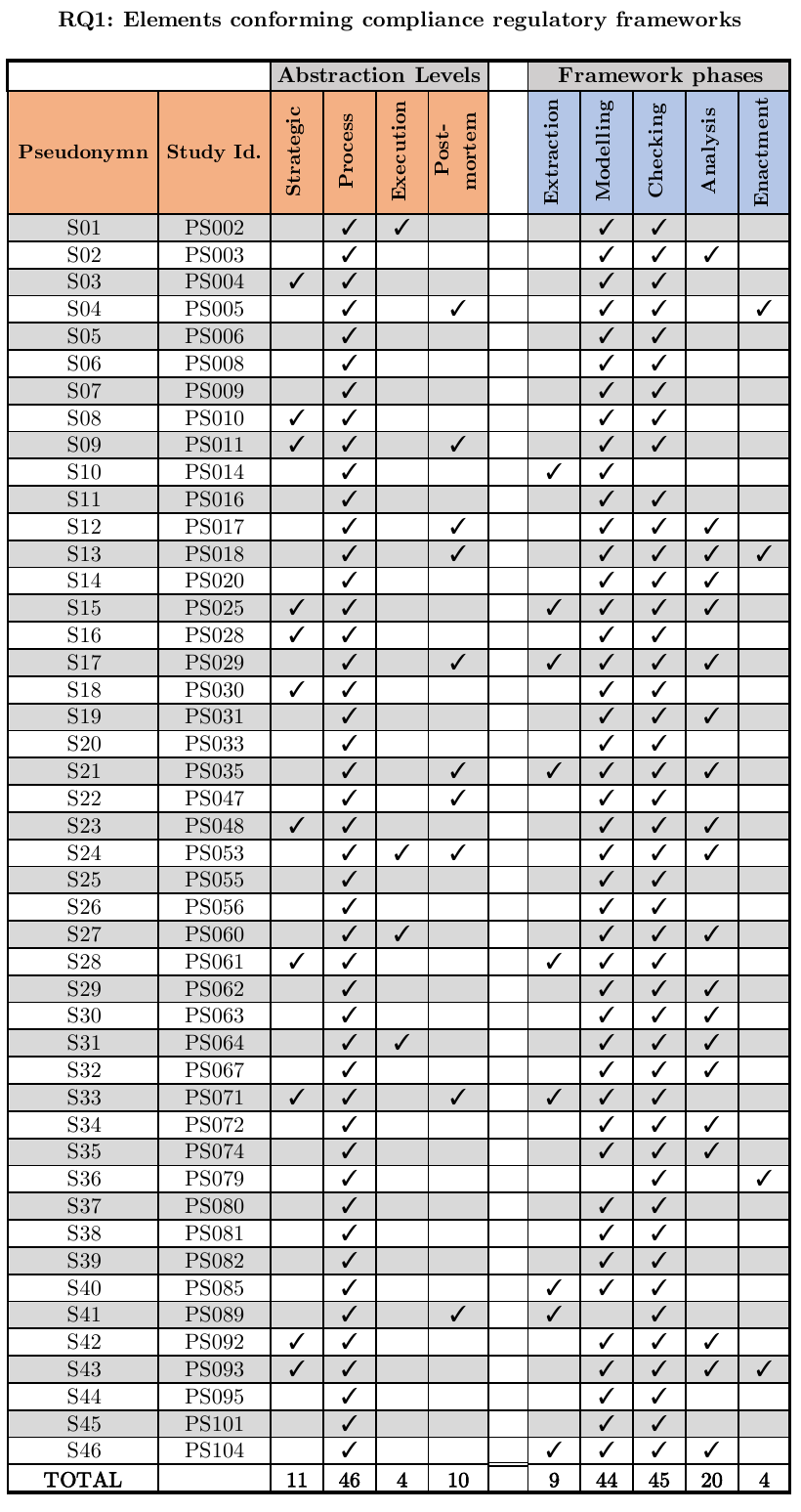}
 \caption{RQ1: Abstraction levels and phases in RQF}
 \label{fig:results-rq1}
     \end{figure}
% \end{wrapfigure}

\subsection{Research Question RQ2: How do current technologies formalize regulatory documents?} \label{sec:rq2-answers}

This research question is divided into several questions:

\subsubsection{    Q2.1: What formal languages are used to extract,
    represent, and organize requirements from regulations?}

We list the languages used for these purposes in
Table~\ref{table:summary-regulatory-languages}. We have classified the languages
by major types: Logic-based, graphical-based, and Domain-Specific Languages. 

Among the \textbf{logic-based languages}, we find two major classes: temporal logics on
the one hand, and combinations and variations of deontic and/or defeasible logic on
the other. Temporal logics emphasize statements about time---hence the
name---and straightforwardly expresses requirements that other things are required
or forbidden to happen in the past or future. Deontic logics~\cite{von1951deontic} formalize notions
of obligations and permissions, and defeasible logic~\cite{nute2001defeasible} formalizes the interaction
of conflicting rules. Other logical languages used for the expression of compliance rules include First Order Logic~\cite{fitting2012first}, Temporal Action Theory \cite{giordano2013reasoning}, a combination of Answer Set Programming \cite{bonatti2010answer} and Dynamic Linear Time Temporal Logic (DLTL) \cite{henriksen1999dynamic}, the Compliance Request Language (CRL) \cite{elgammal2010formal},  Computational Tree Logic (CTL) \cite{clarke1981ctl}, Deontic logic \cite{ribino_modeling_2017}, Extended Compliance Rule Graphs (eCRG) \cite{knuplesch_visual_2013},  Formal Contract Logic (FCL) \cite{governatori_journey_2009}, Linear Temporal Logic (LTL) \cite{LTL},  and the Process Compliance Language (PCL) \cite{governatori_norm_2010}.

We note that, while LTL seems to be the most popular specification language for compliance rules, there is ongoing debate about the appropriateness of temporal logic (in particular LTL) for modeling compliance. The argument against LTL can be summarised as LTL lacking primitive notions corresponding to the central concepts of ``permissions'', (defeasible) ``obligations'', and ``prohibitions''; this lack means that modeling with LTL becomes awkward, and can lead to ``obvious'' models being unsound for the intended
application~\cite{governatori_thou_2014,governatori_no_2015}.

Among the \textbf{graphical languages}, we find as major variants those based on
Petri Net semantics and temporal logic semantics. The different Petri Net semantics used in compliance models include Colored Petri Nets \cite{ColouredPetriNetsJensen}, Norm Nets
  \cite{Jiang:2013:NCC:2484920.2485101}, Workflow Nets     \cite{van1997verification} and Data-aware Petri Nets. Other graphical-based languages include Compliance Descriptors \cite{koetter2014integrating}.	For temporal logic graphical languages, the idea is to reduce the cognitive load on domain specialists not proficient in formal modeling by providing a graphical representation that may be amenable for the user, hiding the underlying semantics. Some examples of these languages include Business Property Specification Language (BPSL) \cite{xu2008bpsl}, a graphical language generating LTL and CTL specifications, and Multi-perspective Declare~\cite{burattin_conformance_2016}. The Extended Compliance Rule Graphs (eCRG) \cite{knuplesch_visual_2013} introduces a formal language to express compliance rules over events used in compliance monitoring.
  % operational semantics at \cite{knuplesch2017operational}.
  Another way of combining the languages is by using them as underlying semantics for runtime verification, as proposed in  BPMN-Q~\cite{awad_efficient_2008}. 

  Finally, some \textbf{domain-specific languages} have been crafted to express compliance properties. Schleicher et al. \cite{schleicher2009maintaining} define a language to specify compliance descriptors that can be used to check the BPEL specifications. Another DSL used is the coordination protocol language (CPG)~\cite{kopp2008model}, a platform-independent choreography language that describes coordination protocol and then generates BPEL code implementing them. Finally, the Extended Privacy Definition Tool (ExPDT)~\cite{sackmann_expdt:_2008} is a DSL extending the Web Ontology Language developed to formalize privacy preferences and process compliance rules.

\begin{table}[t]\footnotesize \centering
\begin{tabular}{m{7cm}rl}
\toprule
\textbf{Compliance Language} & \textbf{N}& \textbf{Primary Study} \\ \midrule
\multicolumn{3}{l}{\textbf{Logic-based languages}} \\ \midrule
\rowcolor[rgb]{ .867,  .922,  .969}
Linear Temporal Logic (LTL) &7& PS055, PS056, PS060, PS061, PS063, PS064, PS093  \\ 
Formal Contract Logic (FCL) &5& PS014, 	PS072, PS080, PS085, PS104  \\
\rowcolor[rgb]{ .867,  .922,  .969}
Process Compliance Language (PCL)  &5& PS006, PS009, PS074, PS081, PS009  \\
Computational Tree Logic (CTL) &2&  PS020, PS031\\
\rowcolor[rgb]{ .867,  .922,  .969}
First Order Logic (FOL) &1& PS048\\
Semantic Constraints (CTL, FCL) &1& PS005\\
\rowcolor[rgb]{ .867,  .922,  .969} 
  Deontic Logic &1& PS010\\
Defeasible Logic &1& PS101 \\\midrule
\multicolumn{3}{l}{\textbf{Graphical-based languages}} \\ \midrule
\rowcolor[rgb]{ .867,  .922,  .969}
Petri Net variants  &5& PS029, PS011, PS016, PS017, PS092  \\
Compliance Request Language (CRL) &3& PS062, PS053, PS002 \\
\rowcolor[rgb]{ .867,  .922,  .969}
Extended Compliance Rule Graphs (eCRG) &2&PS018, PS047\\
  Coordination Protocol Graph (CPG)&1&    PS004\\
\rowcolor[rgb]{ .867,  .922,  .969}
BPMN-Q &1&  PS003\\
Business Property Specification Language (BPSL) &1& PS067 \\ %\midrule
%\multicolumn{2}{l}{\textbf{Pattern-based languages}} \\ \midrule
\midrule 
\multicolumn{3}{l}{\textbf{Domain-Specific Languages}} \\ \midrule
\rowcolor[rgb]{ .867,  .922,  .969}Compliance Templates &2& PS028, PS030 \\
Violation Patterns &1& PS079 \\
\rowcolor[rgb]{ .867,  .922,  .969}Natural Language Patterns &1&PS025 \\ %\midrule 
%\midrule \multicolumn{2}{l}{\textbf{Others}} \\ \midrule
\rowcolor[rgb]{ .867,  .922,  .969}
Temporal Action Theory (TAT) &1& PS008 \\
Any rule modeling language &1& PS071 \\
\rowcolor[rgb]{ .867,  .922,  .969}ExPDT &1& PS082 \\
Inference Rules  &1& PS095 \\
\rowcolor[rgb]{ .867,  .922,  .969}
N/A&3& PS033, PS035, PS089 \\\bottomrule
\end{tabular}
\caption{Regulatory Languages}
 \label{table:summary-regulatory-languages}
\end{table}

\subsubsection{  Q2.2: What \textbf{case studies} of \textbf{formal
    regulatory compliance frameworks} exists?}

\paragraph{Regulations Studied}

We can identify a range of legal instruments used as input for the identification of compliance rules. They all differ in specificity and purpose. The instruments collected include legal codes, acts, guidelines, and bylaws. A \emph{legal code}
is a broad collection of statutes that cover extensive areas of law systematically, whose scope provide a detailed and organized set of statutes that govern a wide range of legal issues within a jurisdiction. 
A \emph{legal act} includes specific laws passed by legislative bodies addressing particular issues or subjects. they have a specific focus and are often narrower in scope compared to legal codes. They are authoritative and binding within the jurisdiction that enacted them.
A \emph{guideline} is a non-binding recommendations intended to assist in the application and interpretation of laws and regulations.  They are not legally binding but aim to influence behavior and practices. They are often used to interpret and implement laws and regulations. Finally, 
\emph{bylaws} are rules governing the internal affairs of a local authority or organization, applicable within a limited jurisdiction, they are limited in scope to the specific entity or area they govern and are enforceable within that jurisdiction or organization.

\textbf{Legal Codes} 
A legal code is a systematic collection or comprehensive written statement of laws, rules, or regulations organized and compiled by a governing body or legal authority. These codes are designed to cover specific areas of law, providing clear guidelines for conduct, procedures, and the administration of justice. Compliance rules extracted from legal codes include a)  access control and separation of duties, for instance, the ISO Code of Practice for Information Security Management \cite{ISO17799} Section 10.1.3 was studied in PS063. The ISO 27002 Code of Practice for information security controls \cite{iso27002}, Section 10.1.3, was studied in PS002, PS060, PS061, PS062, and PS063. Other codes studied include the Code of Conduct of the German Insurance Association (GDV)~\cite{dataHandling-Germany}, a set of principles and guidelines designed to ensure ethical and fair behavior among insurance companies operating in Germany. The GDV was studied in PS071. Other code studied was The European Customs Code (ECC) and its Implementation Provisions \cite{EUCustomsCode} form the regulatory framework that governs the import, export, and transit of goods within the European Union (EU). This framework aims to facilitate trade, ensure the security and safety of goods entering and leaving the EU, and streamline customs procedures across member states. The ECC code was studied in PS011.  Finally, the last code studied is the Australian Telecommunications Customer Protection (TCP) Code~\cite{communicationsCode-Australia}, a regulatory framework designed to safeguard the interests of consumers and ensure fair treatment by telecommunications service providers in Australia. The TCP code was studied in PS072, PS006, PS074.
% \begin{itemize}
%     \item ISO Code of practice for Information Security Management \cite{ISO17799}. Section 10.1.3 (Segregation of Duties)  has been studied in PS063.
%     \item ISO 27002 Code of Practice for information security controls \cite{iso27002}. Section 10.1.3, which establishes guidelines on separation of duties, has been studied in PS002, PS060, PS061, PS062, PS063. 
%     \item Code of Conduct of the German Insurance Association (GDV) \cite{dataHandling-Germany}. It was studied in PS071.   
%     \item Australian Telecommunications Customer Protection Code \cite{communicationsCode-Australia}. This was studied in PS072, PS006, PS074.

%     \item A subset of the European Customs Code and Implementation Provisions\cite{EUCustomsCode} has been studied by PS011.
% \end{itemize}

\textbf{Acts}
A legal act is a formal written law enacted by a legislative body, such as a parliament or congress. It is a primary source of law that creates new legal obligations, rights, or regulations, modifies existing ones, or repeals old laws. Legal acts are authoritative and binding within the jurisdiction of the legislative body that enacts them. At this level, most of the legal acts studied by the primary studies included privacy and data protection. They included the EU Directive directive on the protection of data~\cite{EU95-privacy}, studied in PS002, PS025, PS048, PS060, PS061 and PS062. The German Federal Data Protection Act (BDSG)~\cite{germanDataProtectionAct}, studied in PS071, the German Federal Data Protection Act \cite{germanISecDirective}, studied in PS025. The Privacy Principles (Schedule 1) of the Australian Privacy Amendment Act \cite{AustralianPrivacyAct} was analyzed in PS072. The last privacy act studied was the Japanese Act on the Protection of Personal Information \cite{gdpr-japan}, studied in PS031. Other acts include Section 8 of the Australian National Consumer Credit Protection Act \cite{NCCPA} in PS074.     Public sector laws have also been considered. In particular, the Finnish Administrative Procedure Act (APA) \cite{finnishAdminProc}, which regulates proceedings of government agencies when handling requests submitted by citizens and businesses exercising their lawful rights, was studied in PS004. Finally, section 404 famous Sarbanes-Oxley Act~\cite{sox} protecting investors by improving the accuracy and reliability of corporate disclosures made according to the securities laws, has been studied by (PS053, PS060, PS061, PS063).

    % \item  EU information society Directive on privacy and electronic communications 2002/58/EC. This directive has been studied in PS025, by modeling partial texts from Articles 4 and 5 and correlating them with Article 17 of 95/46/EC, and with  Sections 5, 9, and Annex from the FDPA national law German Federal Data Protection Act (FDPA), that implement EU 2002/58/AC.

% \begin{itemize}
%     % \item German Federal Data Protection Act (BDSG).  \cite{germanDataProtectionAct}. It was studied in PS071.
    
%     % \item The German Federal Data Protection Act \cite{germanISecDirective}. It has been studied by PS025.
    
%     \item Finnish Administrative Procedure Act (APA) \cite{finnishAdminProc} (PS004). It regulates proceedings of government agencies when handling requests submitted by citizens and businesses exercising their lawful rights.
    
%     \item Sarbanes-Oxley Act to protect investors by improving the accuracy and reliability of corporate disclosures made according to the securities laws, and for other purposes\cite{sox}.  The section 404 has been studied by (PS053, PS060, PS061, PS063).
%     \item Japanese Act on the Protection of Personal of Personal Information \cite{gdpr-japan}. This was studied in PS031.

%     % \item Australian National Consumer Credit Protection Act \cite{NCCPA}. Section 8 of this act was considered by PS074.
%     % \item Australian Privacy Amendment (Enhancing Privacy Protection) Act \cite{AustralianPrivacyAct}. An adaptation of Schedule 1 (Privacy Principles) was analysed by PS072.
% \end{itemize}

\textbf{Guidelines}
A legal guideline is an advisory document issued by government agencies, regulatory bodies, professional organizations, or other authoritative entities to provide direction and recommendations on how to comply with laws, regulations, or standards. While not legally binding like statutes or regulations, guidelines are designed to assist individuals, organizations, and entities in understanding and adhering to legal requirements and best practices. We can divide the types of guidelines into money-laundry, privacy and data protection, and public service. With regard to money laundering, the  Malaysian Guidelines on Anti-Money Laundering and Counter Financing of Terrorism~ \cite{guidelines-moneyLaundry} was studied in PS003. Also, the Jamaican Guidelines on Anti-Money Laundering and Counter Financing of Terrorism \cite{moneyLaunderingAct} was studied in PS093. Finally, one primary study analyzed the Chinese rules for anti-money laundering by financial institutions~\cite{moneyLaunderingChina}. Finally, regarding regulatory guidelines in the public sector,  the work in PS095 studied the Luxembourguish legislation regarding the disbursal of family allowances. This was presented in PS095 and consisted of a compendium of five main legal frameworks, as well as two international agreements.  The work in PS029 covered the Dutch legislation governing building permissions.

% \begin{itemize}
%     % \item Malaysian Guidelines on Anti-Money Laundering and Counter Financing of Terrorism - PS003 \cite{guidelines-moneyLaundry}. This was confirmed by personal communication with the author.
    
%     % \item Jamaican Guidelines on Anti-Money Laundering and Counter Financing of Terrorism \cite{moneyLaunderingAct}. This was studied in PS093. The original citation in the paper does not longer exist. \hugo{Perhaps ask the author for confirmation on this one}
    
%     % \item Chinese rules for anti-money laundering by financial institutions \cite{moneyLaunderingChina}

%     \item Luxembourguish legislation regarding the disbursal of family allowances. This was presented in PS095 and consisted of a compendium of five main legal frameworks, as well as two international agreements. There was not enough information to establish which parts of the legal frameworks considered were formalized.
    
%     \item Dutch legislation governing building permissions. This has been described in PS029. \hugo{Not enough information is provided to identify which legislations or the extent in their formalization was used. We could ask authors}.

% \end{itemize}

\textbf{Bylaws } 
A bylaw is a rule or regulation adopted by an organization, local government, or corporation to govern its internal operations and manage specific aspects of its functioning. These rules are typically established to provide clarity, order, and structure within the organization or jurisdiction. Bylaws can cover a wide range of topics, including the roles and responsibilities of members, meeting procedures, election of officers, and the management of property and finances. Few works covered bylaws as their source of compliance documents, including Maintenance Service Contracts (PS085), Clinical Guidelines (PS-16, PS017), Bank Internal Policies (PS017) and sets of business constraints (PS014, PS064).

\paragraph{Business Processes Studied}

% \usepackage{booktabs}
\begin{table}[t] \footnotesize \centering
\begin{tabular}{p{7cm}l}
\toprule
\textbf{Type of Process} & \textbf{Primary Study} \\ \midrule
\multicolumn{2}{l}{\textbf{Financial/Banking Sector}} \\ \midrule
\rowcolor[rgb]{ .867,  .922,  .969}Bank account opening & PS003, PS067, PS093 \\
Loan origination and approval & PS002,  PS017, PS028, PS053, PS060, PS061,  PS062, PS063 \\
\rowcolor[rgb]{ .867,  .922,  .969}Financial Instruments & PS008 \\
Loan/overdraft application & PS089 \\
\rowcolor[rgb]{ .867,  .922,  .969}Order-to-delivery & PS018 \\\midrule
\multicolumn{2}{l}{\textbf{E-Commerce}} \\ \midrule
\rowcolor[rgb]{ .867,  .922,  .969}Internet order processing, invoicing, cash receipting, delivery & PS002, PS031 \\
Purchase handling in an electronic shop & PS079, PS082 \\
\rowcolor[rgb]{ .867,  .922,  .969}Internet Reseller & PS061, PS064 \\
Order and Purchase Process & PS093 \\\midrule
\multicolumn{2}{l}{\textbf{Healthcare sector}} \\ \midrule
\rowcolor[rgb]{ .867,  .922,  .969}Patient's register \& specialised treatments & PS016, PS093 \\
Drug Administration Process & PS005 \\
\rowcolor[rgb]{ .867,  .922,  .969}Laboratory examinations, radiological services, therapeutic treatments, and surgeries & PS047, PS055, PS056 \\
Intensive care treatment & PS017 \\\midrule
\multicolumn{2}{l}{\textbf{Public Sector}} \\ \midrule
\rowcolor[rgb]{ .867,  .922,  .969}Management and handling of citizen complaints & PS006 \\
International Trade of Goods & PS011 \\
\rowcolor[rgb]{ .867,  .922,  .969}Processing and granting of building permits & PS029 \\
Approval of citizens' residence change & PS035 \\
\rowcolor[rgb]{ .867,  .922,  .969}Attribution of family allowances and parental leave & PS095 \\\midrule
\multicolumn{2}{l}{\textbf{Insurance Sector}} \\ \midrule
\rowcolor[rgb]{ .867,  .922,  .969}Claim Management & PS071, PS092 \\ \midrule
\multicolumn{2}{l}{\textbf{Procurement Sector}} \\ \midrule
\rowcolor[rgb]{ .867,  .922,  .969}Purchase-to-Pay & PS104 \\\bottomrule
\end{tabular}
\caption{Business Processes considered}
 \label{table:summary-processes}
\end{table}

The table \ref{table:summary-processes} summarize our case studies. We proceed by categorizing processes studied by application sector:

\textbf{Financial and banking sectors}. The bank account opening process is explored in PS067, where a hypothetical process is modelled based on the Malaysian anti-money laundering guidelines as presented in PS003 and further studied in PS093. This process must adhere to anti-money laundering regulations as per \cite{moneyLaunderingAct}, using Linear Temporal Logic (LTL) to encode compliance rules, which are then employed in process synthesis techniques to generate compliant process models. The EU 7FP ICT project COMPAS provided a simplified, yet realistic BPMN model for the loan origination and approval process, referenced in multiple studies including PS002, PS053, PS060, PS061, PS062, and PS063, and exemplified by a hypothetical scenario in PS028. The loan origination process documented in PS017 is part of the BPI Challenge of 2012\footnote{\url{http://www.win.tue.nl/bpi/doku.php?id=2012:challenge&redirect=1id=2012/challenge}}. Financial instruments are examined in a case study in PS008, which presents a business process in investment firms using YAWL to personalize customer portfolios. The realistic loan and overdraft application process from a Dutch financial institute\footnote{\url{http://dx.doi.org/10.4121/uuid:3926db30-f712-4394-aebc-75976070e91f}} is studied in PS089, with an event log available from the 2012 BPI Challenge describing the process behavior. Additionally, an order-to-delivery process is detailed in PS018, highlighting the necessity of incorporating multiple perspectives—time, data, resource, and session dependencies—into process monitoring, which is further analyzed using the eCRG language as described in \cite{knuplesch_visual_2013}.

\textbf{E-Commerce sector}: In PS002, a wide variety of business processes, including Internet order processing, invoicing, cash receipting, and delivery, were covered. PS031 presented a case study based on a hypothetical goods store, where the interactions between users and the system were formalized using a transition system. PS079 illustrated a fragment of a buying process in an electronic shop, focusing on situations where execution compliance rules are violated. This study defined violation patterns to match and modify the execution of processes by adding or removing parts of the business process model. The internet reseller scenario was featured in PS064 as part of the COMPAS running scenarios, describing online product systems, compliance patterns, and methods for root-cause analysis of compliance constraint violations. PS061 further explored the internet reseller scenario, detailing processes from order processing to ledger maintenance and identifying 52 compliance requirements from SOX, ISO/IEC 27000, and internal policies. These requirements led to the definition of 118 compliance controls and the generation of compliance rules in LTL. PS082 showcased a simple purchase handling process to demonstrate the expressive power of compliance policies in ExPDT, highlighting the definition of intricate compliance rules for permissions, obligations, and sanctions. Finally, PS093 presented compliance rules for an order and purchase process, used to synthesize BPMN process models that adhere to these rules.

\textbf{Healthcare sector:}
The work by Ramezani in PS016 defined compliance rules derived from a hospital's internal policies and clinical guidelines, which dictated the behavior in various business processes, including patient registration and specialized treatments like X-rays, MRIs, and CT scans. The Seaflows compliance framework was employed in PS005 to verify an abstract drug administration process, demonstrating the necessity of considering compliance rules at all stages of business process management. PS047 analyzed six healthcare process collections created from real cases at the University Hospital Ulm, covering laboratory examinations, radiological services, treatments, and surgeries. These collections explored multiple perspectives on compliance rules using eCRG, with one process also used in PS055 and PS056 to introduce algorithmic techniques for verifying compliance rules involving multiple parties through rule graphs and LTL formulae. PS093 presented a modified version of the patient admittance process originally detailed by van der Aalst, extending the model to incorporate explicit data flows. PS017 investigated the compliance of an event log from a Dutch hospital's intensive care unit against a medical guideline on tube feeding nutrition, analyzing compliance rules from multiple perspectives such as data and time. This technique was tested on real-life logs (1207 traces) and provided statistical information on deviations for domain specialists.

\textbf{Public sector:}
A study on the complaint handling process of a local land and property management authority in Australia (PS006) presented a set of policy rules from real legislative documents, covering various types of policies (e.g., permission, obligation, compensation) and their temporal and durational aspects. These policies were checked against the complaint handling process specification from LPMA, NSW Australia, and a compliance analysis was performed on six business processes regarding consumer complaint management against Section 8 of the Australian Telecommunication Consumers Protection Code 2012 using the Regorous compliance framework. In PS011, compliance with international trade processes was studied against the European Customs Community Code and its Implementing Provisions~\cite{EUCustomsCode}, formalizing six regulations in Norm Nets and verifying their consistency through seven event traces. PS029 examined building permit processes across five Dutch municipalities, showing variations in the process while ensuring compliance with Dutch legislative regulations. Compliance patterns were formalized using Petri net patterns and checked against event logs using the Process Mining Toolkit (Prom) in a study involving 1408 cases. In PS035, the citizen migration process model for registering residence changes by Italian citizens was specified in BPMN and mapped onto Petri nets for conformance and performance analysis, though no legislation was tested for compliance. Lastly, PS095 analyzed the attribution of family allowances and parental leave in Luxembourg, involving more than 160,000 individuals. This study focused on the mental processes governing applications and their relationship with legislation, using an inference system to calculate benefits rather than presenting a traditional process model for compliance checking.

\textbf{Insurance sector:}
 The work in PS071 presents a case study in claim management for the German insurance industry. The process is thought to automatically process damage claims from a customer from their origination up to their resolution, and it is governed by the Code of Conduct of the German Insurance Association (GDV) and the German Federal Data Protection Act (BDSG).  The case study presented in PS092 extends the claim handling process with a fraud detection check provided by external services. The compliance framework uses Petri Net artifacts to model both business processes and compliance rules, and verifies that processes are compliant by merging such artifacts.

\textbf{Procurement sector:}
A purchase-to-pay scenario has been presented in PS104 as an illustration of the use of FCL in the definition of control (e.g.: compliance) rules for the business process, as well as the annotation of compliance rules to specific activities in the business process.

\subsubsection{Q2.3: What \textbf{process modeling languages} have been used for achieving 
    \textbf{conformance} against \textbf{law}?} 
\begin{figure*}[t] 
 \centering 
 \includegraphics[width=\textwidth]{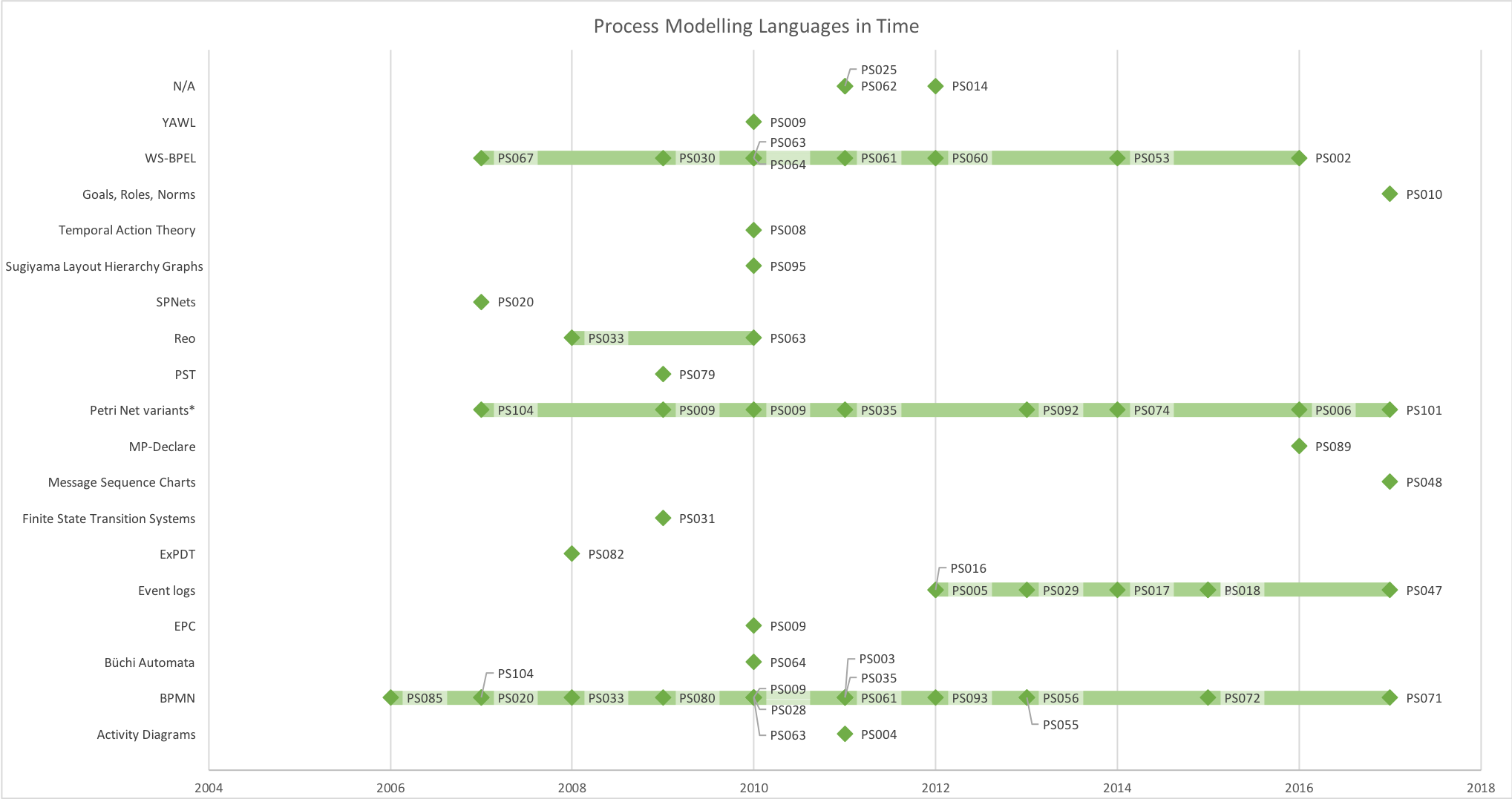}
 % We omit the comparison of process languages against regulatory languages. If needed it is in the graphic Appendixes/RQ23
 \caption{RQ2.3: Process specification languages in RQF}
 \label{fig:results-rq23}
\end{figure*}

We list the process modeling languages used in Table~\ref{table:summary-process-languages}, and we plot the graphically to indicate their frequency over time in Figure~\ref{fig:results-rq23}. We acknowledge four categories: graphical modeling languages such as BPMN and Petri Nets are the most common representations, however, process representations can be found also in terms of programming language specifications, logical specifications, or, in the case of conformance checking, do not exist at all (the input in this case are event logs). The distribution appears to be roughly comparable to the popularity of process modeling notations in general, e.g., Petri Net variants and BPMN by far outweighing other formalisms. It is interesting to note that WS-BPEL appears to have seen a fair share of work on compliance up to as recently as 2016.

% We briefly provide references for the definition of selected modeling languages:

% \begin{enumerate}
%   %\item Activity Diagrams PS004
% %\item Any, transformation to Process Structure Trees (PST): PS079
% %\item BPMN: PS003, PS028, PS020, PS072, PS080, PS085, PS104, PS055, PS056, PS061, PS063, PS093, PS033, PS035, PS009
% %\item Büchi Automata: PS064
% \item Event-Driven Process Chains (EPC) \cite{epc-event-process-chains} %: PS009
% %\item Event logs: PS018, PS047, PS029, PS011, PS016, PS092, PS017
% %\item Event Traces: PS005
% %\item ExPDT: PS082
% %\item Message Sequence Charts: PS048
% %\item MP-Declare: PS089
% %\item Non-deterministic Finite State Transition Systems: PS031
% %\item N/A: PS062, PS014, PS025
% %\item Parametric, BPMN: PS071
% %\item Petri Net variants*: PS006, PS009, PS074, PS081, PS101, PS104, PS035, PS092
% \item Reo \cite{arbab2004reo} %: PS063, PS033
% \item Semantic Process Networks (SPNets) \cite{ghose_auditing_2007} %: PS020
% \item Sugiyama Layout Hierarchy Graphs \cite{sugiyama1981methods} %: PS095
% %\item Temporal Action Theory: PS008
% %\item Tuples of Goals, Roles, Norms: PS010
% %\item WS-BPEL: PS030, PS067, PS053, PS002, 
% %PS060
% \item YAWL \cite{van2005yawl} %: PS009
% \end{enumerate}

%See Figure \ref{fig:results-rq23}

\begin{table}[t] \centering\footnotesize
\begin{tabular}{m{6cm}rm{8cm}}
\toprule
\textbf{Process Language} & \textbf{N}& \textbf{Primary Study} \\ \midrule
\multicolumn{3}{l}{\textbf{Graphical Languages}} \\ \midrule
\rowcolor[rgb]{ .867,  .922,  .969}BPMN   & 15&PS003, PS009,  PS020, PS028, PS033, PS035, PS055, PS056, PS061, PS063, PS072, PS080, PS085,  PS093, PS104 \\ 
 Petri Net variants& 8&   PS006, PS009, PS035, PS074, PS081, PS092, PS101, PS104\\
\rowcolor[rgb]{ .867,  .922,  .969}Activity Diagrams& 1& PS0004    \\
Process Structure Trees & 1& PS079 \\
\rowcolor[rgb]{ .867,  .922,  .969}Event-Driven Process Chains (EPC) \cite{epc-event-process-chains}  & 1&  PS009 \\
Semantic Process Networks (SPNets)~\cite{ghose_auditing_2007} & 1& PS020\\
\rowcolor[rgb]{ .867,  .922,  .969}  Sugiyama Layout Hierarchy Graphs  \cite{sugiyama1981methods}  & 1&  PS095\\
Message Sequence Charts & 1& PS048 \\
\rowcolor[rgb]{ .867,  .922,  .969}Parametric, BPMN & 1& PS071  \\
YAWL~\cite{van2005yawl} & 1& PS009\\\midrule
\multicolumn{3}{l}{\textbf{Programming Languages}} \\ \midrule
\rowcolor[rgb]{ .867,  .922,  .969}WS-BPEL  & 5& PS002, PS030, PS053, PS060, PS067  \\
 Reo \cite{arbab2004reo}  & 2&  PS063, PS033\\
\rowcolor[rgb]{ .867,  .922,  .969} ExPDT & 1&  PS082 \\\midrule
\multicolumn{3}{l}{\textbf{Formal Languages}} \\\midrule 
\rowcolor[rgb]{ .867,  .922,  .969}Temporal Action Theory (TAT) & 1& PS008 \\
Goals, Roles, Norms & 1& PS010 \\
\rowcolor[rgb]{ .867,  .922,  .969} Non-deterministic Finite State Transition Systems & 1& PS031 \\
Büchi Automata & 1& PS064 \\ \midrule
\multicolumn{3}{l}{\textbf{Others}} \\ \midrule
\rowcolor[rgb]{ .867,  .922,  .969}Event logs   & 8& PS005,  PS011,  PS016,  PS017, PS018,  PS029, PS047,  PS092  \\
N/A& 3& PS062, PS014, PS025
\\\bottomrule
\end{tabular}
\caption{Process Modelling Languages}
 \label{table:summary-process-languages}
\end{table}

\begin{figure*}[t] 
 \centering 
 \includegraphics[width=\textwidth]{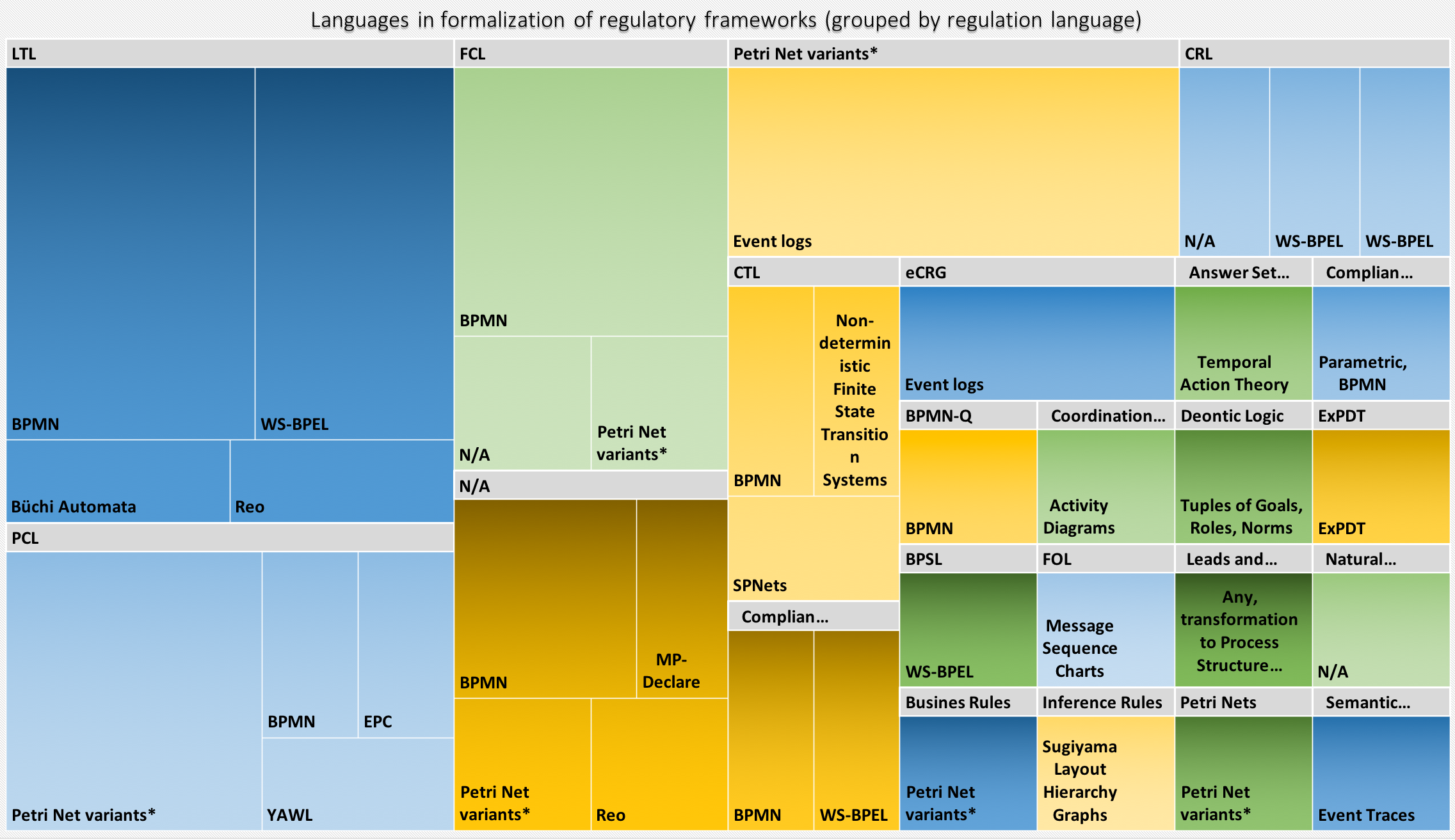}
 \caption{RQ2.1: Compliance Languages versus Process Representations: the outermost category (in dark grey) represents the regulation language used in the approach, and the colored boxes denoting process representation languages, if any}
 \label{fig:results-rq21}
\end{figure*}

All the work on compliance we have surveyed employs (in principle) two models:
one modeling legislation or rules
(Table~\ref{table:summary-regulatory-languages}), and one modeling
business process which we wish to be compliant with that legislation or those
rules (Table~\ref{table:summary-process-languages}). It is interesting also to
consider which formalisms are used \emph{together}. To this end, we depict in
Figure~\ref{fig:results-rq21} shows the correspondence between modeling languages for
regulations and modeling languages for processes. We note that BPMN and
Petri-nets, the major classes of process description languages, have been
used with many regulatory languages, including the major ones. 

Altogether, it seems that in general, the choice of modeling language for regulations does not seriously constrain the choice of modeling language for processes. This is perhaps unsurprising: key work on ``business process compliance'' requires only the process modeling language to have trace semantics for the methods to apply~\cite{hashmi_normative_2016,governatori_business_2013}. 

  \subsubsection{Q2.4: What are the \textbf{competencies} required in compliance frameworks?} \label{q2.4Competencies}
\begin{figure}[t] 
 \centering 
 \includegraphics[width=0.7\columnwidth]{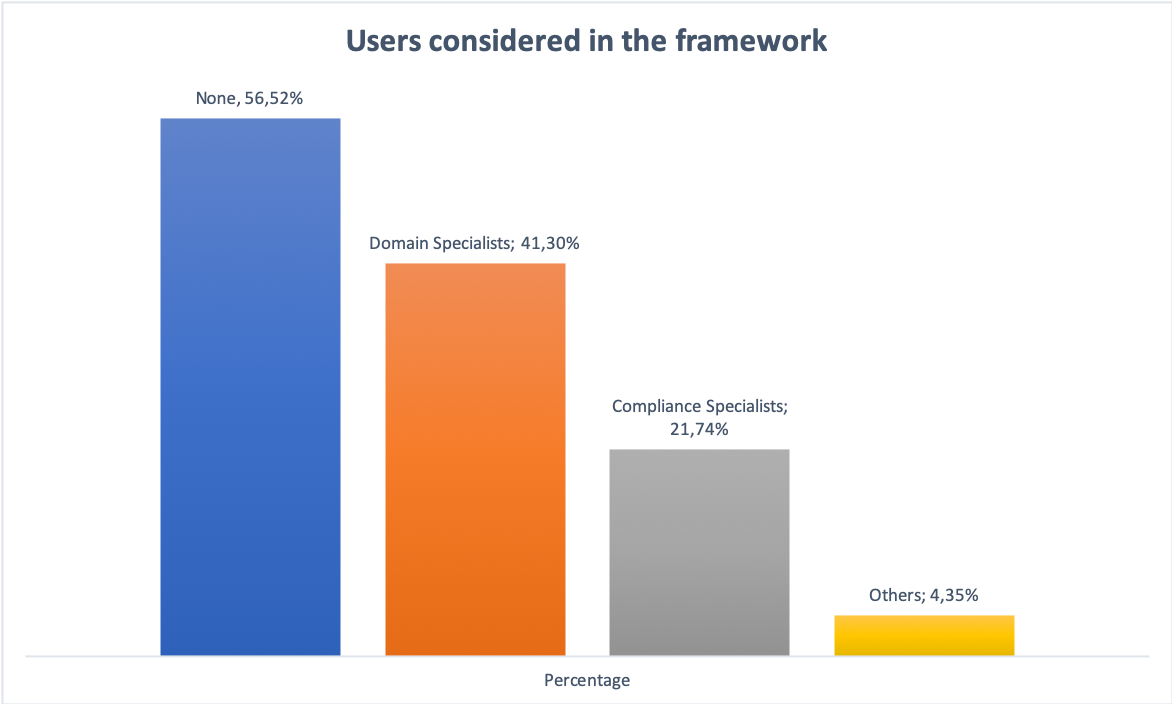}
 \caption{RQ2.4: Users considered in RQF}
 \label{fig:results-rq24}
\end{figure}

This question was answered by gathering evidence on how each framework is used and what competencies are required. A total of 26 studies (56,52\% of the total) did not exhibit evidence of an actual use of the compliance framework proposed. The remaining 43,48 \% mostly considered interactions with domain specialists (41,30 \% of the total) and compliance specialists (21,74 \% of the total). Finally, a single study considered the interaction with other user typologies, namely the interaction with IT security specialists. The results are summarized in Fig. \ref{fig:results-rq24}.

The level of competencies required for using a compliance framework was answered from the evidence collected regarding the manual stages present in the framework. Here we can divide the skills into four: First, the translation from compliance requirements into a compliance specification, second,  the construction of a formal process specification, third, the interpretation of results, and last, other manual activities.

\paragraph{Competencies involved in the formalization of regulatory texts}

From the set of primary studies, 41 (resp. 89,13\%) works considered the formalization of regulatory texts an activity requiring manual effort. 
The process executed to elaborate formal specifications from legal requirements was similar in most frameworks. It includes:
\begin{itemize}
    \item \textbf{Competence L1: Selection of Regulation Documents}: It involves the selection of paragraphs in a regulation that will be good candidates for formalization. 
    \item \textbf{Competence L2: Identify and Solve Risk Management Considerations}: Violating compliance requirements impacts organizations differently. Some compliance requirements may be hard and should never be violated, while others are soft and are valued depending on a cost/benefit scenario. This stage quantifies the cost of a compliance violation in an organization, for instance, by defining control objectives, risk, and internal controls (e.g. PS104).
    
    \item \textbf{Competence L3: Refine Regulatory Documents}: Regulations come in the form of natural language descriptions. An agreement on how terms in legislation are related, and how legal terms map to actions in a business process requires a solution before applying any compliance verification technique. Regulation alignment is typically addressed via terminology matching (e.g.: PS031) and ontological frameworks (e.g.: PS005). This mapping can be abstract, and link terms into groups of activities from the same domain (e.g.: PS005). It also involves the identification of what legal requirements will not be possible to be checked automatically.
    % Not all compliance requirements will be candidates for formalization, and they might require the introduction of manual checks (i.e.: "ensure that the documents are securely stored in locked cabinets").  
    Compliance refinement may lead to controls that map rights (e.g.: see the COSO framework \cite{COSOframework} in PS104 and PS061). This refinement can be done incrementally via abstract business processes and refinement methods (e.g.: PS028).

    \item \textbf{Competence L4: Formalize Refined Documents}: It involves translating compliance requirements into a (formal) specification language. 
    % For instance, the work in PS016 dealt with process auditing using compliance templates. It involved checking relevant regulations and eliciting respective compliance requirements, and,  for each requirement, identifying which available compliance patterns expressing the requirement can be built using Petri-nets. Other works may diverge on the formalism used, but the formalization step remains constant independently of the framework. 
    Since compliance requirements are given in natural language, support for the traceability of specifications to their sources is useful. Also, specification languages must be understandable to compliance specialists not trained in formal methods. Techniques for traceability of CTL formulae have been presented in PS031 using case frames \cite{caseFrames}, and for formulae in LTL using the CompRM tool in  PS061. Natural language patterns were used in PS025 to trace security requirement concepts such as goals, actors, tasks, and resources. Few works have approached the understandability of graphical specification languages: In PS071, the alignment of regulations and formulae is visualized via compliance connectors that link legal paragraphs to sets of logical formulae.
    % \hugo{Mention works on eCRG and BPMN-Q}.

    \item \textbf{Competence L5: Solve Ambiguous, Inconsistent or Incomplete specifications}: Translating legal requirements into formal specifications requires dealing with completeness and consistency issues. As observed in PS085 (and \cite{milosevic2002expressing}), regulatory documents may have explicit and implicit parts and a logical system may need to derive the implied clauses. An inconsistent formalization may be triggered when contradictory conclusions can be deducted from the specification. Business processes may need to comply with multiple compliance rules, but their combination may generate inconsistencies as well~\cite{awad_iterative_2012}. In addition, there might be behavioral incompleteness problems where an alternative situation is possible, but the specification has failed to specify what should happen.  Consistency and completeness considerations have not been at the core of compliance languages, but consistency has been approached in PS093. Similarly, the works in PS085 and PS093 have studied ambiguity resolution techniques in compliance requirements. Finally, consistency checking, loophole identification, and deadlock analysis for laws have been considered complementary activities that may take place in pre-processing steps and resolved in manual (e.g. PS085) or automated (e.g.: PS009, PS011) manners. 
    
\end{itemize}

 \paragraph{Competencies involved in the formalization of business processes}
 Compliance can only be verified if there is a clear account of the possible traces accepted by a business process. 29 primary studies (63,04\%) considered the formal modeling of business processes as a manual activity in their frameworks. Formalizations could be done via standard modeling techniques (c.f. \cite{dumas2018fundamentals}), and process discovery approaches (c.f.~\cite{van2012process}). Even when process discovery can be seen as a semi-automated technique, preparing and extracting behavior as event logs requires a large human effort. Some of the other manual steps in the modeling process for compliance involve the following:
\begin{itemize}
    
    \item \textbf{Competence P1: Define Activity Effects:} The execution of an activity in the process generates changes in the rights, obligations, and prohibitions in place. Keeping track of how the execution of an activity modifies the compliance rules in effect is important for maintaining the correspondence compliance rules. Semantic annotations are a popular technique where each event is annotated with logical formulae (PS009, PS020, PS080, PS081) or with security and access control policies (PS025, PS048, PS092) that hold after execution. The account for effects in process auditing takes the opposite direction: being a data-based process, it is important to capture the \emph{scope} of compliance rules, that is, identify when the occurrence of an activity (or a set of activities) triggers a compliance rule (e.g.: PS017).
    
    % \item \textbf{Security:} As regulations are emphasizing the treatment and management of sensitive information, identifying what are the security requirements derived from regulatory documents becomes more important than ever. The work in PS025 aims at eliciting security requirements (e.g.: secrecy, integrity, authenticity) from regulatory documents; and at tracing these requirements throughout the development stages to ensure that the design indeed supports the required laws and regulations.  The works in PS048 and PS092  consider \emph{access control} policies: In PS048, legal and IT experts collaborate on identifying parts of the regulation that can be formalized, building a general access control policy. This policy is later instantiated and checked against the traces that the IT system can reproduce.  This approach will answer the question on whether the concrete access control policy is compliant with the formalization of the regulation, and provide non-compliant authorization queries that are allowed by the design. In PS092, a compliant-by-design artifact model is extended with role-based access control policies, allowing them to express constraints that focus on which agent may execute which action. Such constraints express separation of duty policies like the four-eyes principle being used to control accounting frauds as in SOX. Nevertheless, the stages in these frameworks related to formalization, generation of access control policies, and creation of authorization queries are performed manually.

     \item \textbf{Competence P2: Define Variability}: Since regulations dictate minimal requirements for processes to be considered valid, more than one business process implementation may be compliant with a rule. 
     % This allows for flexibility in the way processes can be specified. 
     A possibility is that of creating abstract business processes, as well as the identification of variability points where alternative activities and subprocesses can replace original activities and still be compliant (PS030). 
     % The work in PS030 defines variability descriptors for business processes: first, an abstract business process is defined. This process contains compliant activities that realize a regulation, and variability descriptors that present alternative activities (or subprocesses) that concretize an abstract activity.
     % An algorithmic solution ensures that compliance constraints are not compromised by the instantiation of abstract activities.

    \item \textbf{Competence P3: Refine Processes} Sometimes laws define constraints over the interactions among multiple participants, as is the case of contracts. Laws ruling interactions need to be checked against global interaction models (so-called \emph{choreographies})~\cite{carbone2011logic}. However, choreographies only present a partial specification of the behavior of each role, and the process typically contains local business processes for each role involved. While algorithmic techniques to check the conformance between choreographies and implementations have been put forward in PS055, the mapping between local and global specifications remains largely a manual process.
    
    \item \textbf{Competence P4: Compose Processes} Few works have aimed at tackling compliance checking compositionally. Compositional approaches make use of \emph{compliance templates} (also called compliance fragments). These are connected process structures that can be instantiated and composed as building blocks to fulfill a compliance rule, ensuring compliance under composition using model-checking techniques (c.f. PS063, PS092). While compliance templates allow for a reusable design of compliance rules, their creation is done manually. 
    % A compositional approach is also used in some of the compliance-by-design works (e.g.: PS092).  Composition allows for flexibility: the generation of a process model from the compliant rules can be repeated when rules are added, removed, or changed. 

    \item \textbf{Competence P5: Align Terminology} 
    % The alignment of the terminology used within a process model and a compliance model describing the legal requirements is required.
    It is unrealistic that laws and business processes will always be constructed in synch, simply because of their different lifecycles, stakeholder groups, and purposes. This is crucial since regulations (and their encodings into logical formulae) represent general terms (e.g.: the right to be forgotten in GDPR) that may not appear explicitly in a process model. The study in PS104 presents a way to describe the alignment between control objectives in regulation and tasks in a process model. In the case of process auditing (e.g.: PS016) events in an event log will require a mapping to task names in a compliance rule. These alignments are performed manually.
   
\end{itemize}

\paragraph{Interpretation of results} Part of the success criteria for compliance technologies relies on the ability to integrate their output given in further iterations of the process specification. The most used technique in current compliance frameworks is model checking, whose outputs can be in some cases, not providing a lot of support to the user.  Several frameworks approach understandable explanations via \textit{counter-examples} for violating compliance rules (e.g.: PS031, PS055, PS056, PS063, PS064), or root cause analysis techniques (e.g.: PS002, PS060, PS064) that identify critical activities generating violations. In the case of compliance auditing \& monitoring, outputs such as \emph{distance analysis} quantify the degree the execution of a business process has deviated from a compliant execution (e.g. PS020), and in some cases even predict possible changes to enact the fulfillment of compliance rules (e.g.: PS104).

  \subsubsection{ Q2.5: What are the \textbf{gaps} not covered in current formal regulatory compliance frameworks?}

To answer this question, we looked at each primary study and collected information regarding future work. The information was parsed and aggregated in major research trends. 
Research directions considering specific tool improvements that could not be generalized were filtered from the data collection. Table \ref{tab:gaps} summarises what the primary studies have considered main areas of future research. The most frequently occurring topic for future research was empirical studies. Hereafter comes Language expressiveness, with data compliance and timing being the most frequent areas of future study. Finally, runtime monitoring, resource assignment, and usability each occurred as topics in five papers.

\begin{table}[t] \footnotesize \centering
\begin{tabular}{m{9cm}r} \toprule
\textbf{Topic}                                                            & \textbf{Occurrence} \\
\midrule
\rowcolor[rgb]{ .867,  .922,  .969} Empirical Validation                                             & 10         \\
Language Expressiveness: Data Compliance                         & 7          \\
\rowcolor[rgb]{ .867,  .922,  .969} Runtime Monitoring                                               & 7          \\
Resource Assignment                                              & 5          \\
\rowcolor[rgb]{ .867,  .922,  .969} Language expressiveness: General                                 & 5          \\
Usability/UI                                                               & 5          \\
\rowcolor[rgb]{ .867,  .922,  .969} Ontological Reasoning, domain modelling                          & 4          \\
Language Expressiveness: Timing constraints                      & 3          \\
\rowcolor[rgb]{ .867,  .922,  .969}Runtime Adaptation                                               & 3          \\
Scalability (State explosion problem)                            & 3          \\
\rowcolor[rgb]{ .867,  .922,  .969} Process Expressiveness: Loops                                     & 3          \\
Legal Consistency                                                & 2          \\
\rowcolor[rgb]{ .867,  .922,  .969} Natural Language Extraction                                      & 2          \\
Implementation                                                   & 2          \\
\rowcolor[rgb]{ .867,  .922,  .969} Language Expressiveness: Communication and Choreographies                                 & 2          \\
 Diagnostics                                                      & 2          \\
\rowcolor[rgb]{ .867,  .922,  .969} Process Changes                                                  & 2          \\
Optimization Techniques                                          & 2          \\
\rowcolor[rgb]{ .867,  .922,  .969} Context Changes                                                  & 1          \\
Relation between high-level and low-level policies               & 1          \\
\rowcolor[rgb]{ .867,  .922,  .969}Compliance against multiple processes                            & 1          \\
 Law changes                                                      & 1          \\
\rowcolor[rgb]{ .867,  .922,  .969} Multi-dimensional constraints                                    & 1          \\
Industrial Scale validation                                      & 1          \\
\rowcolor[rgb]{ .867,  .922,  .969} Dynamic Compliance Rules (choices between alternative rules)     & 1          \\
Counterexamples management                                       & 1          \\
\rowcolor[rgb]{ .867,  .922,  .969} Risk Management (wrt non-compliance)                             & 1          \\
Compliance metrics                                               & 1          \\
\rowcolor[rgb]{ .867,  .922,  .969} Compliance wrt mined data                                        & 1          \\
Language Expressiveness: Advanced Legal Notions (Mandate, Power) & 1          \\
\rowcolor[rgb]{ .867,  .922,  .969} Relative Expressiveness among different compliance languages     & 1          \\
 Tractable complexity classes of compliance checking              & 1          \\
\rowcolor[rgb]{ .867,  .922,  .969} Automatic Resolution Techniques                                  & 1          \\
Language Expressiveness: Sanctions                               & 1          \\
\rowcolor[rgb]{ .867,  .922,  .969} Compliance Drift                                                 & 1      \\\bottomrule   
\end{tabular}
\caption{RQ 2.5: Future Research Directions}
\label{tab:gaps}
\end{table}

\subsection{Research Question RQ3: How do formal regulatory compliance frameworks improve the behavior of business processes?}
 To answer this question, we subdivided the question into aspects regarding the plausibility, generalizability, and potential benefits of the technologies employed. In particular, we decompose RQ3 into the following questions.

\begin{itemize}
    \item \textbf{Question Q3.1:} \textit{How realistic is it to use existing RQF to solve problems in business process regulatory compliance?} To answer this question, we collected information on whether primary studies have engaged in the elicitation of specifications from real legal documents, or if they only worked inspired in legal texts.
    \item \textbf{Question Q3.2:} \textit{What type of compliance technology has been developed?} This question identified potential proxy variables, so even when certain techniques may be under development, their benefits have been documented and can be transferred to this case.
    \item \textbf{Question Q3.3:} \textit{Which compliance verification outputs have been generated?} This question sets focus on the type of documentation generated by compliance frameworks, helping officers navigate over potentially many cases.
    \item \textbf{Question Q3.4:} \textit{Which compliance enactment techniques have been used?}
    This question looked at the existing techniques used in the runtime monitoring and enactment of compliance frameworks.
\end{itemize}

In the rest of the section we will report the data collection generated for each of these questions.

\subsubsection{\textbf{Evidence for Q3.1. Translation of regulatory documents: Actual RQF support}}
\begin{figure}[t]
  \centering
    \includegraphics[width=0.7\textwidth]{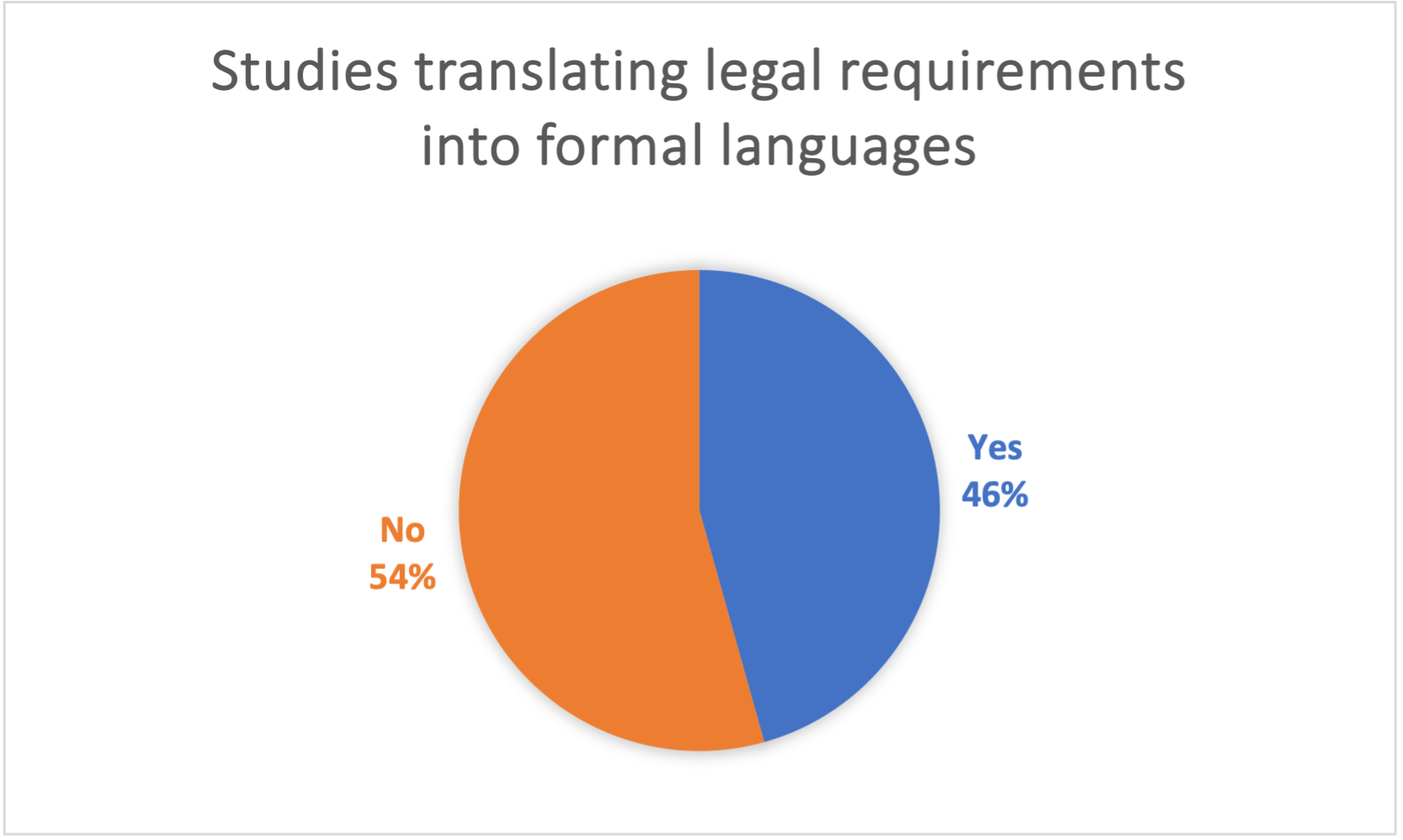}
    \caption{RQ3.1: Translation of regulatory documents.}
    \label{fig:results-rq31}
  \end{figure}

We collected information regarding the regulation used in each study (c.f. Section \ref{sec:rq2-answers}).  Only 21 primary studies (45,65\%)  performed translations of regulatory documents (c.f.: Figure \ref{fig:results-rq31}). The remaining 25 studies either referred generally to laws without providing details of the regulations studied or were theoretical studies.

\subsubsection{\textbf{Evidence for Q3.2 What type of compliance technology has been developed?}}

\begin{figure}[t]
  \centering
    \includegraphics[width=0.7\textwidth]{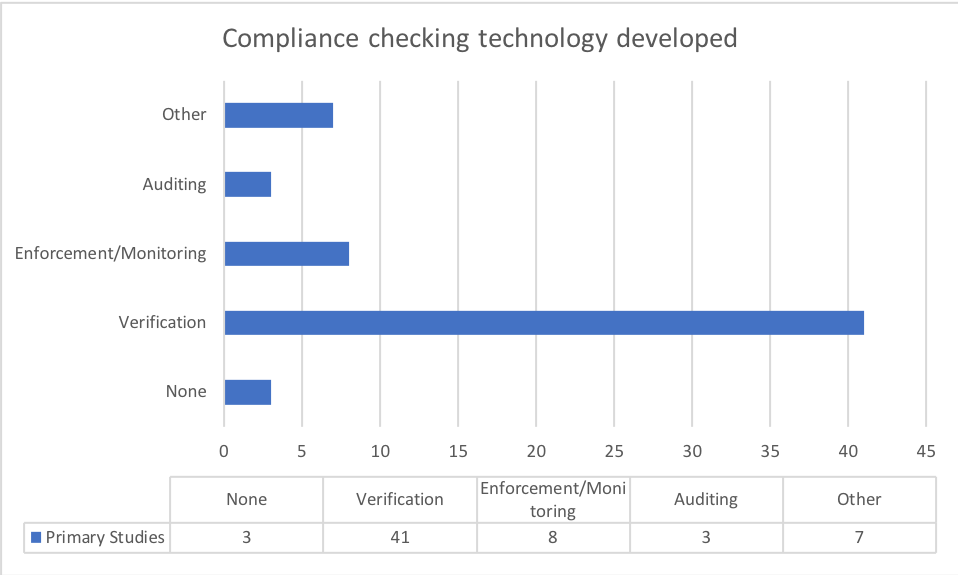}
    \caption{Compliance technologies developed}
    \label{fig:results-rq32}
  \end{figure}

A total of 43 studies (93,47\%) presented evidence for the development of compliance technologies for RQF. Figure \ref{fig:results-rq32} shows a distribution of studies versus the phase where compliance technologies have been applied. A study can apply compliance at more than one stage, for instance, by using the same framework to generate blueprints of compliant processes, that can be used in runtime monitoring. To summarize the types of technologies we will use the categories in Section \ref{def.compliance-checking-technologies}. Table \ref{table:verificationTechniques} provides an overview of each of the major categories used to check compliance verification in each of the frameworks studied. We summarize their most important details.

\begin{table}[t] \centering\footnotesize
\begin{tabular}{m{5cm}rm{7cm}}
\toprule
\textbf{Verification Technique}     &\textbf{Total}                    & \textbf{Study Identifier} \\ \midrule
\rowcolor[rgb]{ .867,  .922,  .969}Model Checking  &13                                & PS002, PS003,  PS008, PS020, PS025, PS031, PS053, PS055, PS056, PS060, PS063, PS064, PS067            \\
Compliance by Design &9                           & PS006, PS009, PS072, PS074, PS080, PS081, PS082, PS092, PS101              \\
\rowcolor[rgb]{ .867,  .922,  .969}Conformance Checking &6                          & PS011, PS016, PS017, PS029, PS035, PS089            \\
Compliance Checking &5                         & PS004,  PS014, PS018, PS030, PS047             \\
\rowcolor[rgb]{ .867,  .922,  .969} Theorem Proving  &2               & PS025, PS093            \\
SAT/SMT solving &2               & PS028, PS048            \\
\rowcolor[rgb]{ .867,  .922,  .969}Constraint Solving &1           & PS005            \\
Practical Reasoning   &1    & PS010            \\
\rowcolor[rgb]{ .867,  .922,  .969}Performance Analysis &1 & PS035            \\
Integrating Approaches &1     & PS071            \\
\rowcolor[rgb]{ .867,  .922,  .969}Massive Simulations                         &1            & PS095            \\
Compliance Distance                 &1           & PS104            \\
\rowcolor[rgb]{ .867,  .922,  .969}N/A                                    &5        & PS033, PS061, PS062, PS079, PS085    \\ \bottomrule
\end{tabular}
\caption{Techniques used in compliance verification }
\label{table:verificationTechniques}
\end{table}

 \textbf{Model Checking} is the most used technique in compliance verification collected in our study, with 28,26\% primary studies applying model-checking techniques. It requires two inputs: regulations encoded as a set of modal logic formulae, and a process model encoded in terms of a transition system. The model checker will exhaustively check that formulae in the specification hold after each transition, or provide counterexamples where properties are violated. The choice of logic may be a limiting factor depending on the kind of properties that can be expressed. Moreover, term alignment between laws and process activities is needed. Finally, limitations on the model-checking technique used may pose limits on the behaviors checked in a process model. Typical limiting factors include infinite state systems or undefined data variables, that can result in a state explosion problem~\cite{clarke2011model}. 

With \textbf{Compliance By Design (CbD)} techniques, the goal is that of generation of executions that do not violate laws.     
    The name "By Design" stresses the fact that the approach is primarily thought at the levels of compliance elicitation and design-time verification: all non-violating processes generate traces that are compliant. Despite their preventive focus, CbD could be used as blueprints of compliant processes, or as input for runtime monitoring (i.e.: PS072). We did not encounter evidence of CbD techniques generating code, or integrated into any Process Aware Information System, thus we consider studies in CbD frameworks only at the level of design phases.  CbD is a general technique, and it can be applied to different frameworks, (i.e.: PENELOPE, Regorous~\cite{governatori_journey_2009}, DECLARE, BPMN-Q, SEAFLOWS, COMPAS, AUDITING, BPC). We will summarize the works in logic-based CbD, and Petri-net CbDs. For the former, 
    % \cite{governatori_journey_2009}
    works using FCL/PCL \& Regorous (e.g.: PS006, PS009, PS072, PS074, PS080, PS081) define compliance as the property of the process to execute while not violating the laws in a regulation. It can be full compliance (the process will never violate at any point) or partial compliance (there is an execution that does not violate the rules).    The compliance checking approach requires 1. Identify the deontic effects of the set of modeled regulations, 2. determine the tasks and the obligations in force for each task, and 3. check whether the obligations have been fulfilled or postponed after the execution of a task. 
    % If there are no pending tasks to be executed and no pending obligations are to be fulfilled, then the process will be considered compliant.
    If this is the case, then the process is compliant.
    In the case of CbD approaches using Petri Nets, laws are not translated into logic, but onto a Petri Net (e.g.: PS101). Compliance is ensured if one can generate a plan that satisfies all rules (akin to logical consistency). The resulting rules will produce a Petri net for each instance. The final step is to merge the nets from each instance into a combined Petri Net for the entire process. Ensuring that such a net can be produced is sufficient to say that the process is compliant with regulations. Finally, data compliance has been modeled on top of Petri Nets. In PS092 artifact models, policies, and compliance rules are encoded into artifact-centric Petri Nets~\cite{lohmann2010artifact}. If the composition of models is weakly terminating, then it is possible to apply a process synthesis technique, thus guaranteeing that executions achieve their goals and preserve compliance.

 \textbf{Conformance checking \cite{rozinat2005conformance}}, a common technique in process mining, has also been used in compliance frameworks. In conformance checking, a reference model is created and compared against an event log. As a result, the checker returns the cost of aligning each trace and the reference model. Applying conformance checking to regulatory compliance requires that the reference model obeys the regulations. This is performed via the definition of imperative process models (e.g.: Petri Nets in PS011, PS016, PS017,  PS029, PS035), compliance patterns, or via declarative process models (e.g.: Multi-perspective DECLARE models in PS089). The event log complies with a rule if each trace in the log can be described by one of the compliance patterns available.      This technique is commonly known as \emph{log replay}. One of the advantages of process conformance is that it provides information regarding where a trace has deviated from the original specification, and, in its aggregated form, can help analyze the cost of adapting a given execution to comply with a regulation.

\textbf{Compliance checking} refers to framework-specific techniques that reduce compliance to a property verification technique where laws are logical properties that a system must satisfy. We will summarize them into protocol verification techniques, monitoring techniques, semantic annotations, and refinement.   The work in PS004 reduces the compliance checking problem to \textit{protocol verification}. Laws and bylaws are encoded as choreographies, while business processes are encoded as end-points. Compliance occurs if all the traces in the choreography are accepted by the parallel product of its endpoints.
    % As a limitation, the choreography language only presents sequences of actions (and no other modalities present in regulatory documents), and models have to be finite (no loops).   
    For \textit{monitoring-based techniques}, the work in PS018 introduces a compliant system as a finite state machine that can be permanently violated, sub-ideally violated, and compliant. Compliance checking examines whether a model starting in a compliant state can execute an activity that leads to a violation state or whether it can move into a state that can be eventually repaired. 
    % The problems of consistency and compliance checking are both resolved via encodings of process models and regulations in Color Petri Nets. 
    % The expressive power of eCRG allows modelers to describe compliance rules combining control flow, resource, data, and time constraints. 
    Regarding \textit{semantic annotations}, the work in PS014  extracts the effects generated by the execution of an activity, pairing them with the compliance checking algorithm of \cite{governatori2008algorithm} to determine whether the effects of executing activities  fulfills all the obligations expressed in the specifications. 
    For works on \textit{refinement}, PS028 and PS030 introduce compliance templates
    % developed according to legal texts, 
    where activities are implemented by a set of variants. In a variant, compliance checking checks that variant business processes do not introduce violations of an abstract business process. 

    \textbf{SAT/SMT and Constraint solving methods} can also provide solutions to compliance checking. A satisfiability problem (SAT) provides answers to the question of whether a boolean expression is satisfiable. 
    % This requires that there is at least one assignment to variables in the boolean expression that makes the whole expression true. 
    The work in PS028 applies \textit{SAT solving} in the following way. First,  it maps compliance rules to formulae in classical logic. These rules describe abstract business processes that may be later refined. 
    % Such a refinement could introduce conflicts due to constraint propagation. 
    Then, an SAT solver takes the (potentially refined) model and checks there are no direct conflicts, meaning that the refinement stages have not violated the compliance requirements. In the case of \textit{constraint programming}, compliance rules define behavioral constraints over traces, describing relations between process activities, subprocesses, and process context data (e.g. PS005).
    Constraints can be satisfied (the formula holds on a trace), violated (the formula does not hold on a trace), or violable (it is neither violated nor satisfied). 
    % This categorization allows to have a verification framework that covers both compliance verification and compliance monitoring. 
    Satisfied and violated constraints can be studied to understand how the process complies before execution, and violable constraints can be used as input for process monitoring tools. 
    Finally, \textit{Satisfiability Modulo Theories (SMT)} solvers have also been applied. 
    % has been paired with an analysis of security-compliant business processes in PS048. 
    In PS048, compliance verification uses a common framework to express security and legal requirements, encoding them into First Order Logic. It then checks whether a concrete Attribute-based Access Control Policy is compliant (e.g.: it is a behavioral refinement) with a general compliance rule, typically denoting an abstract law. 
    
    Other less common techniques have been proposed. We will summarize them below:
\begin{itemize} 
    \item \textbf{Theorem Proving}: 
    % One of the common pitfalls of compliance verification using model checking is its complexity: Deciding satisfiability in LTL is PSPACE-complete \cite{sistla1985complexity}, and the time complexity of the model checking problem is exponential in the size of the formula \cite{clarke1999model}. \hugo{What about complexity in LTL theorem proving? I don't have results}.
    Approaches based on theorem proving take a set of compliance requirements and generate all event traces that satisfy them. The first step is to check the satisfiability of the specification (c.f. Section \ref{q2.4Competencies}). Then, graph-based tableaux methods generate all possible compliant traces (i.e.: PS093). Security-related compliance frameworks (i.e.: PS025) use theorem provers to automatically generate attack sequences that violate the requirements established in a compliance regulation.
    
    \item \textbf{Compliance Distance:} Similar to conformance checking, it quantifies the cost of deviation from regulations at the process model level, not needing event logs. The work in PS104 presents a compliance distance as a quantitative measure of how much a process model may have to be changed in response to a given set of control objectives. For instance, a simplistic compliance distance may take a compliance rule as an input, and calculate the sum of all the checks required to make the rule satisfied, and all actions generated by it. 
        % Compliance distance can work as an indicator to denote how much will be the cost to recover a process instance that has deviated from their compliance objectives, and it can be used in runtime monitoring as a preemptive technique. 

    \item \textbf{Practical Reasoning:}
    The work in PS010 considers the relation of regulatory frameworks with goal-directed systems. Goals provide the reason to do something, and they can be refined to activities in a business process~\cite{lopez2007goal}. Practical Reasoning is an analysis technique directed toward actions \cite{bratman1988plans}. It includes two phases: deliberation - deciding what goals need to be achieved, and means-ends reasoning - determining how to meet them. 
    % To ensure compliance, regulations define the desired state of the world according to the normative context in which the system works. Regulations are composed via a set of logical operators, including deontic operators for permissions, obligations, and violations. 
    Monitoring compliance requires analyzing how pursuing a goal at a given state of the world might lead to a subsequent state that is admissible or not. The advantage of practical reasoning is that guarantees regarding compliance come at a higher level of abstraction (e.g. goals), rather than specific activities in the process.

    % The work in PS035 presents Petri-Net-based techniques for conformance and performance evaluation.
    \item \textbf{Performance Analysis:} derived from conformance checking techniques, performance analysis provides valuable diagnostic information regarding compliance with regulatory requirements that pertain to time constraints of business processes. For example, the work in PS035 uses timestamps to calculate parameters such as the latency between executed activities. The timestamps included in event logs paired with temporal constraints can provide information regarding measurements in implementing a process such as waiting, synchronization, and the sojourn times of a process. 
    % can be calculated by analyzing the time intervals between token production and consumption in the Petri net encoding the process. 

    \item \textbf{Massive Simulation Techniques:} This technique is used to analyze the behavior of a process model against general decision rules, including compliance requirements. The work in PS095 takes a business process model of different process instances by changing the data inputs following using the history of previous instances. Then it analyses how the outcomes of the process comply with the regulations. The result is constituted by a trace, and a final state describing the final point reached by the process instance.  The massive simulation technique collects the analysis of every case to synthesize a statistical answer back to the user.

\end{itemize}

  Finally, one of the challenges in the adoption of compliance frameworks is their technology dependence. The compliance techniques developed for one framework are difficult to port to another. The work in PS071 focuses on providing a language description where different models for regulations can be integrated. These models can come in different specification languages (e.g.: LTL and CTL), and act in different lifecycle phases (e.g.: compliance modeling vs. runtime monitoring). 
  % The results of their independent compliance-checking techniques can be aggregated by the use of compliance descriptors. 
    
\subsubsection{\textbf{Evidence for Q.3.3. Type of verification analysis implemented}} In this section we discuss the type of compliance analysis technology (if any) that has been implemented in the RQF. Notice that by \textit{analysis}, we mean the artifacts given to analysts regarding the compliance analysis. The categories correspond to those defined in Section \ref{def.compliance-analysis-technologies}.

\begin{figure}[t]
  \centering
    \includegraphics[width=0.7\textwidth]{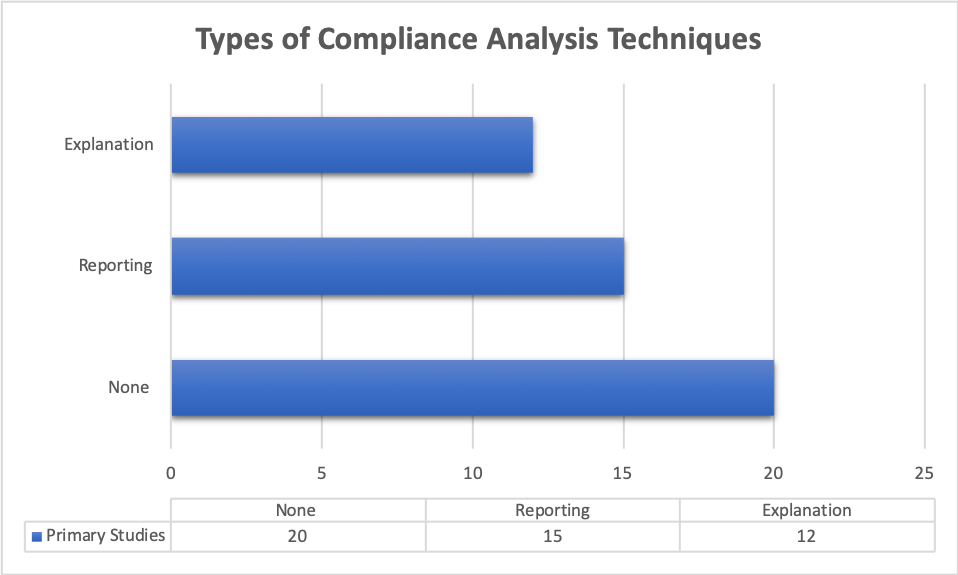}
    \caption{Types of compliance analysis developed}
  \end{figure}

  A significant number of the primary studies (43,47\%) did not present a compliance analysis technique. Among these percentages were works that produced only yes/no answers to compliance checking (for instance, those works based on SAT solving). The remaining 56,52\% mainly provided feedback in the form of counterexamples, root causes, and compliance distance.
  Compliance explanation may come as a by-product of the underlying verification techniques used. In the case of model checking techniques, verification results may come either as \textbf{compliance reports} describing which rules can be violated in a process execution (e.g.: PS017), or it might provide \textbf{counterexamples} showing traces that violate compliance rules (e.g.: PS031, PS053, PS055, PS056, PS063, PS067, PS072, PS074, PS092, PS095). While counterexample-based explanations provided by model checkers may return traces that are possible in theory, some of these traces might not be commonly pursued in reality. The work in PS010 uses \textbf{goal deliberation} \cite{braubach2004goal} to decide which goals should be actively pursued, which ones can be delayed, and which ones can be abandoned. 

  % Other techniques that generate counter-example reports include massive simulations (e.g.: PS095) and conformance checking.
  \textbf{Root-cause analysis }(PS002, PS004, PS060, PS063) is a useful explanation technique to identify the reasons why processes violate compliance rules. While standard compliance techniques report the set of compliance rules that are not satisfied by the business process, the analyst is left to fix all the non-compliant rules alone. Root-cause analysis (originally presented in \cite{elgammal2010root,elgammal2012using} and used in PS002, PS004) builds a constraint taxonomy of compliance rules. The taxonomy maps the violations of compliance rules to patterns that can be observed in the possible event traces. The hierarchical levels in the taxonomy allow the method to identify key causes (the roots) that, if resolved, will efficiently resolve several violations. 
  
  The works based on conformance checking presented different types of reporting strategies. The simplest strategy is the identification of the initial events that violate rules (PS028). Other works apply \textbf{compliance distance} (PS016, PS104). This technique visually represents how existing event logs deviated from a compliance rule. Deviations include log-to-model deviations (e.g.: an event occurred that did not comply with the rule) and model-to-log deviations (e.g.: an event skipped in the log such that the log does not comply with the rule). The work in PS018 presents a set of metrics used to calculate the degree of compliance of an event log. These include the  \emph{compliance rate} and \emph{critical rate}. A compliance rate compares how many rules were satisfied against how many rules were activated in an event log. In contrast, the critical rate only considers \textit{conflicting rules}, that is, those in a process model that can be violated, and those that are active, but not satisfied (i.e.: \emph{pending}). When dealing with timed rules, a violation may occur when an event is executed outside legal time constraints. The works in PS029 and PS104 use compliance distance to present reporting strategies for temporal violations, including \textit{means and maximum times} for violations of compliance rules.

\subsubsection{\textbf{Evidence for Q.3.4: Types of verification enactment implemented}} The monitoring of compliance rules is necessary in the moment models start being used as part of a process-aware information system as execution components. In such cases, a process may deviate from the ideal intent that a compliance rule dictates, and it is not sufficient to only find out whether or not the rule is violated over a process but to provide methods to recover from such a violation. For example, the rule ``every case needs to be answered within 1 week'' may be hard to uphold if the caseworker assigned to the case is overloaded with work.  In this section, we discuss the type of compliance enactment technology implemented in our primary studies, if any. The categories correspond to the ones defined in Section \ref{def.compliance-enactment-technologies}.  Figure \ref{fig:rq34} illustrates the presence of verification enactment techniques. The great majority of the primary studies considered (93,47\%) did not include compliance enactment techniques. We discuss the four techniques considering enactment techniques below:
  
  \begin{figure}[t]
  \centering
  \includegraphics[width=0.7\textwidth]{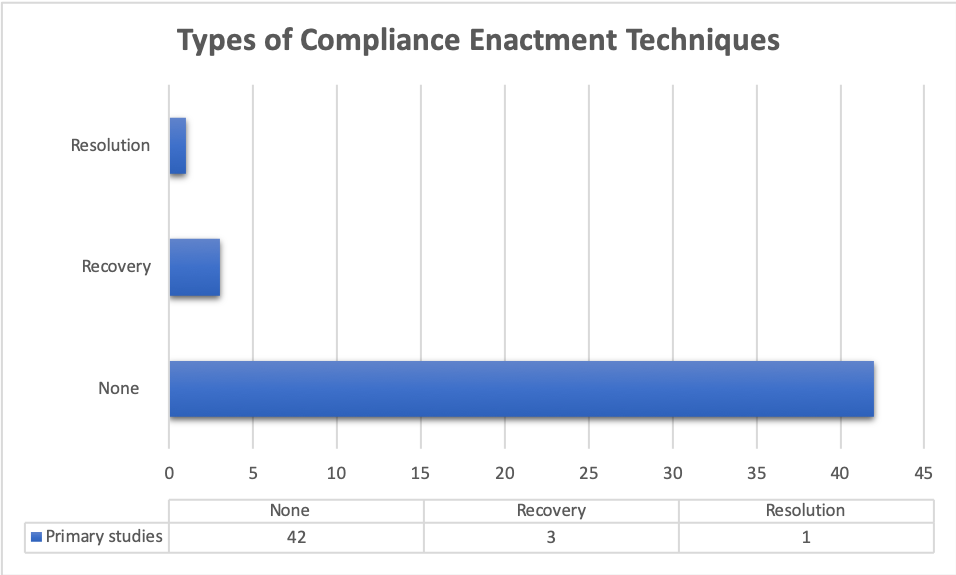}
    \caption{RQ3.4: Types of compliance enactment techniques developed}
    \label{fig:rq34}
  \end{figure}

\textbf{Healable violations}. A violation is healable if and only if a group of process change operations that applied to the current process instance brings the resulting instance back to a non-violation state. Healable violations are introduced in the SeaFlows framework (e.g.: PS005). The work in PS047 considers healable violations for eCRG in the context of monitoring. In their work \cite{knuplesch2012ensuring},
    % To ensure a priori process compliance, eCRG proposes model-checking techniques based on LTL model checking. Formulae derived from eCRG contain compliance and violation states. 
    process instances are checked against a compliance monitor that determines whether a given compliance rule is violated and, in case of violation, whether such a violation can be repaired. 
    % Finally, eCRG implements the compliance auditing technique presented in \cite{van2005process} by applying transformation of LTL rules to generate a  state-based automata, that can be used to check whether the moves in the event log behave according to the automata or not. 
    % The work in PS047 only presents the implementation of process monitoring techniques at the moment.
A similar approach is followed in PS082 using ExPDT as a policy language,
% that will be used to certify the adherence of process models into regulatory documents. Compliance resolution is based on Compliance-by-detection: first, the framework verifies that the policies generated can be enforced. 
that is later fed into runtime monitoring frameworks, such as REALM \cite{giblin2006regulatory}. 

\textbf{Sanction-aware recovery strategies}. Recovery in ExPDT (PS082) takes an approach closer to risk management. Compliance rules are stratified according to their sanctions. While in an ideal context, the execution of a business process does not incur any sanctions, in some contexts it will be necessary to deviate from the ideal flow to maximize a business goal. Such deviation will be penalized with a sanction. In this case, the compliance monitor will include the execution of the sanction as part of the pending events in the process. 

\textbf{Compliance resolution techniques}. Due to their formal definition, the negation of a compliance rule has a set of traces that satisfy it. The collection of such traces is known as a violation pattern. Each compliance rule (precedence rules, implication rules) contains a set of violation patterns, and
% and their existence in a trace will deem the process execution non-compliant. 
each violation pattern has a recovery strategy (add, delete, or move activities in the process) that can modify the behavior of the process model to bring it back to a state where a violated compliance rule is satisfied. As more than one recovery strategy may be possible, a \emph{resolution context} is required to determine the most suitable strategy (PS079).

\subsection{Research Question RQ4: What is the degree of \textbf{flexibility} considered in existing
  \textbf{regulatory compliance frameworks}?}

To answer this question, we considered two types of flexibility. First, the degree of flexibility that comes from frameworks that accept changes in regulations, and second, the degree of flexibility when reasoning about laws with conflicting statements. We will describe them below.
  
  \subsubsection{Adaptation to changing laws}

  Laws are not monolithic structures, they are drafted, approved, and subject to changes. Sometimes the changes are syntactic, but oftentimes they include semantic changes such as the introduction of new rules, the modification of existing ones, or the abrogation of ruling paragraphs. 
  % In compliance, the notion of change may be treated specially, as it can describe a) whether a law paragraph has been modified, or b) whether the outcome of the compliance checking framework changes in the presence of two or more conflicting compliance rules. It is fair to say that none of these two aspects are well supported by the primary studies collected.
  We collected information regarding considerations for law changes in our primary studies, and only 13,04\% supported them. In most cases, adapting a legal change requires an \textbf{adaption process}: the change in the law generates a new specification that needs checking against the existing process. Here the process can continue being compliant or it may require iterative interventions. After the interventions, the modified process will be compliant with the new version of the law. In the case of CbD approaches (PS092, PS093, PS101), a change in regulations generates, modifies, and deletes compliance rules, and the process templates generated from them will be different from the initial processes. No considerations are placed on the adaptation of existing process instances (compliant with previous versions of the law) against new laws.  
  
  An interesting aspect of the legal changes is their \textbf{consistency}, as explored in  PS011. Changes in the regulation generate new compliance rules, and compliance checking considers both the old and new rules. The technique to check for consistency involves merging them into a consolidated model (e.g.: the parallel product of a Petri Net) and determining whether the resulting model is satisfiable (or in the case of a Petri Net, whether it has enabled transitions).

  \begin{figure}[t]
  \centering
    \includegraphics[width=0.7\columnwidth]{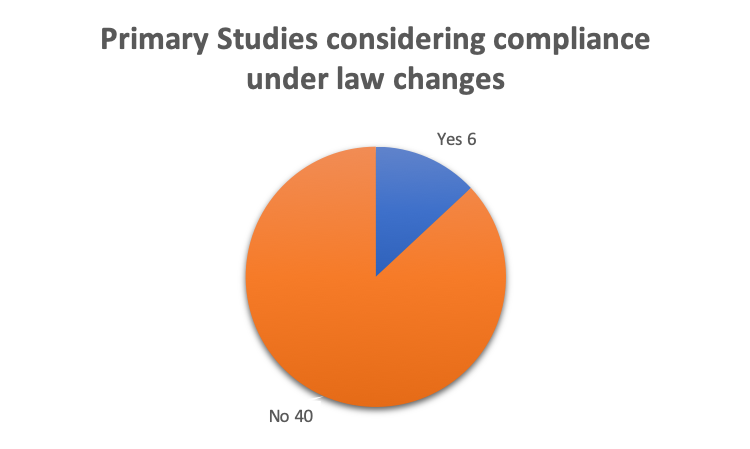}
    \caption{RQ4: Number of frameworks considering changes in law}
    \label{fig:results-rq4}
  \end{figure}
  
  \subsubsection{Conflict Resolution techniques}
\begin{figure}[t]
  \centering
    \includegraphics[width=0.7\columnwidth]{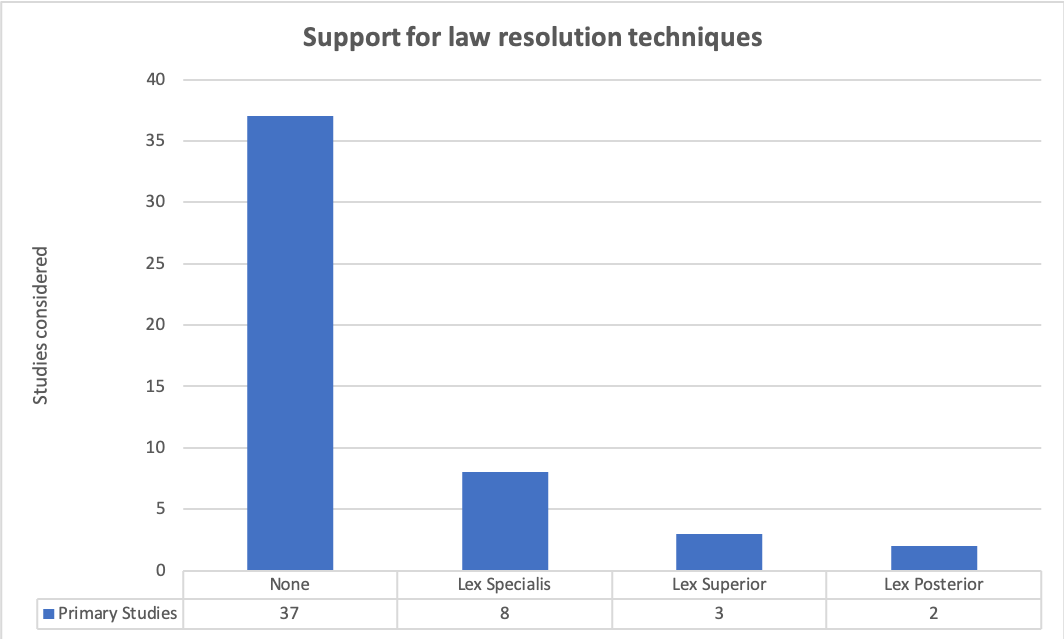}
    \caption{RQ4: Considered Resolution Techniques}
    \label{fig:results-rq4-1}
  \end{figure}

  The second scenario is when multiple rules of \textit{the same version} of the law are inconsistent. This is particularly important when a process is checked for compliance against a set of rules, composed together. Gordon et al. \cite{gordon_rules_2009} present some of the most common problems affecting the consistency of legal systems. Compliance rules $A$ and $B$ might be in conflict if $B$ is an exception of $B$ (or vice-versa), if there is a superiority relation making $A$ rule over $B$, or if $A$ and $B$ have been enacted at different points in time. The term \emph{conflict resolution} is generally used to group the set of techniques that break inconsistencies of legal systems, and they can be divided into: 
  
  \begin{itemize}
      \item \emph{Lex specialis:} If $A$ and $B$ are conflicting compliance rules, and $A$ is more specific than $B$, select $A$ and discard $B$.
      \item \emph{Lex superior:} If $A$ and $B$ are conflicting compliance rules, and $A$ is associated to a superior authority, then select $A$ and discard $B$.
      \item \emph{Lex posterior:} If $A$ and $B$ are conflicting compliance rules, and the time of enactment of $A$ is newer than the time of enactment of $B$, select $A$ and discard $B$.
  \end{itemize}
  Figure \ref{fig:results-rq4-1} shows the distribution of works considering conflict resolution techniques. While most works did not exhibit mechanisms to deliberate about conflicting goals, we have 17,39\%, 8,60\%, and 4,34\% who consider one or multiple scenarios. We will discuss them below:

  \textbf{Resolution on specialized laws (lex specialis)}  Linear Temporal Logic with non-monotonic operators have been introduced in CRL~\cite{baral2007non}. Formulae in this language include operators describing exceptions: $[\langle r \rangle] \phi$ means that $\phi$ is true in most cases, except for $r$. 
  % CRL constructs judgments by describing the main rule (applicable in most cases) linked to its exceptions. 
  % This operator allows us to combine formulae including different exceptional cases without leading to inconsistencies, 
  The work in PS002 shows an example, $[\langle GoldCustomer \rangle] checkCustomerBankPrivileges$ says that all process instances should check for customer bank privileges 
  % (subformula $checkCustomerBankPrivileges$) 
  unless the exceptional case is where the customer has a gold label.
  % (that is, the predicate $GoldCustomer$ is satisfied). 
  Exceptional behavior can be modeled in combinations of Dynamic Linear Temporal Logic and Answer Set Programming.
  % combination of Answer Set Programming (ASP) and Dynamic Linear Temporal Logic to describe exceptional behavior. The approach presented in PS009 uses a 
  % First, non-monotonic rules can be modeled via ASP's default negation.
  For instance, a \emph{persistency law} can be represented as ASP's default negations, as described in PS009: $\square([a]l\leftarrow l, not [a]\lnot l)$ means that if $l$ holds in a state, then $l$ will continue to hold after executing $a$, only if it can be assumed that $\lnot l$ does not hold in the resulting state. Stratified clauses can also represent exceptions. Works based on FCL (e.g.: PS014, PS080, PS081) use this approach.
  A rule in FCL is written as $r \colon A_1, \ldots, A_n \Rightarrow_O P_o \otimes P_r \otimes P_f$ and says that under facts $A_1, \ldots A_n$ the specification should fulfill the primary obligation  $P_o$. If this cannot be fulfilled (for instance, because $P_o$ is violated, or because $P_o$ is not done), then $P_r$ becomes a replacement obligation. If none of these can be fulfilled, then $P_f$ becomes obligatory.
   The syntax of rules in ExPDT (PS082) allows users to define the degree of freedom in the observance of rule actions. Users are always allowed the permitted actions, but they can opt to disregard obligations. If that is the case, the language allows the definition of compensation obligations to take effect, which can be disregarded based on the type of compensation.

  \textbf{Resolution based on jurisdictions (Lex Superior)} Studies: PCL and FCL (PS009, PS080, PS081) also include a defeasible component to resolve conflict between conflicting rules. The composition of rules $r_1 \colon A \Rightarrow_O C$ and $r_2 \colon A \Rightarrow_O \lnot C$ does not lead to a contradiction (i.e.: the logic is \emph{skeptical}), and in the case there is an ordering relation $\prec$ between rules such that  $r_1 \prec r_2$, then $C$ is concluded.

  \textbf{Resolution based on temporal aspects (Lex Posterior)}: Only two studies considered timing in laws. In  PS010, each norm has a state that compares against the state of the world. Part of the considerations regarding norm states include the time of validity. Laws can be \emph{deontically contradictory} when the obligations imposed by two given rules contradict each other. In contrast, FCL implements lex posterior by adding timestamps to propositions in the language, adopting the persistence mechanism in \cite{governatori2005temporalised} to treat temporal rules. This reasoning involves determining whether a rule can be concluded at the given time instant, whether there are newer rules that make the rule invalid, or whether the rule is permanently valid.

\subsection{Research Question RQ5: How \textbf{mature} is the method/tooling available, from the
    evidence present in research literature?}
    
\begin{figure}[t]
    \centering
    \includegraphics[width=0.7\columnwidth]{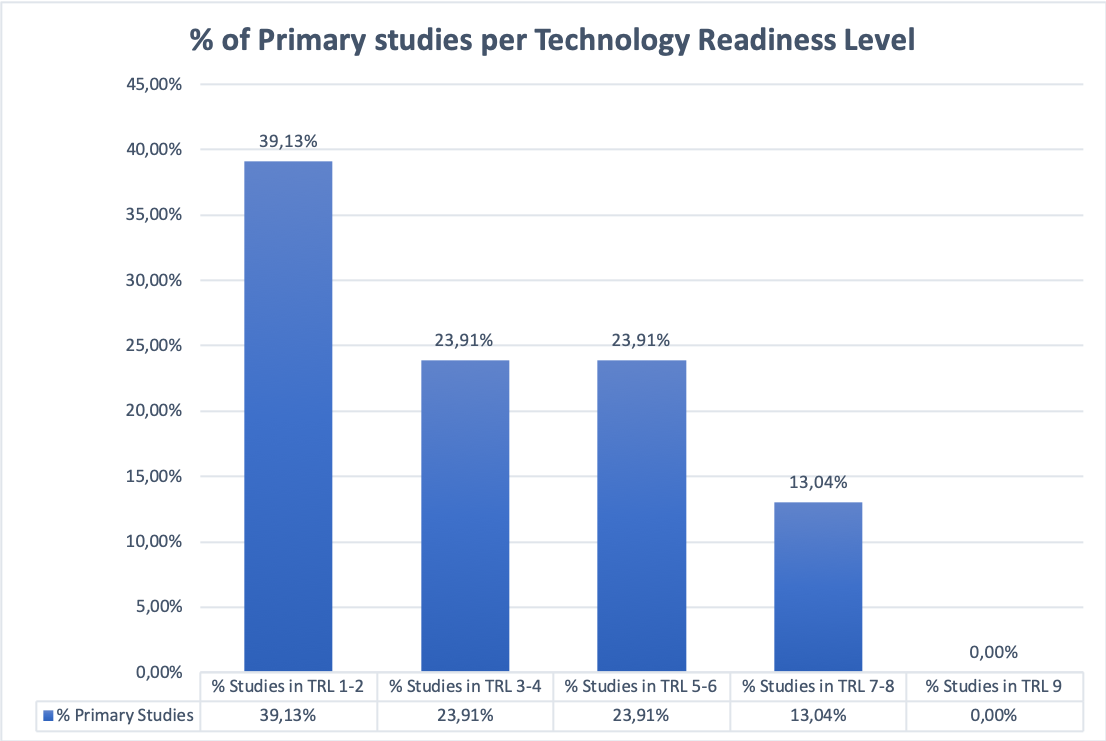}
    \caption{Technology Readiness Levels: Studies aggregated per TRL level}
    \label{fig:TRL}
\end{figure}

    In this analysis, we limited our results to the execution of the research protocol, which does not include either grey literature, or consider versions of the frameworks after publication. Table \ref{tab:TRL} lists each study according to their technology readiness level. Figure \ref{fig:TRL} provides an aggregated view: Most studies collected (39,13\% of the total) considered only theoretical frameworks, and, from the evidence collected, we cannot deduce that any of the proposed frameworks supports organizations in their compliance tasks, although some effort has been placed in moving from theory to practice and empirical evaluations (23,91\% of the total).

\begin{table}[t] \centering \footnotesize
\begin{tabular}{m{1.2cm}rm{1cm}} 
\toprule
\textbf{Pseudonym} & \textbf{Primary Study}     &\textbf{TRL} \\ \midrule
\rowcolor[rgb]{ .867,  .922,  .969}
S01  &	PS002  & TRL 5-6\\
S02	 &	PS003& TRL 3-4\\
\rowcolor[rgb]{ .867,  .922,  .969}S03	 &	PS004& TRL 3-4\\
S04	 &	PS005& TRL 3-4\\
\rowcolor[rgb]{ .867,  .922,  .969}S05	 &	PS006& TRL 7-8\\
S06	 &	PS008& TRL 3-4\\
\rowcolor[rgb]{ .867,  .922,  .969}S07	 &	PS009& TRL 1-2\\
S08	 &	PS010& TRL 1-2\\
\rowcolor[rgb]{ .867,  .922,  .969}S09	 &	PS011& TRL 5-6\\
S10	 &	PS014&TRL 1-2 \\
\rowcolor[rgb]{ .867,  .922,  .969}S11	 &	PS016& TRL 7-8\\
S12	 &	PS017& TRL 7-8\\
\rowcolor[rgb]{ .867,  .922,  .969}S13	 &	PS018& TRL 5-6\\
S14	 &	PS020&TRL 1-2 \\
\rowcolor[rgb]{ .867,  .922,  .969}S15	 &	PS025& TRL 5-6\\
S16	 &	PS028& TRL 1-2\\
\rowcolor[rgb]{ .867,  .922,  .969}S17	 &	PS029& TRL 7-8\\
S18	 &	PS030&TRL 1-2 \\
\rowcolor[rgb]{ .867,  .922,  .969}S19	 &	PS031& TRL 1-2\\
S20	 &	PS033&TRL 1-2 \\
\rowcolor[rgb]{ .867,  .922,  .969}S21	 &	PS035& TRL 3-4\\
S22	 &	PS047&TRL 5-6 \\
\rowcolor[rgb]{ .867,  .922,  .969}S23	 &	PS048& TRL 3-4\\
S24	 &	PS053& TRL 5-6\\
\rowcolor[rgb]{ .867,  .922,  .969}S25	 &	PS055& TRL 1-2\\
S26	 &	PS056& TRL 3-4\\
\rowcolor[rgb]{ .867,  .922,  .969}S27	 &	PS060& TRL 5-6\\
S28	 &	PS061& TRL 5-6\\
\rowcolor[rgb]{ .867,  .922,  .969}S29	 &	PS062& TRL 1-2\\
S30	 &	PS063&TRL 3-4 \\
\rowcolor[rgb]{ .867,  .922,  .969}S31	 &	PS064& TRL 3-4\\
S32	 &	PS067&TRL 3-4 \\
\rowcolor[rgb]{ .867,  .922,  .969}S33	 &	PS071& TRL 7-8\\
S34	 &	PS072&TRL 5-6 \\
\rowcolor[rgb]{ .867,  .922,  .969}S35	 &	PS074& TRL 5-6\\
S36	 &	PS079&TRL 1-2 \\
\rowcolor[rgb]{ .867,  .922,  .969}S37	 &	PS080& TRL 1-2\\
S38	 &	PS081&TRL 1-2 \\
\rowcolor[rgb]{ .867,  .922,  .969}S39	 &	PS082& TRL 1-2\\
S40	 &	PS085&TRL 1-2 \\
\rowcolor[rgb]{ .867,  .922,  .969}S41	 &	PS089& TRL 5-6\\
S42	 &	PS092&TRL 1-2 \\
\rowcolor[rgb]{ .867,  .922,  .969}S43	 &	PS093& TRL 3-4\\
S44	 &	PS095&TRL 7-8 \\
\rowcolor[rgb]{ .867,  .922,  .969}S45	 &	PS101& TRL 1-2\\
S46	 &	PS104  &TRL 1-2 \\\bottomrule
\end{tabular}
\caption{Technology Readiness Level: List of Studies}
\label{tab:TRL}
\end{table}

%%% Local Variables:
%%% mode: latex
%%% TeX-master: "main"
%%% End:

\section{Interpretation of the Results}
\label{sec:interpretation}

Considering the results gathered in the previous section, we propose the following statements as conclusions for each question. 

\subsection{Regarding RQ1: Current structure of Compliance Frameworks} All compliance frameworks analyzed considered the process-level abstraction level, while only $23,9\%$ considered strategic levels, $21,7\%$ considered post-mortem phases, and $8,6\%$ considered the execution phase. Our reading for these results is that compliance as an activity at the design level is well understood, while provenance, execution, and post-mortem phases would benefit from further studies. In particular, we consider particularly worrying the low coverage exhibited at strategic levels. Converting legislation into formal specifications is a difficult task that will be necessary to promote the acceptance of compliance frameworks outside the lab. We believe dual-coding methods and tools where formal specifications can be traced to regulatory documents will be necessary for increasing the trust of non-computer science participants in the framework. Moreover, there may also be a good opportunity to employ process mining technologies more. The field is mature and techniques for process discovery and conformance checking can allow for the alignment of datasets of multiple executions to be checked against normative processes.  Finally, we found process execution particularly absent in the works considered. Since process-aware information systems (BPMS) take process models as input, there is an opportunity to integrate compliance-checking frameworks in BPMS to guarantee compliant executions.

\subsection{Regarding RQ2.1: The technologies used to formalize regulatory documents} We observe that there is no consensus on the best language to formalize regulatory documents. While the use of LTL and Petri-Net variants is common in the community, there is a risk that the formalisms used do not capture the semantics of the laws, that have deontic and defeasible elements. Alternatives such as FCL and PCL seem to be interesting proponents, but it is not clear what their affordances are compared to more established languages. Here the community working in formal compliance frameworks would benefit from comparative expressiveness studies between the different languages, which would enable the creation of automated translation techniques between the diverse specification languages used.

\subsection{Regarding RQ2.2. On Process Modelling Languages Used in Compliance} The results from the primary studies present a dichotomy. First, there is a clear preference for BPMN as the industry standard, with 16 primary studies focusing on its use. This strength reflects research with direct application in the toolset most process modelers are familiar with. On the other hand, with more than 130 notational elements, it may be hard to reason over the full standard, and most of the existing formal semantics consider only fragments of the language~\cite{christiansen2010formal,corradini2022bpmn,dijkman2007formal}. The lack of fully formalized semantics for BPMN makes replicating results between different frameworks difficult. We believe a potential improvement (and simplification of the compliance frameworks) will be the utilization of declarative process models like DCR graphs~\cite{hildebrandt2011declarative} or Declare~\cite{pesic2007declare}. In declarative process models, non-compliant behavior is ruled out already at the design phase, making the alignment phase between process models and logical specifications unnecessary. Possible applications of declarative process models may also include them as abstract business processes. We pursued the use of DCR graphs as a basis for a compliance (and execution) framework in the EcoKnow project.
On the other hand, most techniques in compliance verification disregard the particular modeling language and focus on an abstract model (e.g. a label transition system) and a property. 
The study of parametric frameworks for compliance where particular modeling languages could be instantiated (as proposed in PS071) may foster the adoption of compliance frameworks by not relying on a specific language.

\subsection{Regarding RQ3.1: On the formalization of regulatory documents and their real-world applicability} The information gathered is inconclusive in this aspect. While we could identify the use cases with references to legal codes, acts, guidelines, and bylaws, it was not clear the degree of formalization of each of the legal artifacts, and whether such formalizations were constructed in agreement with legal professionals. In multiple papers, we found the mention of legal documents as inspiration for a formalization, but the reasons behind the choice of particular paragraphs were missing. Normally, a legal text is tens to hundreds of pages long and contains different types of information (definitions, provisions, rules, supplementary material).  What aspects of the legal texts can be formalized now and to which degree? This is an open question after this review. 

A second aspect we considered was the role of ambiguity. Legal documents are complex to parse even for legal practitioners, and sentences in legal documents sometimes contain ambiguous texts. These ambiguities are not errors but design choices where a legal practitioner allows multiple interpretations in a text~\cite{franceschetti2023characterisation}. In formal verification, ambiguities may generate inconsistent specifications and are rejected in formal frameworks, as they generate contradictions that invalidate logical reasoning. A way of reconciling these two different views will be required for compliance practitioners to adopt formal regulatory compliance frameworks.

Concerning the processes studied, we see compliance as an area that has captured interest in Knowledge-Intensive Processes~\cite{di2015knowledge}. Sectors like healthcare, banking, and public services depend on discretionary decisions by participants in the process, where externalities may affect the decisions taken. Surprisingly enough, most process modeling languages used for representing these processes, such as BPMN, EPC, or WS-BPEL are imperative, which is known to be not as well suited to processes with a high degree of variance and need for flexibility~\cite{fahland2009declarative}.

\subsection{Regarding RQ3.2.: The competencies required in compliance frameworks} The results of Q2.4 shed light on a set of new competencies that stakeholders involved in the formal use of regulatory compliance frameworks need to exhibit to benefit from these techniques. First, compliance officers, with their prior background in law and compliance, are ideal candidates to perform the selection of regulation documents, identify risk management considerations, and refine regulatory documents to set compliance controls. However, they may benefit from formal training in mathematical and logical modeling to be able to contribute in the formalization of controls into a set of formulae, as well as the bug-finding process involved when ambiguities, incompleteness, and inconsistencies in the formalization occur. For the domain specialist, basic skills in business process modeling are necessary. To do formal compliance analysis using the present tools, additional skills in identifying process variants, refining global/local business processes, model checking, and composing multiple interconnected processes are also needed. Some activities such as terminology alignment, formalization, identification of activity effects, and solving incomplete or ambiguous specifications will require an interdisciplinary team of lawyers and computer scientists.

While Q2.4 illustrated the manual activities involved in this process, it also opened opportunities for emerging technologies to support some of these skills. In particular, recent advances in Large Language Models for laws similar to LegalBERT~\cite{chalkidis2020legal} may help to speed up the extraction of formulae from legal requirements. Techniques such as \cite{DBLP:conf/icail/ZinNSSN23} are particularly promising, however, they need to be further evaluated, as other studies exhibit negative results \cite{schrack2022can,lopez2021challenges}. Other tasks that can benefit from approaches in NLP include the identification of paragraphs and ambiguities and the annotation of possible anomalies in textual descriptions~\cite{caspary2023does}. Also from the perspective of modeling, possible cross-fertilization with theories of programming and modeling languages may help, as choreographies and their endpoint projectability have been amply studied in the last decade~\cite{carbone2013deadlock}, with new developments allowing the generation of local process models that correspond to their choreographic views \cite{hildebrandt2019declarative,hildebrandt2023declarative}.

\subsection{On the gaps not covered in current compliance frameworks}
The most interesting conclusion that we can draw from the collection of future work is the need for further empirical studies to test the application of compliance frameworks with actual users and real laws. This also coincides with the early phases in the technology readiness levels for compliance frameworks. At a language expressiveness level, there is an identified need for extending compliance policies with data, time, and other legal and process-level notions, such as communications between participants, loops,  sanctions, or power relations. In terms of technologies, there seems to be a growing interest in implementation levels of compliance, where runtime monitoring may help to uncover the state of running cases.
We have pursued these open gaps in the EcoKnow project, doing empirical understandability studies with users~\cite{ABBADANDALOUSSI2023120924} and extending the capabilities of the DCR Graph formalism to handle data, DMN decision modelling~\cite{10.1007/978-3-030-94343-1_28} and time following the ISO 8601 standard. This has taken the technology to TRL 9 and is being used widely in governmental institutions in Denmark.

%\hugo{Comment on current enablers and technologies for object-centric process mining (for data and time relations), runtime monitoring. Mention Ecoknow as one example of high-TRL projects evidencing the impact of compliance technologies}

\subsection{Regarding RQ3.3.: The impact of formal regulatory compliance frameworks} Here the works coincide on the potential benefits of regulatory compliance frameworks based on formal methods. Their potential benefits are to produce business process models that align with the requirements coming from legislation, to monitor that process executions adhere to compliance policies, and to suggest changes in existing processes to make them compliant. Data from question Q.3.1. shows that there is still some gaps between the work in compliance and real application in actual laws, as more than half of the works have yet to model a legal document. Regarding the possible impact of existing frameworks, question Q.3.2 shows that most of the technologies developed operate in the design space, and less so at runtime phases. This hints at a possible application of compliance frameworks in the design of correct process-aware information systems. Technologies such as model checking, compliance by design, compliance checking and SMT solving may particularly be used here. Some of them (e.g.: model checking, SAT/SMT/constraint solving) are mature technologies that can scale extremely well in systems with finite states. However, when looking at future work, we see trends requiring careful thought on the application of these technologies. Extensions to cover compliance against data, time and communications may easily blow up the state space~\cite{clarke2011model}. Here it may be useful to consider techniques for symbolic model checking~\cite{burch1992symbolic} to reduce the number of considered states. Data in Q.3.2 also evidenced the uptake of conformance checking in compliance, and we anticipate further developments in conformance checking techniques in declarative languages such as Declare and DCR graphs, that can map directly legislation instead of normative process models. An important consideration is that the underlying technique in conformance checking, process alignments~\cite{dunzer2019conformance}, is an expensive technique to compute, thus further development in heuristics and approximations may be needed. Outputs need to be further considered: almost half of the works considered only showcased yes/no answers, which does not scale well when a business process has multiple executions. At the design level, techniques such as compliance by design may benefit from further studies in root-case analysis, while at the execution level, studies on conformance checking may consider reports at the trace level, but also in terms of aggregated traces. Here, we foresee three aspects to have more attention in the future: first, the role of data and temporal constraints in object-centric conformance checking, second, the interrelation between multiple process instances, and third, the need for the identification of possible non-conforming traces in real-time. For enactment, we see that very few works (9,53\% of the studies) considered enactment technologies. This is a call for further development in this field, which in our opinion is one of the strongest needs for the industrial application of compliance technologies. Here techniques such as online conformance checking \cite{van2019online} will become particularly useful. Moreover, providing proactive information about the possible healable violations may help identify not only the cases at risk of non-compliance but also the set of strategies that participants in risky traces may perform to achieve compliance. 

\subsection{Regarding RQ4. The flexibility in formal regulatory compliance frameworks?} Our data discussed two types of sources of flexibility, those from adopting changes in laws, and the second on dealing with conflicting rules. We see that 9,53\% of the works still consider compliance as a one-shot activity, which places risk in the adoption of compliance frameworks. Modeling a law in terms of a formal specification requires a considerable amount of resources, and there is a risk that some of the initial assumptions may have changed at the end of the formalization. Here traceability techniques in the elicitation and alignment between laws and models may support the robustness of compliance frameworks. The second aspect we discussed was that of conflict resolution techniques.  Here we see that 80,43\% are yet to consider that legal frameworks have implicit sources of inconsistencies and that logical frameworks need to account for them. Further work on supporting conflict resolution techniques in specialized laws, jurisdictions, and temporal aspects would be necessary at the strategic, analysis, and design phases. For the enactment phase, we suggest supporting users with tools that allow them to work discretionary (e.g. by choosing the next activity being performed according to one of the multiple rules in place) would be better than constraining them onto one given choice.

\subsection{Regarding RQ5. How mature are the methods/tools in compliance?}

Our data shows that compliance has been studied from the point of view of multiple laws at different levels: some of the implications outside an organization  (e.g. legal codes and regulatory acts) and others that may represent internal controls defined by each organization (e.g.: guidelines and bylaws). The use cases evidenced some sectors particularly interested in compliance, namely finance, e-commerce, healthcare, and the public sector. Concerning the maturity of the tools and projects, we see that 63,04\% of frameworks are at levels A-D of the technology readiness levels, which means conceptual solutions with artificial examples. At the same time, we saw only 13,4\% of the studies providing evidence on the application of the frameworks in operational settings, with none of them providing long-term analysis of their impact on the adopting organizations. This challenge seems to correlate with the need for further empirical studies in future works.

\subsection{Some additional Research Challenges}
During the creation of this SLR we identified some additional challenges not foreseen in the beginning, we will discuss them below:

\textbf{Where are the lawyers?} A marked absence of evidence of the use of existing compliance frameworks by compliance specialists (lawyers, consultants, etc). This might be because most of the existing compliance frameworks require compliance officers to know a type of mathematical logic, which is far from their background, and whose training takes time. There is also little efforts in integrating views that are common for legal practitioners (laws) and formal models. In addition, very few primary studies included lawyers as co-authors, which suggests the frameworks yet need to be validated by peers with more expertise in legal systems.

\textbf{How to test a framework?} We lack a benchmark dataset that can be used for comparison among different compliance frameworks. For replicability studies we will need a. the identification of laws that are of interest to a larger community, b. a set of examples of violations and accepting cases, c. models and traces in sectors where these laws are necessary, and d. the expected consequences for compliant/non-compliant behavior. Excepting processes in the banking sector, there is almost no process that has been repeated. Even the examples in the banking sector lack uniformity (the process has variants). Then it is difficult to replicate the same experiments and perform comparisons. We believe that collecting such cases and performing a challenge such as the ``BPI Challenge" or similar will be beneficial. Here the General Data Protection Regulation could help as a baseline, as is a law affecting a large market, and processes in multiple sectors are affected by it. Secondly, the rulings for compliance and violating behavior already exist in the proceedings, as well as the consequences in terms of fines for large infringements.

% \begin{itemize}
%     % \item A marked absence of evidence of the use of existing compliance frameworks by compliance specialists (lawyers, consultants, etc). This might be because most of the existing compliance frameworks require compliance officers to know a type of mathematical logic, which is far from their background, and whose training takes time. There is also little efforts in integrating views that are common for legal practitioners (laws) and formal models.
%     \item We lack a benchmark dataset that can be used for comparison among different compliance frameworks. Excepting processes in the banking sector, there is almost no process that has been repeated. Even the examples in the banking sector lack uniformity (the process has variants). Then it is difficult to replicate the same experiments and perform comparisons. We believe that collecting such cases and performing a challenge such as the ``Process Matching Challenge" or similar will be beneficial.
% \end{itemize}

%%% Local Variables:
%%% mode: latex
%%% TeX-master: "main"
%%% End:

\section{Threats to Validity}
\label{sec:threatsValidity}

The validity of this study can be affected by different aspects, we can categorize them into internal bias and threats to external validity:

\paragraph{Internal bias}

Given the broad spectrum of the primary studies considered, we foresee
a selection bias in the study. All the authors have expertise in formal methods, and primary studies reporting advances in this area were likely to be included in the results. We limit the bias by including all the studies that fulfilled the selection criteria, annotating in the data extraction forms contributions in both the theory and the practice of compliance. Authors were careful not to run into a conflict of interest with the results presented by applying the same criteria to their papers, and as a result, no primary study included the authors' works. 

% \subsubsection{Internal validity}

% The search process has been validated through the construction of a
% Quasi-Gold Standard \cite{zhang2011identifying}. In the selection of
% of sources in Section \ref{sources} we have paid attention to including
% diverse journals including publications in Computer Science,
% Information Systems, Business Processes and Law. The selection of the
% papers included in the QGS complies with the inclusion/exclusion
% criteria defined in this document and has been verified by more than
% one partner. To test the validity of the search we will use
% the recall measure:
% \[
% \textsl{Recall} = \frac{\text{Number of relevant studies in the QGS
%     found by automated search}}{\text{Number of papers in the QGS}}
% \times 100\%
%   \]

% In the construction of the pilot, we have extracted 2666 primary
% studies. Out of our QGS dataset, we have obtained 100\% Recall,
% validating the search query. 

\paragraph{Threats to external validity}
The primary search method selected in this SLR is automatic search,
including explicit terms such as ``formal method'', ``formal model'',
formalism, or ``formal language''. This decision limits the search
space, potentially excluding studies that belong to formal
verification but make not explicit use of these words. This exclusion
may affect the generalizability of our results.  A second aspect is the time window selected. This study collected data published between 1986 -- 2017, and no further updates have been included.  

% A study would be considered
% high-quality if the reviewer has deemed the study to
% have strong internal validity based on the researchers’
% trial quality criteria listed in their coding protocol. In
% Millar et al. (2006), for example, greater weight was
% accorded to studies that established experimental
% control and allowed a reliable investigation of the
% cause-effect relationship between AAC intervention
% and speech production.

%%% Local Variables:
%%% mode: latex
%%% TeX-master: "main"
%%% End:

\section{Related Work}
\label{sec:relatedWork}

Before conducting our SLR we performed a general query\footnote{our query
  ``(compliance or conformance or accordance or coherence or   enforcement) 
and (business process* or workflow* or ``case management'') and (law* or legislation* or policy or regulation*) and (verification or checking or certification or validation)''} in a
general search engine (c.f.: Google Scholar). From the output, no secondary study was identified. The QGS dataset included two secondary studies~\cite{ly_compliance_2015,el_kharbili_business_2012}. In \cite{el_kharbili_business_2012} the selection strategy used in this literature review was not as exhaustive as the one applied in SLRs, but a very focused strategy that required prior knowledge about which primary studies to
assess. The work in \cite{ly_compliance_2015} restricts its scope to consider only works in compliance monitoring of business processes. 

During the screening process, an additional set of secondary studies
was identified. We proceed to comment on them: 
Sackmann et al. \cite{sackmann_classification_2008} propose a classification scheme for existing approaches that address different aspects of automating compliance, along with criteria for a policy language to enable the automation of business process compliance within IT systems. Pourmirza et al. \cite{pourmirza_systematic_2017} add to this body of work, while Shamsaei et al. \cite{shamsaei_systematic_2011} conducted a systematic literature review of goal-oriented compliance management, using Key Performance Indicators (KPIs) to measure organizational compliance levels, and found that few tools effectively represent compliance results via dashboards or similar mechanisms. Ghanavati et al. \cite{ghanavati_systematic_2011} also performed a systematic review, focusing on goal-oriented legal compliance in business processes, with contributions centering on the extraction of legal requirements, modeling with goal languages, and integrating them into business processes. Hashmi et al. \cite{hashmi_are_2018} reviewed the state of the art in compliance management, selecting 79 papers from 183 and identifying three key categories: design-time, run-time, and auditing. Their review also introduced evidence of additional strategic phases in compliance management. Ly et al. \cite{ly_compliance_2015} reviewed the state of the art in compliance monitoring, presenting a framework for benchmarking compliance monitoring approaches in Business Process Management (BPM). While their scope focused on monitoring, they defined core functionalities such as support for compliance rules involving time and data constraints, multiple instances, and compliance explanations, which align with some of the most requested features. Becker et al. \cite{becker_generalizability_2012} examined the generalizability of compliance-checking approaches, comparing 26 compliance and model-checking approaches. Their review discussed the complexity of compliance patterns and highlighted the relevance of model-checking techniques, particularly BPMN, with an emphasis on design techniques rather than the entire compliance lifecycle. Casanovas et al. \cite{casanovas_legal_2017} introduced a survey on Compliance by and through Design, identifying limitations in existing frameworks and suggesting that the selection of formal languages should be context-dependent, considering the nature, complexity, and source of compliance requirements. They also noted that compliance checking might not be sufficient, as legal compliance extends beyond adhering to textual requirements, involving conceptual models and a policy-driven knowledge acquisition process. Hashmi et al. \cite{hashmi_norms_2017} conducted a review on the expressiveness of compliance frameworks, comparing the language expressiveness of regulatory languages against a taxonomy of normative requirements, laying the groundwork for benchmarking the 21 regulatory languages identified in their study. Fellmann et al. \cite{fellmann_state---art_2014} presented a mapping study that characterized compliance frameworks up to 2013, offering an alternative method to Hashmi et al. for comparing compliance languages. Lastly, Zaguir et al. \cite{zaguir_challenges_2024} identified challenges and enablers for GDPR compliance, particularly in the context of IoT and blockchain technologies. Their review highlighted that formalizing rights and obligations, as well as transparency in implementation, remain key challenges in regulatory compliance frameworks based on formal methods.

\section{Conclusions}
\label{sec:conclusions}

In this Systematic Literature Review, we captured the components of regulatory compliance frameworks based on methods. Our review identified 21 types of regulatory languages, 18 process representations, twenty types of processes, and compliance rules stemming from legal codes, acts, guidelines, and bylaws. Moreover, it highlighted a maturity in the use of verification technologies for compliance at the level of design of business processes and opportunities for further research in the implementation and enactment phases. We also identified multiple venues for research that currently impede the adoption of compliance frameworks, including extraction of formal specifications, process refinements, the alignment of terminology, and runtime monitoring. Overall, this work sheds light on some overarching challenges of compliance frameworks: the absence of empirical evaluations, the lack of interdisciplinary approaches, and the absence of studies considering flexibility when laws change. We hope that the results of this mapping study contribute to important research aspects that are necessary for the successful adoption of formal methods in legal and compliance reasoning.

\paragraph*{Acknowledgments:} We would like to thank Søren Debois for his valuable input during earlier versions of this work. This work was supported by the research grant ``Center for Digital CompliancE (DICE)'' (VIL57420) from VILLUM FONDEN.

%%% Local Variables:
%%% mode: latex
%%% TeX-master: "main"
%%% End:

% \begin{sloppypar}
%   \paragraph{Acknowledgments.} This work is supported by the EcoKnow
%   project (XXX)
% \end{sloppypar}

\bibliography{IncludedItems,QGS,SecondaryStudies,biblio,case-studies,Languages}
\bibliographystyle{plain}

\appendix
\newpage

% \input{Appendixes/ManualPilotExtraction}
% \section{Initial pilot extraction results} \label{appendix:pilot}
% \includepdf[scale=0.8,pages={1-3},pagecommand=\section{Initial pilot
%   extraction results} \label{appendix:pilot}
% \vspace{8cm}]{Appendixes/ManualPilotExtraction.pdf}
\onecolumn 

% \section{How to answer the questions}
% This section is a guide on how to get the data required for answering the questions:

% \hugo{The following subsection explains how to navigate the data and get each of the primary studies, the subsection will be removed once the results are plotted}

% \begin{itemize}
%     \item The data containing the included items, and how each research question is answered per the groups of articles is located in the shared Google Drive folder. The file name is "Included Items.xlm" and is directly available from \url{https://drive.google.com/open?id=1iIn2-h9fV7TmfpPVTe9ARFB3MWbbTU4_}. Each article has a study identifier.
%     \item The data sheets per each of the articles are located in a separate file: "Aggregated Studies.xlsx". It is available directly from the following link \url{https://drive.google.com/open?id=1e_8KylJiwfjKxivLZAnYQPYbRHx5w6UP}. Each page has the name of the study identified and referenced in Included Items. 
%     \item To get access to the physical copy of the paper to analyze, you can use the data in aggregated studies. Normally the title of the paper is enough. You can copy and paste that in the Zotero search functionality to get a highlighted copy of the paper in question. 
% \end{itemize}

\includepdf[pages=1,pagecommand=\section{Initial pilot extraction results} \label{appendix:pilot}, offset=0 -3cm]{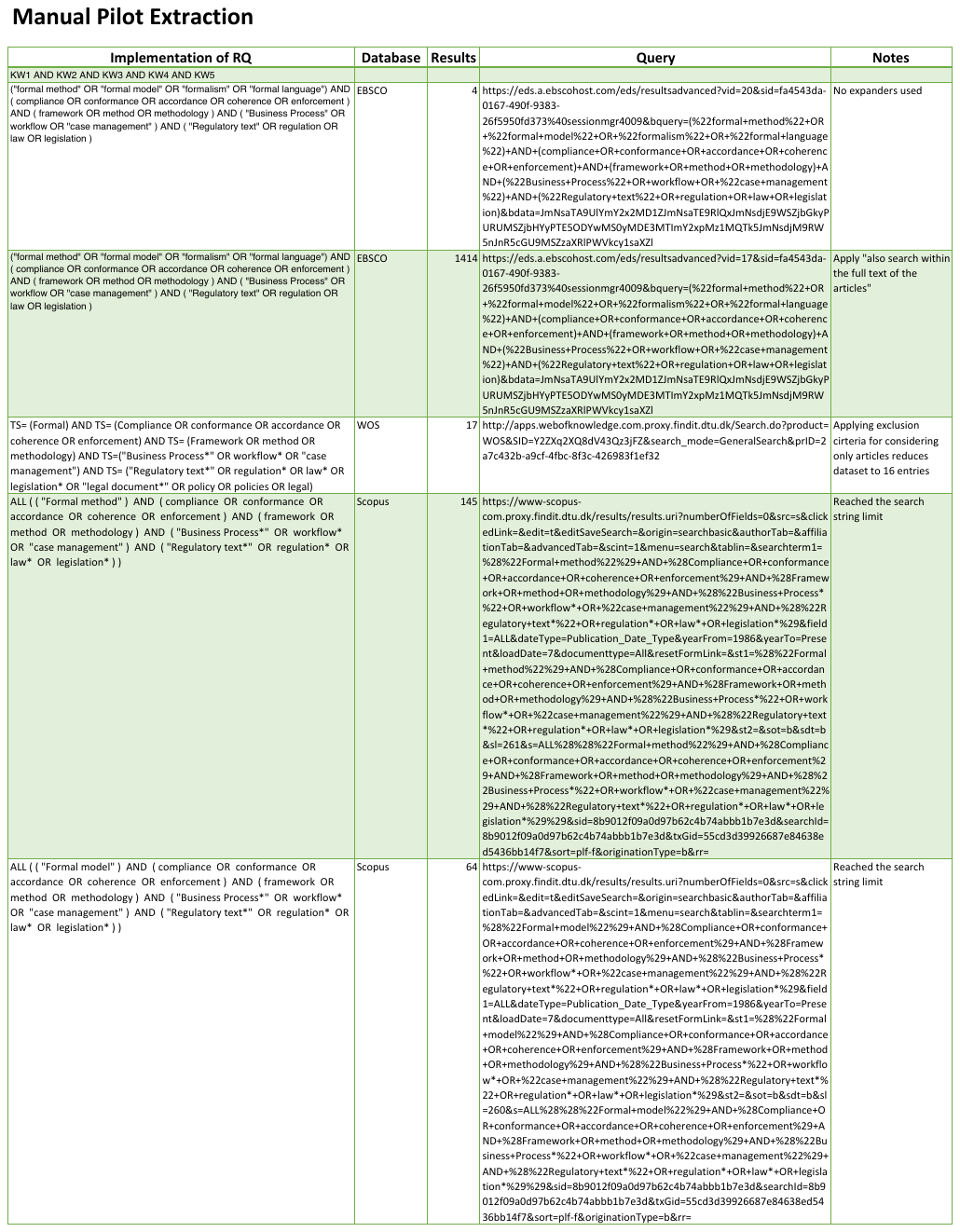}
\includepdf[pages=2-,pagecommand={}, offset=0 -3cm]{Appendixes/ManualPilotExtraction.pdf}

\section{Results from the Quasi-Gold-Standard Dataset}
\label{Appendix:QGS}

Figure \ref{Table:QGS} lists the results of the QGS dataset.

\begin{table}[t]
    \centering \scriptsize
    \begin{tabular}{|lm{7cm}|} \hline
\rowcolor[rgb]{ .867,  .922,  .969} QGS1&                                          \bibentry{liu_static_2007}\\
QGS2&      \bibentry{awad_iterative_2012}\\
\rowcolor[rgb]{ .867,  .922,  .969} QGS3&      \bibentry{lohmann_compliance_2013}\\
QGS4&      \bibentry{letia_compliance_2013}\\
\rowcolor[rgb]{ .867,  .922,  .969} QGS5&      \bibentry{governatori_detecting_2008}\\
QGS6&      \bibentry{elgammal_formalizing_2016}\\
\rowcolor[rgb]{ .867,  .922,  .969} QGS7&      \bibentry{wang_how_2014}\\
QGS8&      \bibentry{governatori_law_2010}\\
\rowcolor[rgb]{ .867,  .922,  .969} QGS9&      \bibentry{sadiq_modeling_2007}\\
QGS10&      \bibentry{hashmi_norms_2017}\\
\rowcolor[rgb]{ .867,  .922,  .969} QGS11&      \bibentry{ly_enabling_2012}\\
QGS12&      \bibentry{jiang_regulatory_2015}\\
\rowcolor[rgb]{ .867,  .922,  .969} QGS13&      \bibentry{governatori_journey_2009}\\
QGS14&      \bibentry{neskovic_using_2011}\\
\rowcolor[rgb]{ .867,  .922,  .969} QGS15&      \bibentry{awad_visually_2011}
    \end{tabular}
    \caption{Quality Gold Standard: initial pilot of primary studies} \label{Table:QGS}
  \end{table}

% \input{examples}
% \newpage
% \input{proofs}

\end{document}